\documentclass[12pt]{article}
\usepackage{graphicx}
\usepackage{latexsym,amsmath,amsfonts,amssymb}
\newcommand{\be}{\begin{equation}}
\newcommand{\ee}{\end{equation}}
\newcommand{\beq}{\begin{equation}}
\newcommand{\eeq}{\end{equation}}
\newcommand{\bea}{\begin{eqnarray}}
\newcommand{\eea}{\end{eqnarray}}

\newcommand{\ba}{\begin{eqnarray}}
\newcommand{\ea}{\end{eqnarray}}
\begin{document}
\baselineskip=15.5pt
\pagestyle{plain}
\setcounter{page}{1}
%--------+---------+---------+---------+---------+---------+---------+
%Body

% Ofer's definitions

\def\del{{\partial}}
\def\vev#1{\left\langle #1 \right\rangle}
\def\cn{{\cal N}}
\def\co{{\cal O}}
%\newfont{\Bbb}{msbm10 scaled 1200}     %instead of eusb10
%\newcommand{\mathbb}[1]{\mbox{\Bbb #1}}
\def\IC{{\mathbb C}}
\def\IR{{\mathbb R}}
\def\IZ{{\mathbb Z}}
\def\RP{{\bf RP}}
\def\CP{{\bf CP}}
\def\Poincare{{Poincar\'e }}
\def\tr{{\rm tr}}
\def\tp{{\tilde \Phi}}

\newcommand{\bA}{{\bf A }}
\newcommand{\bN}{{\bf N }}

\def\TL{\hfil $\displaystyle{##}$ }
\def\TR{$\displaystyle{{}##}$ \hfil}
\def\TC{\hfil $\displaystyle{##}$\hfil}
\def\TT{\hbox{##}}
\def\HLINE{\noalign{\vskip1\jot}\hline\noalign{\vskip1\jot}}
\def\seqalign#1#2{\vcenter{\openup1\jot
  \halign{\strut #1\cr #2 \cr}}}
\def\lbldef#1#2{\expandafter\gdef\csname #1\endcsname {#2}}
\def\eqn#1#2{\lbldef{#1}{(\ref{#1})}%
\begin{equation} #2 \label{#1} \end{equation}}
\def\eqalign#1{\vcenter{\openup1\jot
    \halign{\strut\span\TL & \span\TR\cr #1 \cr
   }}}
\def\eno#1{(\ref{#1})}
\def\href#1#2{#2}
\def\half{{1 \over 2}}

%-------------------Ioannis macros-----------------------------------
%%%%%%%%%%%%%%%%%%%%%%%%%%%%%%%%%%%%%%%%%%%%%%%%%%%%%%%%%%%%%%%%%%%%%%%%%%%%%
%                  CUSTOM DEFINITIONS
%%%%%%%%%%%%%%%%%%%%%%%%%%%%%%%%%%%%%%%%%%%%%%%%%%%%%%%%%%%%%%%%%%%%%%%%%%%%%

% Shortcuts

\def\NO{\nonumber}

\def\bea{\begin{eqnarray}}
\def\eea{\end{eqnarray}}

\def\beqx{\begin{displaymath}}
\def\eeqx{\end{displaymath}}

\newcommand{\bmat}{\left(\begin{array}}
\newcommand{\emat}{\end{array}\right)}

\def\half{\frac{1}{2}}

% New environments

\newtheorem{definition}{Definition}[section]
\newtheorem{theorem}{Theorem}[section]
\newtheorem{lemma}{Lemma}[section]
\newtheorem{corollary}{Corollary}[section]

% Abbreviations for Greek letters

\def\a{\alpha}
\def\b{\beta}
\def\c{\chi}
\def\d{\delta}
\def\e{\epsilon}
\def\f{\phi}
\def\g{\gamma}
\def\h{\eta}
\def\i{\iota}
\def\j{\psi}
\def\k{\kappa}
\def\l{\lambda}
\def\m{\mu}
\def\n{\nu}
\def\o{\omega}
    \def\om{\omega}
\def\p{\pi}
\def\q{\theta}
    \def\th{\theta}
\def\r{\rho}
\def\s{\sigma}
\def\t{\tau}
\def\x{\xi}
\def\z{\zeta}
\def\D{\Delta}
\def\F{\Phi}
\def\G{\Gamma}
\def\J{\Psi}
\def\L{\Lambda}
\def\O{\Omega}
    \def\Om{\Omega}
\def\P{\Pi}
\def\Q{\Theta}
    \def\Th{\Theta}
\def\S{\Sigma}
\def\U{\Upsilon}
\def\X{\Xi}

% Varletters

\def\ve{\varepsilon}
\def\vr{\varrho}
\def\vs{\varsigma}
\def\vq{\vartheta}
    \def\vth{\vartheta}
\def\tvf{\tilde{\varphi}}
\def\vf{\varphi}
    \def\vphi{\varphi}

% Calligraphic letters

\def\ca{{\cal A}}
\def\cb{{\cal B}}
\def\cc{{\cal C}}
\def\cd{{\cal D}}
\def\ce{{\cal E}}
\def\cf{{\cal F}}
\def\cg{{\cal G}}
\def\ch{{\cal H}}
\def\ci{{\cal I}}
\def\cj{{\cal J}}
\def\ck{{\cal K}}
\def\cl{{\cal L}}
\def\cm{{\cal M}}
\def\cn{{\cal N}}
\def\co{{\cal O}}
\def\cp{{\cal P}}
\def\cq{{\cal Q}}
\def\car{{\cal R}}
\def\cs{{\cal S}}
\def\ct{{\cal T}}
\def\cu{{\cal U}}
\def\cv{{\cal V}}
\def\cw{{\cal W}}
\def\cx{{\cal X}}
\def\cy{{\cal Y}}
\def\cz{{\cal Z}}

% Accents and foreign (in text):

% Fonts

\def\Sc#1{{\hbox{\sc #1}}}      % script for single characters in equations
\def\Sf#1{{\hbox{\sf #1}}}      % sans serif for single characters in equations
\def\mb#1{\mbox{\boldmath $#1$}}% bold math character

% Math symbols

\def\slpa{\slash{\pa}}                         % slashed partial derivative
\def\slin{\SLLash{\in}}                                 % slashed in-sign
\def\bo{{\raise-.3ex\hbox{\large$\Box$}}}               % D'Alembertian
\def\cbo{\Sc [}                                         % curly "
\def\pa{\partial}                                       % curly d
\def\de{\nabla}                                         % del
\def\dell{\nabla}                                       % hi ho the dairy-o
\def\su{\sum}                                           % summation
\def\pr{\prod}                                          % product
\def\iff{\leftrightarrow}                               % <-->
\def\conj{{\hbox{\large *}}}                            % complex conjugate
\def\ltap{\raisebox{-.4ex}{\rlap{$\sim$}} \raisebox{.4ex}{$<$}}   % < or ~
\def\gtap{\raisebox{-.4ex}{\rlap{$\sim$}} \raisebox{.4ex}{$>$}}   % > or ~
\def\face{{\raise.2ex\hbox{$\displaystyle \bigodot$}\mskip-2.2mu \llap {$\ddot
        \smile$}}}                                   % happy face
\def\dg{\dagger}                                     % hermitian conjugate
\def\ddg{\ddagger}                                   % double dagger
\def\trans{\mbox{\scri T}}                           % T for transposition
\def\>{\rangle}                                      %right angle
\def\<{\langle}                                      %left angle

% Math stuff with one argument

\def\sp#1{{}^{#1}}                                   % superscript (unaligned)
\def\sb#1{{}_{#1}}                                   % sub"
\newcommand{\sub}[1]{\phantom{}_{(#1)}\phantom{}}    % subscript in ( )
\newcommand{\supt}[1]{\phantom{}^{(#1)}\phantom{}}    % superscript in ( )
\def\oldsl#1{\rlap/#1}                               % poor slash
\def\slash#1{\rlap{\hbox{$\mskip 1 mu /$}}#1}        % good slash for lower case
\def\Slash#1{\rlap{\hbox{$\mskip 3 mu /$}}#1}        % " upper
\def\SLash#1{\rlap{\hbox{$\mskip 4.5 mu /$}}#1}      % " fat stuff (e.g., M)
\def\SLLash#1{\rlap{\hbox{$\mskip 6 mu /$}}#1}       % slash for no-in sign
\def\wt#1{\widetilde{#1}}                            % big tilde
\def\Hat#1{\widehat{#1}}                             % big hat
\def\lbar#1{\ensuremath{\overline{#1}}}              % big bar
\def\VEV#1{\left\langle #1\right\rangle}             % < >
\def\abs#1{\left| #1\right|}                         % | |
\def\leftrightarrowfill{$\mathsurround=0pt \mathord\leftarrow \mkern-6mu
        \cleaders\hbox{$\mkern-2mu \mathord- \mkern-2mu$}\hfill
        \mkern-6mu \mathord\rightarrow$}        % <--> double differential
\def\dvec#1{\vbox{\ialign{##\crcr
        \leftrightarrowfill\crcr\noalign{\kern-1pt\nointerlineskip}
        $\hfil\displaystyle{#1}\hfil$\crcr}}}           % <--> accent
\def\dt#1{{\buildrel {\hbox{\LARGE .}} \over {#1}}}     % dot-over for sp/sb
\def\dtt#1{{\buildrel \bullet \over {#1}}}              % alternate "
\def\der#1{{\pa \over \pa {#1}}}                        % partial derivative
\def\fder#1{{\d \over \d {#1}}}                         % functional derivative
\def\tr{{\rm tr \,}}                                    % trace
\def\Tr{{\rm Tr \,}}                                    % Trace
\def\diag{{\rm diag \,}}                                % diagonal
\def\preal{{\rm Re\,}}                                  % Real part
\def\pim{{\rm Im\,}}                                    % Imaginary part

% Math stuff with more than one argument

\def\partder#1#2{{\partial #1\over\partial #2}}        % partial derivative of
\def\parvar#1#2{{\d #1\over \d #2}}                    % variation of
\def\secder#1#2#3{{\partial^2 #1\over\partial #2 \partial #3}}  % second "
\def\on#1#2{\mathop{\null#2}\limits^{#1}}              % arbitrary accent
\def\bvec#1{\on\leftarrow{#1}}                         % backward vector accent
\def\oover#1{\on\circ{#1}}                             % circle accent

% Physics related

\def\Deq#1{\mbox{$D$=#1}}                               % Dimension
\def\Neq#1{\mbox{$cn$=#1}}                              % SUSY
\newcommand{\ampl}[2]{{\cal M}\left( #1 \to #2 \right)} % Scattering amplitude

% Abbreviations for journals

\def\NPB#1#2#3{Nucl. Phys. B {\bf #1} (19#2) #3}
\def\PLB#1#2#3{Phys. Lett. B {\bf #1} (19#2) #3}
\def\PLBold#1#2#3{Phys. Lett. {\bf #1}B (19#2) #3}
\def\PRD#1#2#3{Phys. Rev. D {\bf #1} (19#2) #3}
\def\PRL#1#2#3{Phys. Rev. Lett. {\bf #1} (19#2) #3}
\def\PRT#1#2#3{Phys. Rep. {\bf #1} C (19#2) #3}
\def\MODA#1#2#3{Mod. Phys. Lett.  {\bf #1} (19#2) #3}

% Miscellaneous

\def\norder{\raisebox{-.13cm}{\ensuremath{\circ}}\hspace{-.174cm}\raisebox{.13cm}{\ensuremath{\circ}}}
\def\bz{\bar{z}}
\def\bw{\bar{w}}
\def\-{\hphantom{-}}
\newcommand{\dd}{\mbox{d}}
\newcommand{\scr}{\scriptscriptstyle}
\newcommand{\scri}{\scriptsize}
\def\rand#1{\marginpar{\tiny #1}}               % Randbemerkung
\newcommand{\rstar}{\rand{\bf\large *}}
\newcommand{\rup}{\rand{$\uparrow$}}
\newcommand{\rdown}{\rand{$\downarrow$}}

%%%%%%%%%%%%%%%%%%%%%%%%%%%%%%%%%%%%%%%%%%%%%%%%%%%%%%%%%%%%%%%%%%%%%%%%%%%%

%--------+---------+---------+---------+---------+---------+---------+
%Hirosi's macros:
\def\ads{{\it AdS}}
\def\adsp{{\it AdS}$_{p+2}$}
\def\cft{{\it CFT}}

\newcommand{\ber}{\begin{eqnarray}}
\newcommand{\eer}{\end{eqnarray}}

\newcommand{\beqar}{\begin{eqnarray}}
\newcommand{\cN}{{\cal N}}
\newcommand{\cO}{{\cal O}}
\newcommand{\cA}{{\cal A}}
\newcommand{\cT}{{\cal T}}
\newcommand{\cF}{{\cal F}}
\newcommand{\cC}{{\cal C}}
\newcommand{\cR}{{\cal R}}
\newcommand{\cW}{{\cal W}}
\newcommand{\eeqar}{\end{eqnarray}}
\newcommand{\tht}{\thteta}
\newcommand{\lm}{\lambda}\newcommand{\Lm}{\Lambda}
\newcommand{\eps}{\epsilon}

%--------+---------+---------+---------+---------+---------+---------+

\newcommand{\nonu}{\nonumber}
\newcommand{\oh}{\displaystyle{\frac{1}{2}}}
\newcommand{\dsl}
  {\kern.06em\hbox{\raise.15ex\hbox{$/$}\kern-.56em\hbox{$\partial$}}}
\newcommand{\id}{i\!\!\not\!\partial}
\newcommand{\as}{\not\!\! A}
\newcommand{\ps}{\not\! p}
\newcommand{\ks}{\not\! k}
\newcommand{\dv}{d^2x}
\newcommand{\Z}{{\cal Z}}
\newcommand{\N}{{\cal N}}
\newcommand{\Dsl}{\not\!\! D}
\newcommand{\Bsl}{\not\!\! B}
\newcommand{\Psl}{\not\!\! P}
\newcommand{\eeqarr}{\end{eqnarray}}
\newcommand{\ZZ}{{\rm \kern 0.275em Z \kern -0.92em Z}\;}
%--------------------------------Alfonso's definitions%%%%%%%%%%%%%

% DEFINITIONS
                                                                                                    
\def\del{{\delta^{\hbox{\sevenrm B}}}} \def\ex{{\hbox{\rm e}}}
\def\azb{A_{\bar z}} \def\az{A_z} \def\bzb{B_{\bar z}} \def\bz{B_z}
\def\czb{C_{\bar z}} \def\cz{C_z} \def\dzb{D_{\bar z}} \def\dz{D_z}
\def\im{{\hbox{\rm Im}}} \def\mod{{\hbox{\rm mod}}} \def\tr{{\hbox{\rm Tr}}}
\def\ch{{\hbox{\rm ch}}} \def\imp{{\hbox{\sevenrm Im}}}
\def\trp{{\hbox{\sevenrm Tr}}} \def\vol{{\hbox{\rm Vol}}}
\def\rl{\Lambda_{\hbox{\sevenrm R}}} \def\wl{\Lambda_{\hbox{\sevenrm W}}}
\def\fc{{\cal F}_{k+\cox}} \def\vev{vacuum expectation value}
\def\nodiv{\mid{\hbox{\hskip-7.8pt/}}}
\def\ie{{\em i.e.}}
\def\ie{\hbox{\it i.e.}}

\def\CC{{\mathchoice
{\rm C\mkern-8mu\vrule height1.45ex depth-.05ex
width.05em\mkern9mu\kern-.05em}
{\rm C\mkern-8mu\vrule height1.45ex depth-.05ex
width.05em\mkern9mu\kern-.05em}
{\rm C\mkern-8mu\vrule height1ex depth-.07ex
width.035em\mkern9mu\kern-.035em}
{\rm C\mkern-8mu\vrule height.65ex depth-.1ex
width.025em\mkern8mu\kern-.025em}}}
                                                                                                    
\def\RR{{\rm I\kern-1.6pt {\rm R}}}
\def\NN{{\rm I\!N}}
\def\ZZ{{\rm Z}\kern-3.8pt {\rm Z} \kern2pt}
\def\IB{\relax{\rm I\kern-.18em B}}
\def\ID{\relax{\rm I\kern-.18em D}}
\def\II{\relax{\rm I\kern-.18em I}}
\def\IP{\relax{\rm I\kern-.18em P}}
\newcommand{\CS}{{\scriptstyle {\rm CS}}}
\newcommand{\CSs}{{\scriptscriptstyle {\rm CS}}}
\newcommand{\rc}{\nonumber\\}
\newcommand{\bear}{\begin{eqnarray}}
\newcommand{\eear}{\end{eqnarray}}
\newcommand{\W}{{\cal W}}
\newcommand{\LL}{{\cal L}}
                                                                                                    
\def\mani{{\cal M}}
\def\calo{{\cal O}}
\def\calb{{\cal B}}
\def\calw{{\cal W}}
\def\calz{{\cal Z}}
\def\cald{{\cal D}}
\def\calc{{\cal C}}
\def\to{\rightarrow}
\def\ele{{\hbox{\sevenrm L}}}
\def\ere{{\hbox{\sevenrm R}}}
\def\zb{{\bar z}}
\def\wb{{\bar w}}
\def\nodiv{\mid{\hbox{\hskip-7.8pt/}}}
\def\menos{\hbox{\hskip-2.9pt}}
\def\dr{\dot R_}
\def\drr{\dot r_}
\def\ds{\dot s_}
\def\da{\dot A_}
\def\dga{\dot \gamma_}
\def\ga{\gamma_}
\def\dal{\dot\alpha_}
\def\al{\alpha_}
\def\cls{{closing}}
\def\vev{vacuum expectation value}
\def\tr{{\rm Tr}}
\def\to{\rightarrow}
\def\too{\longrightarrow}

% Umut likes:

\def\a{\alpha}
\def\b{\beta}
\def\c{\gamma}
\def\d{\delta}
\def\e{\epsilon}           % Also, \varepsilon
\def\f{\phi}               %      \varphi
\def\vf{\varphi}  \def\tvf{\tilde{\varphi}}
\def\vp{\varphi}
\def\g{\gamma}
\def\h{\eta}
\def\i{\iota}
\def\j{\psi}
\def\k{\kappa}                    % Also, \varkappa (see below)
\def\l{\lambda}
\def\m{\mu}
\def\n{\nu}
\def\o{\omega}  \def\w{\omega}
\def\q{\theta}  \def\th{\theta}                  %     \vartheta
\def\r{\rho}                                     %     \varrho
\def\s{\sigma}                                   %     \varsigma
\def\t{\tau}
\def\u{\upsilon}
\def\x{\xi}
\def\z{\zeta}
\def\pt{\tilde{\varphi}}
\def\tt{\tilde{\theta}}
\def\lab{\label}  
\def\6{\partial}
\def\wg{\wedge}
\def\atanh{{\rm arctanh}}
\def\bpsi{\bar{\psi}}
\def\bt{\bar{\theta}}
\def\bvf{\bar{\varphi}}

%
% FONTS
                                                                                                    
%\newfont{\headfont}{cmbx10 scaled 1440}
\newfont{\namefont}{cmr10}
%\newfont{\initialfont}{cmr10 scaled 1200}
\newfont{\addfont}{cmti7 scaled 1440}
\newfont{\boldmathfont}{cmbx10}
%\newfont{\figfont}{cmr7 scaled 1200}
\newfont{\headfontb}{cmbx10 scaled 1728}
\renewcommand{\theequation}{{\rm\thesection.\arabic{equation}}}
\begin{titlepage}

\begin{center} \Large \bf Comments on the String dual to $N=1$ SQCD

\end{center}

\vskip 0.3truein
\begin{center}
Carlos Hoyos \footnote{c.h.badajoz@swansea.ac.uk}, 
Carlos 
N\'u\~nez \footnote{c.nunez@swansea.ac.uk} and 
Ioannis 
Papadimitriou\footnote{i.papadimitriou@swansea.ac.uk}
\vspace{0.7in}\\
\vskip 0.2truein
 \it{Department of Physics\\ University of Swansea, Singleton 
Park\\
Swansea SA2 8PP\\ United Kingdom.}
\vspace{0.3in}
\end{center}
\vskip 1cm
%\begin{center}
\centerline{{\bf Abstract}}
We study the String dual to $N=1$ SQCD deformed by a quartic 
superpotential in the quark superfields.
We present a unified view of the previous results in the literature and 
find new {\it exact} solutions and new asymptotic solutions. Then we study 
the Physics encoded in these backgrounds, giving among other things a 
resolution to an old puzzle related to the beta function and a sufficient 
criteria for screening. We also extend our results to the $SO(N_c)$ case 
where  we present a candidate for the Wilson loop in the spinorial 
representation. Various aspects of this line of research are critically 
analyzed.
%
%\end{center}

\vskip1truecm
\vspace{0.1in}
\end{titlepage}
\setcounter{footnote}{0}
\tableofcontents
%--------+---------+---------+---------+---------+---------+---------+
%Body

\section{Introduction}
\setcounter{equation}{0}
In the last decade, the AdS/CFT conjecture originally proposed by 
Maldacena \cite{Maldacena:1997re}
and refined in \cite{Gubser:1998bc, Witten:1998qj} have shown to be one of 
the most 
powerful analytic tools to study strong coupling effects in gauge 
theories. 
The project of extending the original duality to 
theories with a renormalization-group flow was initiated in 
\cite{Itzhaki:1998dd} and many different lines of research were proposed 
to compute non-perturbative effects in quantum field theories with small 
amount of Supersymmetry (SUSY).

In this paper we will focus on the type of set-ups called 
``wrapped branes models''. The first example of these models was presented 
by Witten in \cite{Witten:1998zw} and consists of a set of $N_c$ D4 branes 
that wrap a circle with SUSY breaking periodicity conditions, giving 
at low energies an effective theory with a massless vector field. This 
String background may be thought as  capturing  strongly coupled 
aspects of a version of Yang-Mills theory  completed at slightly higher 
energies 
by a 
set of extra states. This type of ideas have been applied to a variety 
of models, preserving different amounts of SUSY. For example in 
\cite{Gauntlett:2001ps} a String background dual to an $N=2$ Super 
Yang-Mills theory was given. In the paper \cite{Maldacena:2000yy}, the 
dual to a version 
of $N=1$ Super Yang-Mills was presented 
\footnote{The solution in \cite{Maldacena:2000yy}, 
 is the one found in a 4-d gauged Supergravity context by the authors of 
\cite{Chamseddine:1997nm}. There are 
actually a whole family of solutions dual to N=1 SYM with an UV 
completion. See Section 8 of \cite{Casero:2006pt}.}.

The models described above involve a (large) number $N_c$ of ``color 
branes'' usually wrapping calibrated cycles inside CY folds.
In this paper we will focus on duals to field theories encoding the 
dynamics of adjoint fields in interaction with
fundamental matter. Fundamental fields (quarks) are added, following the 
results of the paper \cite{Karch:2002sh}, 
with ``flavor branes''. These  are  branes  sharing the 
Minkowski directions with the ``color branes'' and extended over 
non-compact calibrated manifolds inside the CY fold. For the case of 
wrapped brane set-ups dual to a version of $N=1$ SYM theory 
\cite{Maldacena:2000yy}, the addition 
of flavors was first considered in the limit 
$\frac{N_f}{N_c}\to 0$ (similar  to the quenched 
approximation in the Lattice) in \cite{Nunez:2003cf}.
In the papers \cite{Casero:2006pt} and \cite{Casero:2007jj} 
the full dynamics of fundamentals was taken into account, by working in 
the so called Veneziano scaling, that is considering 
$x=\frac{N_f}{N_c}$ fixed, in the large $N_c$ limit. The field theory dual 
to the backgrounds in  
\cite{Casero:2006pt} and \cite{Casero:2007jj} is a version of $N=1$ SQCD 
on which we will elaborate below.

The rest of this paper is organized as follows: In Section \ref{stft}, we 
will review the field theory and the String dual(s) presented 
in \cite{Casero:2006pt} and \cite{Casero:2007jj}.
Then, in Section \ref{generalbackground} we will present a unified 
treatment of the 
different string duals. This will allow us to systematically classify and find
new solutions (done in Section \ref{newsolutions}). We will then dedicate 
Section 
\ref{Physics} to the study of 
field theory aspects that can be read from the string dual; most notably 
we will 
provide a sufficient condition for screening of the Wilson and other loops, 
solve an old puzzle related to the beta function computation and study 
domain walls, Wilson, 't Hooft and dyon loops. In Section \ref{sons}, we 
will 
present a dual version to the field theory described in Section \ref{stft} 
for the 
case of orthogonal gauge group $SO(N_c)$ and present an object that can be 
associated with the Wilson loop in the spinor representation. Finally, in 
Section \ref{gcc} we 
close with general criticism to this line of research and some 
conclusions. Various Appendixes with technical details complement our 
presentation, trying to 
make the main part of the paper more readable.

\section{Comments on the Field Theory and String dual}\label{stft}
\subsection{Field Theory aspects}\label{qft}
Let us start by commenting on the field theory at weak coupling. Before 
the 
addition of fundamental fields, this is a four 
dimensional theory preserving $N=1$ SUSY obtained via a twisted 
compactification of six dimensional Super Yang-Mills with sixteen 
supercharges. It is precisely the twisting in the compactification that 
preserves only four supercharges. The weakly coupled massless spectrum and 
multiplicities were 
studied in detail in \cite{Andrews:2006aw}. The theory contains a massless 
vector 
multiplet plus a tower of massive chiral and massive vector multiplets, 
usually called ``Kaluza-Klein (KK) modes''. 
The massive chiral superfields, 
denoted below by $\Phi_k$ have  masses (in 
units of the size of the inverse $S^2$ radius) given by $M_{\Phi_k }^2
= k(k+1)$ and degeneracy $g=(2k+1)$. The massive 
vector superfields, denoted below by $V_k$, have masses $M_{V_k}^2= k^2$ 
and degeneracy $g=4k$. 

Generically, the Lagrangian describing 
the weakly coupled theory (see \cite{Andrews:2006aw} for 
the quadratic part of the Lagrangian),
consists of a massless vector multiplet plus an infinite set of KK 
multiplets. 
Denoting the massless vector multiplet and its curvature by 
$(V,\W_\alpha)$, the massive 
vector multiplets by $V_k$ (its curvature by $W_k$) and massive chiral 
multiplets as $\Phi_k$, 
the action is  
\beq
S=\int d^4\theta \sum_k \Phi_k^\dagger e^{V}\Phi_k + \m_k |V_k|^2 
+ \int d^2\theta \Big[ \W_\alpha \W^\alpha + 
\sum_k W_k W_k + \mu_k |\Phi_k|^2 
+ {\cal 
W}(\Phi_k, V_k)\Big].
\label{unflav}
\eeq
On the basis of renormalizability, we will propose that the 
superpotential is at most cubic in the chirals, and we also expect that 
all the 
KK modes interact among 
themselves and that chirals interact with massive vectors
via the kinetic term
\beq
{\cal W}=  \sum_{ijk}z_{ijk} \Phi_i \Phi_j \Phi_k + 
\sum_{k}\hat{f}(\Phi_k)W_{k,\alpha} W_{k}^\alpha.
\eeq
Now, suppose we want to couple this field theory to fundamental matter.

\underline{{\it{Addition of Flavors}}}: 
We now introduce flavors as a pair of chiral multiplets $Q,\bar{Q}$
transforming in the fundamental of $SU(N_c)$. The action will have the usual kinetic 
terms plus interactions between quarks and KK modes of the form
\beq
S_{Q,\bar{Q}}=\int d^4\theta \left(\bar{Q}^\dagger e^{V}\bar{Q} +Q^\dagger 
e^{-V}Q\right) + \int d^2\theta \sum_{p,i,j,a,b} \kappa^{ij}_p 
\bar{Q}^{a,i}\Phi^{ab}_p 
Q^{b,j}.
\label{oooo}
\eeq
Here, $a,b=1,....,N_c$ are indexes in the fundamental and anti-fundamental 
of $SU(N_c)$, while $i,j=1,....,N_f$ belong to the fundamental and 
anti-fundamental of $SU(N_f)$.
Notice that the interaction between the quark superfields and the KK 
fields is such that the $SU(N_f)_V$ global symmetry is broken to 
$U(1)^{N_f}$. This reflects in the String solution via 
a smearing procedure that will be applied. With these interaction terms, 
it is possible to integrate out most of the KK modes at any given range of 
energy, although some may be light and should be kept in the action.

We can look for a configuration along the lines of what was proposed in 
the paper
\cite{Casero:2007jj}, that is, to search for a solution to the F term 
equations 
that is such that the cubic potential has no contribution. After 
integrating out the massive KK fields we find a low energy effective 
action of the form
\beq
S= S_{N=1\;SQCD} + W_{eff},
\label{ourtheory}
\eeq
with a superpotential given by (the color indexes of the quarks are 
contracted and suppressed)
\beq
W_{eff}\sim 
-\sum_{p,i,j}\frac{\kappa_{(p),ij}^2}{2\mu_{p}^2}(\bar{Q_i}Q_j)^2
\sim \frac{\kappa^2}{2\mu}M^2.
\label{weff}
\eeq
Let us now explain how this particular field theory, realized in a 
particular vacuum, is encoded in a  String 
background.
\subsection{The String dual}
As it should be clear to the reader, the addition of flavors to a QFT 
using  a dual 
String background 
is achieved by supplementing the putative closed string background with 
open string degrees of freedom \cite{Karch:2002sh}. 

One may decide to neglect the effects of pair creation of 
fundamental fields. This is analogous to neglect the deformation that 
the flavor branes should produce on the closed string background mentioned 
above-for more discussion see Section \ref{gcc}.

On the contrary, if the full quantum dynamics of fundamentals is to be 
considered, it should be encoded in a String background whose equations 
of motion 
are derived from the action
\beq
S= S_{type\;IIB/A} + S_{branes}.
\label{accion}
\eeq
In this paper, we will follow the lead of \cite{Casero:2006pt}
where the closed string background was taken to be the one of 
\cite{Maldacena:2000yy}
supplemented by the dynamics of fundamentals. In the case at 
hand, the 
fundamentals are represented by a  set of $N_f$ D5 branes extended on a 
non-compact two-cycle of a CY3-fold. As mentioned above and explained in 
detail in \cite{Casero:2006pt} a smearing procedure is 
applied so that the eqs. of motion derived from (\ref{accion}) are 
ordinary differential eqs. From this perspective, this smearing is just a 
matter of technical convenience, for more discussion about the effects of 
the smearing, see Section \ref{gcc}.

To be concrete, we will choose coordinates $x^M=[t, \vec{x}_3,\rho, 
\theta,\varphi,\tilde{\theta},\tilde{\varphi},\psi]$. In the present 
situation, our eqs. of motion will describe two 
sets of branes: the $N_c$ D5 color branes that wrap a compact calibrated 
two cycle 
inside a CY3-fold and $N_f$ D5 flavor branes that extend along the same 
`Minkowski' directions as the color branes wrapping a non-compact two 
manifold inside the CY3-fold. 
Besides, these flavor branes are smeared
along the transversal four directions (hence avoiding dependencies on those 
four coordinates denoted by 
$\theta,\tilde{\theta},\varphi,\tilde{\varphi}$). 
The action from which the dynamics follows is (see \cite{Casero:2006pt} 
for a derivation)
\bea
& & S=\frac{1}{2\kappa_{(10)}^2}
\int d^{10}x \sqrt{-g} \left[R-\frac12 \partial_\mu \phi
\partial^\mu \phi-\frac{1}{12}e^{\phi}F_{(3)}^2\right]\nonumber\\
& &
+\frac{T_5 N_f}{4\pi^2} \left(
-\int_{{\cal M}_6} d^{10}x \sin\theta \sin \tilde 
\theta e^{\frac{\phi}{2}}\sqrt {-\hat g_{(6)}}
+ \int Vol({\cal Y}_4) \wedge C_{6}  \right).
\label{actioniib}
\eea
The authors of \cite{Casero:2006pt} proposed a configuration where the 
`topology' of the metric was basically $\mathbb{R}^{1,3}_{x^\mu}\times 
S^2_{\theta,\varphi}\times 
\mathbb{R}_{\rho}\times 
S^3_{\tilde{\theta},\tilde{\varphi},\psi}$ (there is a fibration between 
$S^2$ and $S^3$ preserving four supercharges), a
dilaton field $\phi(\rho)$ depending on the radial coordinate and a RR 
three 
form that indicates the presence of sources via the Bianchi identity. 

Two types of backgrounds have been proposed as possible solutions. In the 
paper 
\cite{Casero:2007jj}, these solutions were referred to as type {\bf A} and 
type {\bf N} 
backgrounds and we will adopt that notation here. We will set the 
constants $\alpha'=g_s=1$ in the following. Then,
$
(2\pi)^5 T_5=1$ and $2\k^2_{10}=(2\p)^7$.

\vskip.2in
\underline{{\it The type {\bf A} backgrounds}}:
are believed to faithfully reproduce non-perturbative physics if the 
relation $N_c\leq N_f$ is satisfied. The metric in the Einstein frame, the RR 
three form and the dilaton read
\bea
ds^2& =& e^{\frac{\phi(\rho)}{2}} \Big[ dx_{1,3}^2 + 4 Y(\rho) d\rho^2 +
H(\rho)
(d\theta^2 + \sin^2\theta d\varphi^2)
+G(\rho) (d\tilde\theta^2 +\sin^2\tilde\theta
d\tilde\varphi^2)\nonumber\\
& & + Y(\rho)
(d\psi +\cos\tilde\theta d\tilde\varphi + \cos\theta d\varphi)^2 \Big],
\nonumber\\
F_{(3)}&=&- \Big[\frac{N_c }{4} \sin\tilde\theta d\tilde\theta \wedge
d\tilde\varphi
+\frac{N_f - N_c}{4}
\sin\theta d\theta \wedge d\varphi \Big] \wedge ( d\psi + \cos
\tilde \theta d\tilde\varphi +\cos\theta d\varphi),\,\,\nonumber\\
\phi&=&\phi(\rho)\, ,
\label{metricA}
\eea
The functions $H(\rho)$, $G(\rho)$, $Y(\rho)$, $\phi(\rho)$ 
satisfy a
set of BPS equations that as usual solve all the Euler-Lagrange 
equations derived 
from (\ref{actioniib}). These first order non-linear equations read,
\ba
H'&=& \frac{1}{2} (N_c-N_f)  + 2 Y,
\label{newbps1} \\
G' &=& -\frac{N_c}{2} + 2Y,
\label{newbps2} \\
Y' &=&-\frac{1}{2}(N_f- N_c) \frac{Y}{H} -\frac{N_c}{2} \frac{Y}{G}
-2Y^2\left(\frac{1}{H}+\frac{1}{G}\right) +4Y,
\label{newbps3}  \\
\phi'&=&-\frac{ (N_c-N_f)}{4H}   + \frac{N_c}{4G}.
\label{newbps4}
\ea
In the paper \cite{Casero:2007jj} these equations where solved (partly 
analytically and partly numerically). The interesting finding was the 
existence of  three types of solutions having very different IR 
behavior 
\footnote{The word IR (infrared) actually refers to the IR of the dual 
QFT. On the string side we mean a solution for small values of the 
radial coordinate $\rho$. Here and in the rest of the paper we will 
adopt this terminology.}. 
They were called  respectively type I, type II, and type III. The three types of IR 
solutions connect smoothly with  solutions at very large 
values of $\rho$ (the UV of the dual QFT). The difference between the 
three types of UVs is in very suppressed terms $\co(e^{-4\rho})$, that can 
be interpreted as different VEVs for operators of dimension six in the 
dual QFT. See \cite{Casero:2007jj} for a clear explanation of this issue. 
For completeness, we will re-analyze these solutions in 
Section \ref{newsolutions}, also finding new behaviors.

\vskip.2in
\underline{{\it The type {\bf N} backgrounds}}: we know much less about 
these backgrounds. The only available solution was originally found 
(in part analytically and in part numerically) in \cite{Casero:2006pt}. The 
configuration of the dilaton, metric (in Einstein frame) and RR three form are 
given by
\ba
ds^2 &=& e^{ \frac{\phi(\rho)}{2}} \Big[dx_{1,3}^2 + e^{2k(\rho)}d\rho^2
+ e^{2 h(\rho)}
(d\theta^2 + \sin^2\theta d\varphi^2) +\nonumber\\
&+&\frac{e^{2 g(\rho)}}{4}
\left((\tilde{\omega}_1+a(\rho)d\theta)^2
+ (\tilde{\omega}_2-a(\rho)\sin\theta d\varphi)^2\right)
 + \frac{e^{2 k(\rho)}}{4}
(\tilde{\omega}_3 + \cos\theta d\varphi)^2\Big], \nonumber\\
F_{(3)} &=&\frac{N_c}{4}\Bigg[-(\tilde{\omega}_1+b(\rho) d\theta)\wedge
(\tilde{\omega}_2-b(\rho) \sin\theta d\varphi)\wedge
(\tilde{\omega}_3 + \cos\theta d\varphi)+\nonumber\\
& & b'd\rho \wedge (-d\theta \wedge \tilde{\omega}_1  +
\sin\theta d\varphi
\wedge
\tilde{\omega}_2) + (1-b(\rho)^2) \sin\theta d\theta\wedge d\varphi \wedge
\tilde{\omega}_3\Bigg]\nonumber\\
&-&\frac{N_f}{4}\sin\theta d\theta \wedge
d\varphi \wedge (d\psi +\cos\tilde{\theta} d\tilde{\varphi}),
\label{nonabmetric424}
\ea
where $\tilde\omega_i$ are the left-invariant forms of $SU(2)$
\bea\lab{su2}
&&\tilde{\omega}_1= \cos\psi d\tilde\theta\,+\,\sin\psi\sin\tilde\theta
d\tilde\varphi\,\,,\rc
&&\tilde{\omega}_2=-\sin\psi d\tilde\theta\,+\,\cos\psi\sin\tilde\theta
d\tilde\varphi\,\,,\rc
&&\tilde{\omega}_3=d\psi\,+\,\cos\tilde\theta d\tilde\varphi\,\,,
\eea
and the BPS equations defining the functions $h,g,k,a,b,\phi$ as presented 
in Appendix B of \cite{Casero:2006pt} and appear quite involved; we will 
not 
quote them here because in the following sections we will present them in a 
unified way with the type {\bf A} BPS equations written above 
eqs.(\ref{newbps1})-(\ref{newbps4}).

It was argued in \cite{Casero:2006pt} and \cite{Casero:2007jj} that the 
type {\bf N} backgrounds faithfully capture non-perturbative physics for 
any value of the number of flavors $N_f>0$. Also, it was proposed that 
backgrounds of type {\bf N} describe vacua of the theory where the gaugino 
condensate is 
non-zero, while type {\bf A} backgrounds describe vacua with 
$<\lambda\lambda>=0$. By the Konishi anomaly relation, this seems to 
indicate that the meson superfield should vanish in type {\bf A} 
backgrounds whilst being non-zero in type {\bf N} solutions.

Many checks or predictions about  aspects of the QFT in 
eq.~(\ref{ourtheory}) using the type 
{\bf A} and type {\bf N} solutions have been presented in 
\cite{Casero:2006pt,Casero:2007jj,varios}.

Let us then start with the main part of this paper, 
describing in a unified fashion the type {\bf A, N} backgrounds
mentioned above.
\section{Unified view of the type A and type N backgrounds}
\label{generalbackground}
\setcounter{equation}{0}
It is clear that the type {\bf A} configurations
are  a special case of  the type {\bf N} ones. Indeed, if the 
functions $a(\rho)=b(\rho)=0$, then there is no difference 
between the type {\bf A} 
and type {\bf N} backgrounds. Hence, the fact that the BPS equations for 
type 
{\bf A} solutions read like in eqs.~(\ref{newbps1})-(\ref{newbps4}) 
suggests that there 
should be some variables where the type {\bf N} first order eqs. will look 
similarly nice.

So, we start by rewriting the type {\bf N} ansatz (\ref{nonabmetric424}) using a more symmetric parametrization as 
\bea\label{metric}
ds^2&=&e^{\phi(\r)/2}\left\{\rule{.0in}{.2in}dx_{1,3}^2 +Y(\r)\left(4d\rho^2+(\om_3+\tilde{\om}_3)^2\right)
+\frac12P(\r)\sinh\t(\r)\left(\om_1\tilde{\om}_1-\om_2\tilde{\om}_2\right)\right.\\
&&\left.+\frac14\left(P(\r)\cosh\t(\r)+Q(\r)\right)(\om_1^2+\om_2^2)
+\frac14\left(P(\r)\cosh\t(\r)-Q(\r)\right)(\tilde{\omega}_1^2+\tilde{\omega}_2^2)\right\},\NO
\eea
\be
F\sub{3}=-d\left\{\s(\r)\left(\om_1\wedge\tilde{\om}_1-\om_2\wedge\tilde{\om}_2\right)\right\}
-\left(\frac{N_f-N_c}{4}\om_1\wedge\om_2+\frac{N_c}{4}\tilde{\om}_1\wedge\tilde{\om}_2\right)
\wedge\left(\om_3+\tilde{\om}_3\right),
\label{3-form}
\ee
where we have defined, in addition to the left invariant forms (\ref{su2}),
$
\om_1=d\th, \quad \om_2=\sin\th d\vf,\quad \om_3=\cos\th d\vf.
$
The relation between these new variables and the functions 
originally parametrizing the type {\bf N} backgrounds in 
eq.~(\ref{nonabmetric424}) is
\be
e^{2h}=\frac14\left(\frac{P^2-Q^2}{P\cosh\t-Q}\right),
\quad e^{2g}=P\cosh\t-Q,\quad e^{2k}=4Y,\quad
a=\frac{P\sinh\t}{P\cosh\t-Q},\quad b=\frac{\s}{N_c}.
\label{changevariables}
\ee
Evaluating the action (\ref{actioniib}) on  
the ansatz  (\ref{metric})-(\ref{3-form}) we obtain the one-dimensional 
effective action
\be
(2\p)^4 {\rm Vol}(\mathbb{R}^{1,3})^{-1}S\equiv S_{1d}=\int d\rho {\cal L},
\ee
where the 1-d effective Lagrangian is,
\be\label{1d-eff-action}
\cl= \frac{\F}{4\sqrt{Y}} 
\left(\left(\frac{\F'}{\F}\right)^2
 -\frac{1}{4}\left(\frac{Y'}{Y}\right)^2
-\frac12\left(\frac{P'+Q'}{P+Q}\right)^2-
\frac12\left(\frac{P'-Q'}{P-Q}\right)^2-\left(\frac{P^2\t'^2+\s'^2}{P^2-Q^2}\right)\right)-V,
\ee
we have defined $\F\equiv(P^2-Q^2)Y^{1/2}e^{2\f}$. 
The potential is given by
\bea\label{potential}
V&=&\frac{\F}{(P^2-Q^2)\sqrt{Y}}\left\{\frac{(8YP)^2+\left((2N_c-N_f)P\cosh\t+N_fQ-2\s P\sinh\t\right)^2}{2(P^2-Q^2)}
\right.\NO\\
&&\left.-4Y(4P\cosh\t-N_f+4Y)+P^2\sinh^2\t+N_c(N_f-N_c)+\s^2\rule{.0in}{.3in}\right\},
\eea

At this point a word of caution is in order. 
Namely, the solutions of the Euler-Lagrange
equations of the one dimensional effective action 
(\ref{1d-eff-action}) are {\em not}
necessarily solutions of the Type IIB plus branes 
equations of motion. This is because we
have obtained (\ref{1d-eff-action}) by inserting 
the ansatz in the supergravity action and not 
in the supergravity equations of motion. It 
turns out, however, that there is
a special class of solutions of the Euler-Lagrange equations 
of (\ref{1d-eff-action}) that satisfy
precisely the BPS equations of the supersymmetric type \bA and type \bN 
backgrounds. Since
it is known that these 
BPS equations imply the second order supergravity 
plus branes equations \cite{Casero:2006pt} (for a general proof see 
\cite{Koerber:2007hd}),
it follows that this  particular class of 
solutions of (\ref{1d-eff-action}) are actually solutions
of the  second order supergravity plus branes equations. 
The reason why we consider 
the one dimensional effective action 
(\ref{1d-eff-action}) at all is that it naturally leads
to a `superpotential' for the 
BPS equations of both  type \bA and type \bN backgrounds.

It will often be convenient 
to write $P$, $Q$, in terms of further new variables $H, G$ as
$
P=2(H+G),\; Q=2(H-G).
$
Note that for the type {\bf A} backgrounds, 
obtained by setting $\t=\s=0$, the variables $H, G$,
defined as above coincide with those in (\ref{metricA})-(\ref{newbps4}). 
 
From the effective action (\ref{1d-eff-action}) 
then we compute the canonical momenta
\bea
&&\p_H=\frac{\pa {\cal L} }{\pa H'}=-\frac14\F 
Y^{-1/2}\frac{H'}{H^2},\phantom{morem}
\p_G=\frac{\pa  {\cal L} }{\pa G'}=-\frac14\F Y^{-1/2}\frac{G'}{G^2},\NO\\
&&\p_Y=\frac{\pa {\cal L}  }{\pa Y'}=-\frac18\F Y^{-5/2}Y',\phantom{moremo}\;
\p_\F=\frac{\pa {\cal L} }{\pa \F'}=\frac12Y^{-1/2}\frac{\F'}{\F},\NO\\
&&\p_\t=\frac{\pa {\cal L} }{\pa \t'}=-\frac12\F Y^{-1/2}\frac{P^2\t'}{P^2-Q^2},\phantom{mo}\;
\p_\s=\frac{\pa {\cal L} }{\pa \s'}=-\frac12\F Y^{-1/2}\frac{\s'}{P^2-Q^2},
\label{momenta}
\eea
as well as the Hamiltonian
\bea\label{hamiltonian}
{\cal H}&=&H'\p_H+G'\p_G+Y'\p_Y+\F'\p_\F+\t'\p_\t+\s'\p_\s- {\cal L}\\
&=&Y^{1/2}\F^{-1}\left(\F^2\p_\f^2-2H^2\p_H^2-2G^2\p_G^2-4Y^2\p_Y^2-(P^2-Q^2)(P^{-2}\p_\t^2+\p_\s^2)\right)+V.\NO
\eea
Using Hamilton-Jacobi theory we can obtain the canonical 
momenta as derivatives of Hamilton's principal function, $\cs$, as
\be
\p_H=\frac{\pa\cs}{\pa H},\quad \p_G=\frac{\pa\cs}{\pa G},\quad \p_Y=\frac{\pa\cs}{\pa Y},\quad \p_\F=\frac{\pa\cs}{\pa \F},\quad
\p_\t=\frac{\pa\cs}{\pa \t},\quad \p_\s=\frac{\pa\cs}{\pa \s},
\label{momentos2}\ee
where $\cs$ is a solution of the Hamilton-Jacobi equation
\be
{\cal H}\left(H,G,Y,\F,\t,\s;\frac{\pa\cs}{\pa H},\frac{\pa\cs}{\pa 
G},\frac{\pa\cs}{\pa Y},\frac{\pa\cs}{\pa \F},\frac{\pa\cs}{\pa \t},\frac{\pa\cs}{\pa \s}\right)
=0.
\label{hamiltonjacobi}
\ee
A particular solution to this equation is,
\bea\label{superpotential}
\cs&=&\frac{\F}{16\sqrt{Y}}\left\{(8Y-N_f)\left(\frac{1}{H}+\frac{1}{G}\right)
+((2N_c-N_f)\cosh\t-2\s\sinh\t)\left(\frac{1}{H}-\frac{1}{G}\right)\right.\NO\\
&&\left.\phantom{moremore}+16\cosh\t\rule{.0in}{.25in}\right\}.
\eea
By equating the canonical momenta (\ref{momenta}) 
to the corresponding ones obtained from (\ref{momentos2}),
this incomplete integral (cf. superpotential) 
leads to a system of first order equations, whose
solutions {\em automatically} satisfy the second 
order equations following from the Lagrangian 
(\ref{1d-eff-action}) -- but, we stress again, {\em not} 
necessarily the Type IIB plus branes equations --
\bea\label{BPS-eqs}
&&H'=\frac14\left\{8Y-N_f+\left((2N_c-N_f)\cosh\t-2\s\sinh\t\right)\right\},\NO\\
&&G'=\frac14\left\{8Y-N_f-\left((2N_c-N_f)\cosh\t-2\s\sinh\t\right)\right\},\NO\\
&&Y'=\frac{Y}{4}\left\{-\left(8Y+N_f\right)\left(\frac{1}{H}+\frac{1}{G}\right)+\left((2N_c-N_f)\cosh\t
-2\s\sinh\t\right)\left(\frac{1}{H}-\frac{1}{G}\right)\right.\NO\\
&&\left.\phantom{moremore}+16\cosh\t\rule{.0in}{.25in}\right\},\NO\\
&&\F'=\frac18\F\left\{(8Y-N_f)\left(\frac{1}{H}+\frac{1}{G}\right)
+((2N_c-N_f)\cosh\t-2\s\sinh\t)\left(\frac{1}{H}-\frac{1}{G}\right)\right.\NO\\
&&\left.\phantom{moremore}+16\cosh\t\rule{.0in}{.25in}\right\},\NO\\
&&\t'=\frac{1}{2(H+G)^2}\left\{(H-G)\left((2N_c-N_f)\sinh\t-2\s\cosh\t\right)-16HG\sinh\t\right\},\NO\\
&&\s'=-4(H-G)\sinh\t.
\eea
Notice that for 
$\tau=\sigma=0$, this system of first order equations reduces to the BPS 
eqs. (\ref{newbps1})-(\ref{newbps4}), as claimed above. 
However, as they stand, (\ref{BPS-eqs}) are not quite
the same as the type {\bf N} BPS equations. 

Let us next introduce the function 
\be
\om\equiv \s-\tanh\t\left(Q+\frac{2N_c-N_f}{2}\right).
\ee
The  first order equations (\ref{BPS-eqs}) can 
then be rearranged in the form
\bea\label{first-order-rearranged}
&&P'=8Y-N_f,\NO\\
&&\pa_\r\left(\frac{Q}{\cosh\t}\right)=\frac{(2N_c-N_f)}{\cosh^2\t}-\frac{2\om}{P^2}(P^2-Q^2)\tanh\t,\NO\\
&&\pa_\r\log\left(\frac{\F}{\sqrt{P^2-Q^2}}\right)=2\cosh\t,\NO\\
&&\pa_\r\log\left(\frac{\F}{\sqrt{Y}}\right)=\frac{16YP}{P^2-Q^2},\NO\\
&&\t'+2\sinh\t=-\frac{2Q\cosh\t}{P^2}\om,\NO\\
&&\om'=\frac{2\om}{P^2\cosh\t}\left(P^2\sinh^2\t+Q\left(Q+\frac{2N_c-N_f}{2}\right)\right).
\eea
It is clear from this form of the first order equations 
that the algebraic constraint
\be\label{constraint}
\om=0 \Leftrightarrow  \s=\tanh\t\left(Q+\frac{2N_c-N_f}{2}\right),
\ee
is consistent with the equations and therefore 
defines a subclass of solutions, eliminating one integration constant. 
It can be shown that the first order equations 
(\ref{BPS-eqs}), {\em together} with the algebraic constraint $\om=0$, are 
equivalent to the BPS equations for the type {\bf N} 
backgrounds, derived in Appendix B of \cite{Casero:2006pt}. 
The subclass of solutions of eqs.~(\ref{BPS-eqs}) that satisfy the constraint (\ref{constraint}) then 
correspond to supersymmetric solutions. 
As stressed above, in the case $\sigma=\tau=0$, eqs. (\ref{BPS-eqs}) 
reduce to the type {\bf A} 
BPS eqs. (\ref{newbps1})-(\ref{newbps4}),
and so these solutions 
correspond to the supersymmetric type {\bf A} solutions. The constraint 
in eq.~(\ref{constraint}) is automatically satisfied in this case.

The form of eqs. 
(\ref{first-order-rearranged}) 
makes it manifest that imposing the constraint in eq.
(\ref{constraint}) leads to a dramatic simplification. 
Firstly, the equation for $\om$ is trivially satisfied, 
while the equation for $\t$ decouples and gives
\be\label{tau}
\sinh\t=\frac{1}{\sinh(2(\r-\r_o))}.
\ee
Given this, one can then integrate the equations for $Q$ and $\F$ to obtain
\be\label{Q}
Q=\left(Q_o+\frac{2N_c-N_f}{2}\right)\cosh\t+\frac{2N_c-N_f}{2}\left(2\r\cosh\t-1\right),
\ee
\be\label{dilaton}
e^{4(\f-\f_o)}=\frac{\sinh^2(2\r_o)}{(P^2-Q^2)Y\sinh^2\t}.
\ee
Note that both $Q$ and the dilaton are given {\em algebraically} in terms of the rest of the functions
parametrizing the backgrounds. 
Here $\r_o$, $Q_o$ and $\f_o$ are constants of 
integration and we have chosen the integration constant in (\ref{dilaton})
such that it admits a smooth 
limit as $\r_o\to -\infty$ (this limit gives $\t=\s=0$ and so corresponds to the type {\bf A} backgrounds).
Finally, $Y$ is determined in terms of $P$ as
\be\label{Y}
Y=\frac18(P'+N_f),
\ee
while the only remaining unknown, the function $P$, then satisfies the 
decoupled second order equation
\be\label{master}
P''+(P'+N_f)\left(\frac{P'+Q'+2N_f}{P-Q}+\frac{P'-Q'+2N_f}{P+Q}-4\cosh\t\right)=0.
\ee
We will refer to this equation as the `master' equation, since once we have a solution of (\ref{master}) all other functions 
are determined via (\ref{tau})-(\ref{Y}).

Some comments are due here. First, note that the number of 
integration constants is indeed
as expected. Namely, (\ref{BPS-eqs}) are 
six first order equations for six variables and
so we expect six integration constants. 
However, the algebraic constraint (\ref{constraint}) eliminates one
of the integration constants. The five integration constants 
are then $\r_o$, $Q_o$, $\f_o$ plus
the two integration constants coming from 
the second order equation (\ref{master}). As we shall discuss in Section 
\ref{Physics}, 
the integration constant $\r_0$ is related to the gaugino 
condensate VEV.
In particular, a finite value of $\r_o$ corresponds 
to the type {\bf N} backgrounds,
while the limit $\r_o\to -\infty$ gives the type {\bf A} backgrounds.
As we will state below, the type {\bf N}/{\bf A} backgrounds are dual to the field 
theory in vacua with/without gaugino bilinear VEV.
\section{New (and old) solutions}\label{newsolutions}
\setcounter{equation}{0}
In this section we will present solutions to the 
first order eqs.~(\ref{BPS-eqs}), which also satisfy the constraint
(\ref{constraint}). As we have just discussed, 
these correspond to the supersymmetric type \bA and type \bN solutions. We will 
write the solutions in terms of the 
new variables $P, Q, Y, \tau,\sigma$ and $e^{2\f}$  as 
given in eqs.~(\ref{changevariables}), with the values of 
$Q,Y, \tau,\sigma$ and $e^{2\f}$ being read from eqs. 
(\ref{tau})-(\ref{Y}) and
obtained after solving for $P$ in eq.~(\ref{master}). 
In order to facilitate easy comparison with the earlier 
literature, however, we will often write the solutions 
in the original variables too.

We will present first some solutions that were already known, just to 
convince the reader that our general treatment of Section 
\ref{generalbackground} is correct (and convenient). Then we will 
present a new exact solution (that we have found for the particular 
case of $N_f=2N_c$). Finally, we will present more general 
solutions that we know only as series expansions for large and small 
values of the radial coordinate, or perturbatively 
in certain integration constants. 
They will describe the UV and IR physics of the dual field theory.

To close the section, we will present a study 
of the first order eqs.~(\ref{BPS-eqs}) as a dynamical system. This will give 
further insight into the dual theory dynamics.

\subsection{Exact solutions that were already known}
\begin{flushleft}
\underline{Unflavored $N_f=0$ solutions:}\vskip.1in
\end{flushleft}

For $N_f=0$ an exact solution of 
(\ref{master}) is the well known unflavored 
solution \cite{Maldacena:2000yy,Chamseddine:1997nm}, which for
$\r_o>-\infty$ (that is type {\bf N} solution) takes the form
\be\label{unflavored}
P=2N_c(\r-\r_o),\quad Q_o=-N_c-2N_c\r_o,\quad \r\geq \r_o,
\ee
or equivalently
\bea\label{nf=0solu}
&&e^{2h}=\frac{N_c}{8 \sinh^2(2(\r-\r_0))}\Big[1-8(\r -\r_0)^2 - 
\cosh(4(\r-\r_0)) -4(\r-\r_0)\sinh(4 (\r-\r_0))   \Big] ,\NO\\
&&e^{2g}=e^{2k}=N_c,\;\;\; a(\r)=b(\r)= 
\frac{2(\rho-\rho_0)}{\sinh(2(\r -\r_0))},\NO\\
&&e^{-2\phi}={2e^{h}\over \sinh(2(\rho-\rho_0))}.
\eea

This is a smooth solution. On the other hand, for $\r_o\to-\infty$ (type 
{\bf A} solution), we have instead
\be
P=2 N_c(\r-\r_*),\quad \r_*\equiv 
-\frac{1}{2N_c}\left(Q_o+\frac{N_c}{2}\right),\quad \r\geq \r_*,
\ee
that gives a background like the one in eq.~(\ref{metricA}) with,
\be
H=N_c(\r-\r_*),\quad G=\frac{N_c}{4},\quad Y=
\frac{N_c}{4},\quad e^{4(\f-\f_o)}=\frac{e^{4\r}}{N_c^3(\r-\r_*)}.
\ee
This solution has a singularity at $\r=\r_*$.
\begin{flushleft}
\underline{The case $N_f=2N_c$:}\vskip.1in
\end{flushleft}

For $N_f=2N_c$ an exact type {\bf A} ($\r_o\to -\infty$) solution of 
(\ref{master}) is known \cite{Casero:2006pt,Casero:2007jj}
\footnote{Another exact solution valid for any 
$\r_o>-\infty$ is $P=-N_c\cosh\t,\quad Q=\pm \sqrt{3}N_c\cosh\t$, 
but it is clearly singular. However, it might be useful 
in finding a related non-singular solution.}.
It reads
\be\label{self-dual-1}
P=N_c+\sqrt{N_c^2+Q_o^2}, \quad Q=Q_o\equiv 
4N_c\frac{(2-\x)}{\x(4-\x)},\quad 0<\x<4,
\ee
or in terms of the background variables described in eq.~(\ref{metricA}) 
\be
H=\frac{N_c}{\x},\quad G=\frac{N_c}{4-\x},\quad Y=\frac{N_c}{4},\quad e^{4(\f-\f_o)}=\frac{\x(4-\x)e^{4\r}}{4N_c^3}.
\ee

\subsection{A new exact solution}
Here we present a new one-parameter family of type {\bf A} solutions for the case $N_f=2N_c$. This solution is a one-parameter
deformation of (\ref{self-dual-1}) for $\x=4/3$ 
or $\x=8/3$, the two values of $\x$ being interchanged under Seiberg duality (Seiberg duality in these 
backgrounds corresponds to the transformation $P\to P,\; Q \to -Q,\; \tau \to \tau,\; Y \to Y,\;N_c \to N_f-N_c$. See
(\ref{seibergr}) below).

The new solution of eq.~(\ref{master}), 
with $\r_o\to-\infty$ (type {\bf A}), takes the form
\be\label{self-dual-2}
P=\frac{9 N_c}{4}+c_+ e^{4\r/3}, 
\quad c_+>0,\quad Q=\pm\frac{3N_c}{4},
\ee
or, in terms of the functions in 
the background described in eq.~(\ref{metricA}),
\bea\label{newexactnf=2nc}
H=\frac{N_c}{16}(9\pm3)+
\frac{c_+}{4}e^{4\r/3}, \quad G=\frac{N_c}{16}(9\mp3)
+\frac{c_+}{4}e^{4\r/3}, 
\quad Y=\frac{N_c}{4}+\frac{c_+}{6}e^{4\r/3},\NO\\
\quad e^{4(\f-\f_o)}=\frac{6e^{4\r}}{\left(3N_c+c_+e^{4\r/3}\right)
\left(\frac{3N_c}{2}+c_+e^{4\r/3}\right)^2},\phantom{moremoremore}
\eea
where $c_+$ is an arbitrary positive constant. 
It can be seen that this solution approaches in the IR 
($\rho\to-\infty$) the solution in 
eq.~(\ref{self-dual-1}) for 
the values $(\x=\frac43,\,\frac83)$, while it drastically differs from
(\ref{self-dual-1}) in the UV ($\rho\to\infty$). The new solution 
leads to an asymptotic UV geometry  of the 
form $\mathbb{R}^{1,3}\times M_6$, where $M_6$ is the 
conifold. Such UV asymptotics was 
anticipated in \cite{Caceres:2007mu} and the solution (\ref{self-dual-2})
is the first exact example possessing this UV behavior. 
The physics of this and 
other solutions with similar UV behavior will be analyzed 
in Section \ref{Physics}. 
Finally, let us mention in passing that it 
may be possible to find solutions like the 
ones above where a black hole (non-extremal deformation) is introduced.

\subsection{Classification of asymptotic solutions}
\label{sec:asymp}
Here we classify the asymptotic solutions of the `master' equation (\ref{master}), both in the UV ($\rho\to\infty$) and in the 
IR ($\rho\to \r_{IR}$), where $\r_{IR}\geq \r_o$ is the minimum value of the radial coordinate in the background geometry
\footnote{We 
stress again that the words UV-ultraviolet- and IR-infrared- 
make reference to the high and low energies (compared to $\Lambda_{QCD}$) 
in the dual field theory. Here we will use these terms indistinctly with 
large and small $\rho$ asymptotics.}.
Note that for the type {\bf A} backgrounds $\r_o\to-\infty$ and so it is possible that $\r_{IR}\to -\infty$. For
the type {\bf N} backgrounds, however, 
we have instead $\r_{IR}\geq \r_o>-\infty$.

\subsubsection{UV asymptotics}
\label{UV-asymptotics}
Let us start by discussing first the UV behavior of solutions to eq.~(\ref{master}). As we have already mentioned,
by UV we mean the asymptotic behavior of the solutions as $\r\to\infty$
\footnote{Note that this excludes solutions
that describe a Landau pole in the dual theory, where the gauge coupling 
diverges at a finite value of the 
radial coordinate, $\r$. }. The first observation one can now make is 
that as $\r\to\infty$, $\t\to 0 $ (see (\ref{tau})) and so 
the leading asymptotic behavior of the type {\bf A} and type {\bf N} solutions of (\ref{master}) is the same. 
This is indeed expected from the holographic interpretation of the type {\bf N} backgrounds, which are believed to 
describe the same theory as the one described by the type {\bf A} backgrounds, but with a non-zero VEV for the 
gaugino bilinear, $\langle \l \l\rangle \neq 0$. At energies much above the energy scale set by the gaugino
condensate the two types of solutions must be identical. 

Secondly, from (\ref{Q}) we see that $Q\to\pm\infty$ as $\r\to\infty$ respectively for $N_f<2N_c$ or
$N_f>2N_c$, while it goes to a constant for $N_f=2N_c$. Moreover, from (\ref{dilaton}) follows that
both $(P+Q)$ and $(P-Q)$ should remain positive as $\r\to\infty$, which 
requires that $P\to+\infty$ as $\r\to\infty$ for 
$N_f\neq 2N_c$, while it can can either asymptote to a finite constant or to infinity for $N_f=2N_c$.

To summarize then, for $N_f\neq 2N_c$ one needs to look for asymptotic solutions of (\ref{master}) 
with $P\to+\infty$ as $\r\to+\infty$, while for $N_f=2N_c$ one should look 
for solutions 
such that $P\to+\infty$ or $P\to constant$ as $\r\to+\infty$. One then finds that there are
two classes of qualitatively different asymptotic solutions. The first class
consists of solutions where $P$ behaves linearly with $\r$ as $\r\to+\infty$,
with the coefficient dependent on the relative values of $N_f$ and $N_c$. These asymptotic
solutions have been studied in \cite{Casero:2006pt,Casero:2007jj}. The second class consists of
solutions for which $P\sim c_+ e^{4\r/3}$ as $\r\to+\infty$, where $c_+$ is an arbitrary constant.
This type of UV asymptotics was anticipated-in the {\it flavorless} case- 
in the paper \cite{Casero:2006pt}, 
where it was pointed out (see Appendix B of \cite{Casero:2006pt})
that the geometry in this case asymptotes to four dimensional Minkowski spacetime times the conifold for the type {\bf A} 
backgrounds ($\r_o\to-\infty$), or times the deformed conifold for for the type {\bf N} backgrounds ($\r_o>-\infty$).

Given this general classification of the UV asymptotics, which we have summarized in Table \ref{UV-classes}, we are now ready to 
construct explicitly the corresponding asymptotic solutions.

\begin{table}
\begin{center}
\begin{tabular}{|c||l|l|}
\hline
$N_f$ & I & II \\
\hline\hline &&\\
$<2N_c$ & $P\sim Q\sim|2N_c-N_f|\r$ \mbox{}&  \\ 
	& $e^{2h}\sim\frac12\left(2N_c-N_f\right) \rho $			&\\ 
	& $e^{2g}\sim N_c$			&\\ 
	& $Y\sim \frac{N_c}{4}$			&\\ 
	& $e^{4(\f-\f_o)}\sim \frac{e^{4 \left(\rho -\rho _o\right)}\sinh^2\left(2 \rho _o\right)}
{2N_c^2\left(2 N_c- N_f\right) \rho }$			&\\ 
	& $a\sim \frac{2}{N_c}\left(2N_c- N_f\right) e^{-2 \left(\rho -\rho _o\right)}\rho     $			&\\ 
	& &\\ \cline{1-2} && \\
$>2N_c$ & $P\sim -Q\sim|2N_c-N_f|\r$ &  $P\sim c_+e^{4\r/3}$  \\ 
	& $e^{2h}\sim \frac{1}{4} \left(N_f-N_c\right)$			& $e^{2h}\sim \frac14 c_+e^{4\r/3}$	\\ 
	& $e^{2g}\sim \frac{1}{2} \left(N_f-2 N_c\right) \rho$			& $e^{2g}\sim c_+e^{4\r/3}$\\ 
	& $Y\sim \frac{1}{4} \left(N_f-N_c\right)$			& $Y\sim \frac16 c_+e^{4\r/3}$\\ 
	& $e^{4(\f-\f_o)}\sim \frac{e^{4 \left(\rho -\rho _o\right)}\sinh^2\left(2 \rho _o\right)}
{2 \left(N_c-N_f\right){}^2 \left(N_f-2 N_c\right) \rho }$			& $e^{4(\f-\f_o)}\sim 1 $	\\ 
	& $a\sim  e^{-2 \left(\rho -\rho _o\right)}\rho $	& $a\sim2 e^{-2(\r-\r_o)} $\\ 
	&&\\ \cline{1-2} && \\
$=2N_c$ & $P\sim N_c+\sqrt{N_c^2+Q_o^2}\sim \frac{8 N_c}{(4-\xi ) \xi }$ &    \\ 
	& $e^{2h}\sim \frac{N_c}{\xi }$			&\\ 
	& $e^{2g}\sim \frac{4 N_c}{4-\xi }$			&\\ 
	& $Y\sim \frac{N_c}{4}$			&\\  
	& $e^{4(\f-\f_o)}\sim e^{4 \left(\rho -\rho _o\right)}\sinh^2\left(2 \rho _o\right)\frac{(4-\xi ) \xi   }
{16 N_c^3}$			&\\ 
	& $a\sim \frac{4}{\xi  }e^{-2 \left(\rho -\rho _o\right)}$			&\\
\hline
\end{tabular}
\caption{The two classes of leading UV behaviors.}
\label{UV-classes}
\end{center}
\end{table}

\begin{flushleft}
\underline{\bf Class I:}
\end{flushleft}

As we mentioned above, this class consists of solutions where $P$ grows at most linearly with $\r$ as $\r\to+\infty$. 
Since $\cosh\t=(1+e^{-4(\r-\r_o)})/(1-e^{-4(\r-\r_o)})=1+\co\left(e^{-4\r}\right)$, it follows that unless one
includes exponentially suppressed terms, the class I UV expansions of the type {\bf A} and type {\bf N} backgrounds
are identical. Namely,

\begin{enumerate}

\item[(i)]  $N_f<2N_c$:
\be\label{UV-e}
P=Q+N_c\left(1+\frac{N_f}{4Q}+\frac{N_f(N_f-2N_c)}{8Q^2}
+\frac{N_f(16N_c^2-19N_cN_f+5N_f^2)}{32Q^3}+\co\left(Q^{-4}\right)\right).
\ee
Using (\ref{tau})-(\ref{Y}) this leads to 
\bea
e^{2h}&=&\left(N_c-\frac{N_f}{2}\right) \rho +\frac{1}{4} \left(N_c+2 Q_o\right)+\frac{N_c N_f}{\left(32 N_c-16 N_f\right) \rho }\NO\\
&&+\frac{N_c N_f \left(-2 N_c+N_f-2 Q_o\right)}{32 \left(-2 N_c+N_f\right){}^2 \rho ^2}+\co(\r^{-3}),\NO\\\NO\\
e^{2g}&=&N_c+\frac{N_c N_f}{\left(8 N_c-4 N_f\right) \rho }+\frac{N_c N_f \left(-2 N_c+N_f-2 Q_o\right)}{8 \left(-2 N_c+N_f\right){}^2 \rho ^2}+\co(\r^{-3}),\NO\\\NO\\
Y&=&\frac{N_c}{4}-\frac{N_c N_f}{\left(64 N_c-32 N_f\right) \rho ^2}+\frac{N_c N_f \left(2 N_c-N_f+2 Q_o\right)}{32 \left(-2 N_c+N_f\right){}^2 \rho ^3}+\co(\r^{-4}),\NO\\\NO\\
e^{4(\f-\f_o)}&=&e^{4 \left(\rho -\rho _o\right)}\left(\frac{1}{\left(4 N_c^3-2 N_c^2 N_f\right) \rho }-\frac{2 N_c+N_f+4 Q_o}{8 \left(N_c^2 \left(-2 N_c+N_f\right){}^2\right) \rho ^2}+\co(\r^{-3})\right)\NO\\
&&\times \sinh^2\left(2 \rho _o\right) ,\NO\\\NO\\
a&=&e^{-2 \left(\rho -\rho _o\right)}\left(\left(4-\frac{2 N_f}{N_c}\right) \rho +\frac{4 N_c-N_f+4 Q_o}{2 N_c}+\frac{4 N_c N_f-N_f^2}{\left(16 N_c^2-8 N_c N_f\right) \rho }\right.\NO\\
&&\left.\phantom{moremoremoremoremoremoremoremoremoremo}+\co(\r^{-2})\rule{0.0in}{.25in}\right).
\eea
It is now obvious that these expansions describe in a unified way both the 
type {\bf A} UV expansions in eq. (3.5) in \cite{Casero:2007jj}, as
well as the type {\bf N} UV expansions in eqs. (4.18)-(4.19) in 
\cite{Casero:2006pt}.

\item[(ii)]  $N_f>2N_c$:
\bea\label{UV-m}
P=-Q+(N_f-N_c)\left(1-\frac{N_f}{4Q}-\frac{N_f(N_f-2N_c)}{8Q^2}
-\frac{N_f(16N_c^2-13N_cN_f+2N_f^2)}{32Q^3}\right.\NO\\
\left.+\co\left(Q^{-4}\right)\rule{.0in}{.24in}\right).
\eea
This expansion is related to the expansion (\ref{UV-e}) by the Seiberg duality transformation $Q\to -Q$, $N_c\to N_f-N_c$.
Using again (\ref{tau})-(\ref{Y}) this leads to the expansions
\bea
e^{2h}&=&\frac{1}{4} \left(-N_c+N_f\right)+\frac{\left(N_c-N_f\right) N_f}{\left(32 N_c-16 N_f\right) \rho }-\frac{\left(N_c-N_f\right) N_f \left(2 N_c-N_f+2 Q_o\right)}{32 \left(-2 N_c+N_f\right){}^2 \rho ^2}\NO\\
&&+\co(\r^{-3}),\NO\\\NO\\
e^{2g}&=&\frac{1}{2} \left(-2 N_c+N_f\right) \rho +\frac{1}{4} \left(-N_c+N_f-2 Q_o\right)+\frac{\left(N_c-N_f\right) N_f}{\left(32 N_c-16 N_f\right) \rho }\NO\\
&&-\frac{\left(N_c-N_f\right) N_f \left(2 N_c-N_f+2 Q_o\right)}{32 \left(-2 N_c+N_f\right){}^2 \rho ^2}+\co(\r^{-3}),\NO\\\NO\\
Y&=&\frac{1}{4} \left(-N_c+N_f\right)+\frac{N_f \left(-N_c+N_f\right)}{\left(64 N_c-32 N_f\right) \rho ^2}+\frac{\left(N_c-N_f\right) N_f \left(2 N_c-N_f+2 Q_o\right)}{32 \left(-2 N_c+N_f\right){}^2 \rho ^3}\NO\\
&&+\co(\r^{-4}),\NO\\\NO\\
e^{4(\f-\f_o)}&=&e^{4 \left(\rho -\rho _o\right)}\left(-\frac{1}{2 \left(\left(N_c-N_f\right){}^2 \left(2 N_c-N_f\right)\right) \rho }+\frac{2 N_c-3 N_f+4 Q_o}{8 \left(2 N_c^2-3 N_c N_f+N_f^2\right){}^2 \rho ^2}\right.\NO\\
&&\left.\phantom{moremoremoremoremoremore}+\co(\r^{-3})\rule{.0in}{.24in}\right)\sinh^2\left(2 \rho _o\right) ,\\ \NO\\
a&=&e^{-2 \left(\rho -\rho _o\right)}\left(1+\frac{N_c-N_f}{\left(4 N_c-2 N_f\right) \rho }-\frac{\left(N_c-N_f\right) \left(2 N_c-N_f+4 Q_o\right)}{8 \left(-2 N_c+N_f\right){}^2 \rho ^2}+\co(\r^{-2})\rule{0.0in}{.25in}\right).\NO
\eea
Clearly, these expansions reproduce in a unified way both the type {\bf A} 
UV expansions of eq.~(3.6) in \cite{Casero:2007jj} and 
the type {\bf N} UV expansions of eqs.~(4.20) in \cite{Casero:2006pt}.

\item[(iii)] $N_f=2N_c$:\vskip.1in
Finally, for $N_f=2N_c$ the type {\bf A} class I UV solution is given by the exact solution (\ref{self-dual-1}), while the type {\bf N} 
class I solution is again given by (\ref{self-dual-1}) up to exponentially suppressed terms. 
Namely
\be\label{UV-c}
P=N_c+\sqrt{N_c^2+Q_o^2}+\co(e^{-4(\r-\r_o)})=\frac{8 N_c}{(4-\xi ) \xi }+\co(e^{-4(\r-\r_o)}), \quad 0<\x<4.
\ee
Using (\ref{tau})-(\ref{Y}) this leads to the expansions
\bea
e^{2h}&=&\frac{N_c}{\xi }+\co(e^{-4(\r-\r_o)}),\NO\\
e^{2g}&=&\frac{4 N_c}{4-\xi }+\co(e^{-4(\r-\r_o)}),\NO\\
Y&=&\frac{N_c}{4}+\co(e^{-4(\r-\r_o)}),\NO\\
e^{4(\f-\f_o)}&=&e^{4 \left(\rho -\rho _o\right)}\sinh^2\left(2 \rho _o\right)\frac{(4-\xi ) \xi   }
{16 N_c^3}\left(1+\co(e^{-4(\r-\r_o)})\right),\NO\\
a&=&\frac{4}{\xi  }e^{-2 \left(\rho -\rho _o\right)}+\co(e^{-4(\r-\r_o)}).
\eea
Note that as $\r_o\to-\infty$, this reproduces the exact solution (\ref{self-dual-1}). 
\end{enumerate}
These asymptotic expansions exhaust all possible UV behaviors of the 
physically accepted solutions of eq. (\ref{master}),
with $P$ growing at most linearly with $\r$ as 
$\r\to\infty$. As we have already mentioned, these expansions 
receive exponentially suppressed corrections, which are of two different types. Firstly, if $\r_o>-\infty$, i.e. we are interested
in the type {\bf N} solutions, then, as we have indicated, 
there are exponentially suppressed corrections due to the presence of
$\cosh\t$ in $Q$ and in eq. (\ref{master}). Moreover, the expansions 
of eqs. (\ref{UV-e}), (\ref{UV-m}) and (\ref{UV-c}) do not involve any
integration constants (except for $Q_o$ and $\r_o$ which appear as parameters in (\ref{master})). As we shall see below, turning on
one of the two integration constants of (\ref{master}) will change the UV asymptotics to that of class II. Turning on the second 
integration constant, however, will introduce certain exponentially suppressed terms. See e.g. \cite{Casero:2007jj} for a discussion
of these terms.

\begin{flushleft}
\underline{\bf Class II:}
\end{flushleft}

The second class of UV asymptotics consists of solutions where $P\sim c_+e^{4\r/3}$ as $\r\to\infty$, independently of the values of the parameters $N_c$, $N_f$, $Q_o$ or $\r_o$. The positive constant $c_+$ here corresponds to one of the integration constants of (\ref{master}),
and it completely determines the leading asymptotic behavior of the solution. Keeping the leading exponentially suppressed corrections
that differentiate the expansions of the type {\bf A} and the type {\bf N} solutions in this case, we have respectively
\bea\label{UV-II-A}
P&=&e^{4\rho/3}\left\{c_++\frac{9 N_f}{8}e^{-4\rho/3}+\frac{1}{64c_+}\left[64(2N_c-N_f)^2\rho^2+128(2N_c-N_f)Q_o\rho+\right.\right.\NO\\
&&\left.\left.9(4N_c-3N_f)(4N_c-N_f)+64Q_o^2\right]e^{-8\rho/3}\right.\NO\\
&&\left.-\frac{1}{64c_+^2}\left[\frac{64}{3}N_f(2N_c-N_f)^2\rho^3+16N_f(2N_c-N_f)(3(2N_c-N_f)+4Q_o)\rho^2\right.\right.\\
&&\left.\left.+\left(3N_f(32N_c(N_c-N_f)+5N_f^2)+32N_f Q_o(2Q_o-3N_f+6N_c)\right)\rho+c_-\rule{0.0in}{.24in}\right]e^{-12\rho/3}\right.\NO\\
&&\left.+\co\left(\rho^3 e^{-16\rho/3}\right)\rule{0.0in}{.24in}\right\},\NO
\eea
for $\r_o\to-\infty$ (type {\bf A} solution), and
\bea\label{UV-II-N}
P&=&e^{4\rho/3}\left\{c_+\left(1-\frac{8}{3}\r e^{-4\rho}+\co(e^{-8\rho})\right)+\frac{9N_f}{8}\left(1+\co(\r e^{-4\rho})\right)e^{-4\rho/3}\right.\NO\\
&&\left.+\frac{1}{64c_+}\left[64(2N_c-N_f)^2\r^2+128(2N_c-N_f)Q_o\r
+144N_c(N_c-N_f)+27N_f^2\right.\right.\NO\\
&&\left.\left.+64Q_o^2\right]e^{-8\rho/3}\right.\NO\\
&&\left.-\frac{1}{64c_+^2}\left[\frac{64}{3}N_f(2N_c-N_f)^2\rho^3+16N_f(2N_c-N_f)(3(2N_c-N_f)+4Q_o)\rho^2\right.\right.\\
&&\left.\left.
+\left(3N_f(32N_c(N_c-N_f)+5N_f^2)+32b Q_o(2Q_o-3N_f+6N_c)\right)\rho+c_-\rule{0.0in}{.24in}\right]e^{-12\rho/3}\right.\NO\\
&&\left.+\co\left(\r^3 e^{-16\rho/3}\right)\rule{0.0in}{.24in}\right\},\NO
\eea
for $\r_o>-\infty$ (type {\bf N} solution), where we have set $\r_o=0$ for 
convenience. Notice that these two expansions involve two integration constants, $c_+>0$ and $c_-$, and are identical 
except for the exponentially suppressed corrections in the first line of (\ref{UV-II-N}), coming from the expansion of $\cosh\t$. 
Using eqs. (\ref{tau})-(\ref{Y}) these expansions give to first subleading 
order
\bea
e^{2h}&=&\frac{1}{4}\left(c_+e^{4\r/3}+(2N_c-N_f)\r+Q_o+\frac{9N_f}{8}+\co\left(\r^2e^{-4\r/3}\right)\right),\NO\\
e^{2g}&=&c_+e^{4\r/3}-(2N_c-N_f)\r-Q_o+\frac{9N_f}{8}+\co\left(\r^2e^{-4\r/3}\right),\NO\\
Y&=&\frac{1}{8}\left(\frac{4c_+}{3}e^{4\r/3}+N_f+\co\left(\r^2e^{-4\r/3}\right)\right),\NO\\
e^{4(\f-\f_o)}&=&1-\frac{3N_f}{c_+}e^{-4\r/3}+\co\left(\r^2e^{-8\r/3}\right),\NO\\
a&=&2e^{-2(\r-\r_o)}\left(1+\frac{1}{c_+}\left((2N_c-N_f)\r+Q_o\right)e^{-4\r/3}+\co\left(\r^2e^{-8\r/3}\right)\right).
\label{gggg}\eea

As it can be seen directly by inserting these asymptotics in the metric (\ref{nonabmetric424}), the geometry asymptotes to the conifold.
These solutions then correspond to turning on an irrelevant operator with coupling $c_+$ that drives us away from
the near horizon geometry of the D5 branes (cf. \cite{Intriligator:1999ai,Skenderis:2006di}). We will give 
arguments supporting this interpretation in Section \ref{Physics}. 

In fact, it turns out that the geometry asymptotes to the undeformed conifold for the type \bA case, while it asymptotes to the deformed conifold in the type \bN case.  In order to see this, however, we need to resum certain terms in 
the asymptotic expansion in eqs. (\ref{UV-II-N}). This can be done as 
follows. Since the powers of $e^{4\r/3}$ in these expansions for 
large $\r$ come with powers of $c_+$, one can, instead of asymptotic solutions for large $\r$, look for
solutions of (\ref{master}) in an expansion for large $c_+$. In Appendix \ref{moduli-expansions} we 
construct systematically both the type \bA and type \bN solutions in the large $c_+$ expansion, showing
that the former asymptote to the conifold, while the latter to the deformed conifold.

\subsubsection{IR asymptotics}
\label{IR-asymptotics}

Let us next analyze the IR behavior of the solutions of (\ref{master}). By infrared we generically mean the minimum value, 
$\r_{IR}$, of the radial coordinate, where the geometry ends. For the type \bN backgrounds $\r\geq \r_o>-\infty$ and so necessarily $\r_{IR}\geq \r_o>-\infty$. For the type \bA backgrounds, however, $\r_o\to-\infty$ and so $\r_{IR}$ is totally unconstrained. 
In contrast to the UV asymptotics where we always have $\r\to\infty$ 
(unless we consider Landau poles), the first
task in determining the IR asymptotics is to determine the location of $\r_{IR}$. This is not too complicated, though, since 
there are two possible inequivalent locations for $\r_{IR}$ in each case. Namely, for the type \bA backgrounds ($\t=0$)
it is either located at $\r_{IR}=-\infty$  or at a finite radial distance $\r_{IR}>-\infty$, which can be taken to be zero by a shift in the radial coordinate. 
For the type \bN case the IR is located either at $\r_{IR}=\r_o$ ($\t(\r_{IR})=\infty$) or at $\r_{IR}>\r_o$ ($\t(\r_{IR})<\infty$), in which case again it can be taken to be zero by a shift of the radial coordinate. 
In summary, then
\bea
\r_{IR}=\left\{\begin{matrix}
                0  & {\rm or} & -\infty, & & {\bf A},\\
		0 & {\rm or} & \r_o, & & {\bf N}.
               \end{matrix}\right.
\eea 

Looking at the metric (\ref{metric}) and the dilaton (\ref{dilaton}) we see that the IR will generically be located where
some of the functions $Y$, $H=(P+Q)/4$, $G=(P-Q)/4$ become zero
\footnote{Note that from (\ref{metric}) we see that
the geometry degenerates when $P\cosh\t\pm Q=0$, while from 
(\ref{dilaton}) we see that the dilaton will 
generically diverge if $P\pm Q=0$ 
(unless $Y$ or $\sinh\t$ go to infinity suitably fast). 
For the type \bA case
these two conditions coincide, but for the type \bN case $P\pm Q$ go to zero necessarily before $P\cosh\t\pm Q$. Hence,
in either case, one should use the condition $P\pm Q=0$ as a 
criterion for the location of the IR.}. Since the dilaton must remain
finite for solutions with acceptable ``good'' singularities, this means 
that at the same time 
some of these functions, or $\sinh\t$, must go to infinity sufficiently
fast.

Another possibility for the location of the IR is that none of the functions $H$, $G$, $Y$ vanish, but instead some 
of them, or $\sinh\t$, go to infinity, forcing  $e^{\phi} \to 0$. However, if either $H$ 
or $G$ goes to infinity, then $P\to\infty$ in the IR. But given that $P'$ should remain finite due to (\ref{Y}), it 
is easy to show that equation (\ref{master}) excludes this possibility. Therefore, $H$ and $G$ must remain finite, 
while $Y$ and/or $\sinh\t$ would have to go to infinity. An easy calculation using the fact that $P'\sim Y$ and eq.~(\ref{master}) 
excludes the possibility $Y\to\infty$ while $P$ remains finite as $\r\to \r_{IR}$. The only possibility of this kind then
would be that all $H$, $G$ and $Y$ remain finite and non-zero, while $\sinh\t\to\infty$. This possibility is also easily
excluded by eq. (\ref{master}). 

We therefore conclude that all possible IRs are classified by the vanishing of some of the functions $H$, $G$ and $Y$. 
One expects that depending on which of these functions vanish, there will be qualitatively different IR behaviors. 
Following \cite{Casero:2007jj} we will call type I the case where $Y\to 0$, type II the case where {\em either} 
$H$ {\em or} $G$ vanish, and type III the case where {\em both} $H$ {\em and} $G$ vanish. In the 
type \bA case and only for $N_c=0$ or $N_f=N_c$ there is one more case where respectively $G$ or $H$ vanishes in addition to $Y$, 
but we will include this in the type I case. As we shall see, the type I backgrounds have $\r_{IR}=-\infty$ in the type \bA case
and $\r_{IR}=\r_o=0$ in the type {\bf N}, whereas the type II and type III backgrounds have $\r_{IR}=0$ in both cases.

\begin{flushleft}
\underline{\bf Type I:}
\end{flushleft}

By definition, these are solutions that have $Y\to 0$ as $\r\to \r_{IR}$. Now there is an exact solution
of eq.~(\ref{master}), valid for arbitrary parameters $N_c$, $N_f$, $Q_o$ 
and $\r_o$, namely
\be\label{bad-solution}
P=-N_f\r+P_o,\quad \r\leq P_o/N_f,
\ee
where $P_o$ is an arbitrary constant. This solution, via (\ref{Y}), leads to $Y=0$ identically, and it is therefore not an
acceptable solution. Since it involves only one integration constant, $P_o$, though, there is one more integration
constant left which can be used to deform this solution to a non-singular solution. Together these two integration constants
parametrize {\em any} solution of (\ref{master}) in the vicinity of $Y=0$. However, since in the solution 
(\ref{bad-solution}) $\r$ is bounded from above, the deformed solution must deviate
strongly from  (\ref{bad-solution}) in order to match with the UV asymptotics described above. 
One can then construct the type I IR expansions by solving (\ref{master}) perturbatively in the second integration
constant around the unphysical solution (\ref{bad-solution}). In Appendix \ref{moduli-expansions} we systematically construct these expansions both in the type {\bf A} and type \bN cases. Since (\ref{bad-solution})
solves eq.~(\ref{master-c+-integrated}) with $c_+=0$, these expansions are expansions in powers of $c_+^3$. 
From these then we obtain the following type I IR expansions. 

For the type \bA backgrounds we find that $\r_{IR}\to-\infty$ and 
\bea
P&=&-N_f\r+P_o+c_+^3e^{4\r}\left(N_c(N_f-N_c)\r^2-\frac12(N_c(N_f-N_c)+N_f P_o+(2N_c-N_f)Q_o)\r\right.\NO\\
&&\left.+\frac14(P_o^2-Q_o^2)+\frac18(N_fP_o+(2N_c-N_f)Q_o+N_c(N_f-N_c))\right)+\co\left(c_+^6e^{8\r}\r^3\right).
\eea
Evaluating the rest of the functions parametrizing the background using (\ref{tau})-(\ref{Y}), we find 
\bea
H&=&\frac{1}{2} \left(N_c-N_f\right) \rho +\frac{1}{4} \left(P_o+Q_o\right)\NO\\
&&+c_+^3e^{4 \rho } \left(\frac{1}{4}  N_c \left(-N_c+N_f\right) \rho ^2+\frac{1}{8}  \left(N_c^2+N_f \left(-P_o+Q_o\right)-N_c \left(N_f+2 Q_o\right)\right) \rho \right.\NO\\
&&\left.+\frac{1}{32}  \left(-N_c^2+N_c \left(N_f+2 Q_o\right)+\left(P_o-Q_o\right) \left(N_f+2 \left(P_o+Q_o\right)\right)\right)\right)+\co\left(c_+^6e^{8\r}\r^3\right),\NO\\
G&=&-\frac{N_c \rho }{2}+\frac{1}{4} \left(P_o-Q_o\right)\NO\\
&&+c_+^3e^{4 \rho } \left(\frac{1}{4}  N_c \left(-N_c+N_f\right) \rho ^2+\frac{1}{8} \left(N_c^2+N_f \left(-P_o+Q_o\right)-N_c \left(N_f+2 Q_o\right)\right) \rho \right.\NO\\
&&\left.+\frac{1}{32}  \left(-N_c^2+N_c \left(N_f+2 Q_o\right)+\left(P_o-Q_o\right) \left(N_f+2 \left(P_o+Q_o\right)\right)\right)\right)+\co\left(c_+^6e^{8\r}\r^3\right),\NO\\
Y&=&c_+^3e^{4 \rho } \left(\frac{1}{2} N_c \left(-N_c+N_f\right) \rho ^2+\frac{1}{4}  \left(-2 N_c Q_o+N_f \left(-P_o+Q_o\right)\right) \rho +\frac{1}{8} \left(P_o^2-Q_o^2\right)\right)\NO\\
&&+\co\left(c_+^6e^{8\r}\r^3\right),\NO\\
e^{4(\f-\f_o)}&=&\frac{8}{c_+^3\left(4 N_c \left(-N_c+N_f\right) \rho ^2+2 \left(-2 N_c Q_o+N_f \left(-P_o+Q_o\right)\right) \rho + \left(P_o^2-Q_o^2\right)\right)^2}\NO\\
&&+\co\left(e^{4\r}\right).
\eea
This asymptotic solution then coincides with the type I IR solution of \cite{Casero:2007jj} (cf. eqs. (3.7)-(3.9)). 
However, when either $N_f=N_c$ and $P_o=-Q_o$ or $N_c=0$ and $P_o=Q_o$, we find a second asymptotic solution, namely
\bea
P=-N_f\r+P_o+c_+^3e^{2\r}\sqrt{-N_f\r+P_o+N_f/4}+\cdots,
\eea
which gives
\bea
H&=&\frac{1}{2} \left(N_c-N_f\right) \rho +\frac{1}{4} \left(P_o+Q_o\right)+\frac{1}{4}c_+^3e^{2\r}\sqrt{-N_f\r}+\co\left(e^{2\r}|\r|^{-1/2}\right),\NO\\
G&=&-\frac{N_c \rho }{2}+\frac{1}{4} \left(P_o-Q_o\right)+\frac{1}{4}c_+^3e^{2\r}\sqrt{-N_f\r}+\co\left(e^{2\r}|\r|^{-1/2}\right),\NO\\
Y&=&\frac{1}{4}c_+^3e^{2\r}\sqrt{-N_f\r}+\co\left(e^{2\r}|\r|^{-1/2}\right),\NO\\
e^{4(\f-\f_o)}&=&\frac{4(-N_f\r)^{-1/2}e^{2\r}}{c_+^3\left(4 N_c \left(-N_c+N_f\right) \rho ^2+2 \left(-2 N_c Q_o+N_f \left(-P_o+Q_o\right)\right) \rho + \left(P_o^2-Q_o^2\right)\right)}+\cdots\NO\\
\eea

For the type \bN backgrounds, we find that $\r_{IR}=\r_o$, which we can take to be zero without loss of generality. 
Moreover we must set $Q_o=-(2N_c-N_f)/2$ in order for a well defined solution of this form to exist.
The asymptotic solution then takes the form
\bea
P=-N_f\r+P_o+\frac43c_+^3P_o^2\r^3-2c_+^3N_fP_o\r^4+\frac{4}{5}c_+^3\left(\frac{4}{3}P_o^2+N_f^2\right)\r^5+\co\left(\r^6\right),
\eea
which leads to 
\bea
e^{2h}&=&\frac{P_o \rho }{2}-\frac{N_f \rho ^2}{2}-\frac{2 P_o \rho ^3}{3}+\co\left(\r^4\right),\NO\\
e^{2g}&=&\frac{P_o}{2 \rho }-\frac{N_f}{2}+\frac{2 P_o \rho }{3}+\frac{2}{3} \left(-2 N_c+c_+^3 P_o^2\right) \rho ^2-\frac{1}{45} \left(\left(8+45 c_+^3 N_f\right) P_o\right) \rho ^3+\co\left(\r^4\right),\NO\\
Y&=&\frac{1}{2} c_+^3 P_o^2 \rho ^2-c_+^3 N_f P_o \rho ^3+\frac{1}{6} c_+^3 \left(3 N_f^2+4 P_o^2\right) \rho ^4+\co\left(\r^5\right),\NO\\
e^{4(\f-\f_o)}&=&1+\frac{4 N_f \rho }{P_o}+\frac{10 N_f^2 \rho ^2}{P_o^2}+\left(\frac{20 N_f^3}{P_o^3}-\frac{8 N_f}{3 P_o}-\frac{8 c_+^3 P_o}{3}\right) \rho ^3+\co\left(\r^4\right),\NO\\ 
a &=&1-2 \rho ^2-\frac{4 \left(-2 N_c+N_f\right) \rho ^3}{3 P_o}+\frac{2 \left(4 N_c N_f-2 N_f^2+5 P_o^2\right) \rho ^4}{3 P_o^2}+\co\left(\r^5\right). 
\label{kxkx}\eea
Identifying then $2c_+^3P_o^2=c_2N_c$ and $c_1=4(2N_c-N_f)/3P_o$ the 
expansion 
(\ref{kxkx}) exactly reproduces the expansion of eq. (4.21) in 
\cite{Casero:2006pt}. Finally, note there is another 
isolated type I solution, namely the exact solution in 
eq.~(\ref{unflavored}).

\begin{flushleft}
\underline{\bf Type II:}
\end{flushleft}

Type II infrared behavior corresponds to $H=0$ or $G=0$. Let us first assume that this behavior occurs when the IR
is located at $\r_{IR}>\r_o$. Without loss of generality we can choose $\r_{IR}=0$. With this choice we then
necessarily have $\r_o<0$. Expanding $Q$ in (\ref{Q}) around $\r=0$ we obtain
\be\label{Q-exp}
Q=b_0+b_1\r+\co(\r^2),
\ee
where
\bea
b_0&=&-\coth(2\r_o)\left(Q_o+\frac{2N_c-N_f}{2}\right)-\frac{2N_c-N_f}{2},\NO\\
b_1&=&-\frac{2}{\sinh^2(2\r_o)}\left(Q_o+\frac{2N_c-N_f}{2}\right)-(2N_c-N_f)\coth(2\r_o),\NO\\
b_2&=&\ldots.
\eea
Looking for IR solutions of (\ref{master}) with $G\to 0$ as $\r\to 0$ we find we must require that $b_0>0$. The
corresponding asymptotic solution then takes the form
\bea\label{G-II-1}
P&=&Q+h_1\r^{1/2}-\frac{1}{6b_0}\left(h_1^2+12b_0(b_1+N_f)\right)\r\NO\\
&&+\frac{h_1}{72b_0^2}\left(5h_1^2+6(5b_1+2N_f)b_0-72b_0^2\coth(2\r_o)\right)\r^{3/2}+\co(\r^2),
\eea
where $h_1$ is an arbitrary constant. Note that this expansion for $P$ admits a smooth limit when $\r_o\to-\infty$ and so it 
is valid for both type \bA ($\r_o\to-\infty$) and type \bN backgrounds. Using then (\ref{tau})-(\ref{Y}) we get
\bea
H&=&\frac{b_0}{2}+\frac{h_1 \rho^{1/2}}{4}+\left(-\frac{h_1^2}{24 b_0}-\frac{N_f}{2}\right) \rho \NO\\
&&+\frac{h_1 \left(30 b_1 b_0-72 \coth(2\r_o) b_0^2+5 h_1^2+12 b_0 N_f\right) \rho ^{3/2}}{288 b_0^2}+\co(\r^2),\NO\\
G&=&\frac{h_1 \rho^{1/2}}{4}+\left(-\frac{b_1}{2}-\frac{h_1^2}{24 b_0}-\frac{N_f}{2}\right) \rho \NO\\
&&+\frac{h_1 \left(30 b_1 b_0-72 \coth(2\r_o) b_0^2+5 h_1^2+12 b_0 N_f\right) \rho ^{3/2}}{288 b_0^2}+\co(\r^2),\NO\\
Y&=&\frac{h_1}{16 \rho^{1/2}}+\frac{1}{48} \left(-6 b_1-\frac{h_1^2}{b_0}-6 N_f\right)\NO\\
&&+\frac{h_1 \left(30 b_1 b_0-72 \coth(2 \rho_o) b_0^2+5 h_1^2+12 b_0 N_f\right) \rho^{1/2}}{384 b_0^2}+\co(\r),\NO\\
e^{4(\f-\f_o)}&=&1+\frac{4 \left(b_1+N_f\right) \rho^{1/2}}{ h_1}\NO\\
&&+\frac{2  \left(18 b_1^2 b_0-b_1 \left(h_1^2-36 b_0 N_f\right)+2 N_f \left(h_1^2+9 b_0 N_f\right)\right) \rho }{3 b_0 h_1^2}+\co(\r^{3/2}),\NO\\
a &=&\frac{\sinh^{-1}(2\r_o)}{1+\coth(2\r_o)}+\frac{\sinh^{-1}(2\r_o) h_1 \rho^{1/2}}{(1+\coth(2\r_o))^2 b_0}+\co(\r).
\eea
Note that we have given the functions $H$ and $G$ here instead of the old variables  $e^{2h}$ and $e^{2g}$. The reason
is that the expansion of these functions takes the form
\bea
e^{2h}&=&-\frac{h_1 \r^{1/2}}{2+\coth(2\r_o)+\tanh(2\r_o)+\co(\r^{1/2})}+\co(\r),\NO\\
e^{2g}&=&-(1+\coth(2\r_o)) b_0-\coth(2\r_o) h_1 \r^{1/2}+\co(\r),
\eea
and so it is clear that the expansion around $\r\to 0$ of $e^{2h}$ does not commute with the 
$\r_o\to -\infty$ (type {\bf A}) limit. In particular, to evaluate this limit one needs to
keep the subleading $\co(\r^{1/2})$ term in the denominator of $e^{2h}$. The same is true
for $a$, whose expansion, keeping the subleading term in the denominator, takes the form
\be
a=\frac{b_0}{b_0e^{2\r_o}+\cosh(2\r_o)h_1\r^{1/2}+\cdots}+\cdots
\ee 
In the form we have given the asymptotic solution, however, one can directly take the
the limit $\r_o\to -\infty$, except in $a$ which vanishes in this limit once one takes into account
the subleading corrections. In this limit this asymptotic solution reduces to the 
asymptotic solution in eq.~(3.11) of  \cite{Casero:2007jj}.\footnote{Note that $h_1$ here differs by a factor of 4 from
$h_1$ in \cite{Casero:2007jj}.} The expansion here 
gives the generalization to the type \bN backgrounds.

Similarly, looking for IR solutions of (\ref{master}) with $H\to 0$ as $\r\to 0$ we find we must require that $b_0<0$ and then,
\bea\label{H-II-1}
P&=&-Q+h_1\r^{1/2}+\frac{1}{6b_0}\left(h_1^2+12b_0(b_1-N_f)\right)\r\NO\\
&&+\frac{h_1}{72b_0^2}\left(5h_1^2+6(5b_1-2N_f)b_0-72b_0^2\coth(2\r_o)\right)\r^{3/2}+\co(\r^2),
\eea
where $h_1$ is an arbitrary constant. From  (\ref{tau})-(\ref{Y}) then we get
\bea
e^{2h}&=&-\frac{h_1 \r^{1/2}}{-2+\coth(2\r_o)+\tanh(2\r_o)}\NO\\
&&-\frac{\left(6 (-1+\coth(2\r_o))  b_o(b_1-N_f)+h_1^2 (1+\coth(2\r_o)+\tanh(2\r_o))\right) \rho }{6 b_o(-1+\coth(2\r_o))^2 }+\co(\r^{3/2}),\NO\\
e^{2g}&=&(-1+\coth(2\r_o)) b_o-\coth(2\r_o) h_1 \r^{1/2}\NO\\
&&+\left(\coth(2\r_o)\left(2N_f-b_1-\frac{ h_1^2}{6 b_o}\right)+2 b_o\sinh^{-2}(2\r_o) -b_1\right) \rho +\co(\r^{3/2}),\NO\\
Y&=&\frac{h_1}{16 \r^{1/2}}+\frac{1}{48} \left(6 b_1+\frac{h_1^2}{b_o}-6 N_f\right)\NO\\
&&+\frac{h_1 \left(30 b_1 b_o-72 \coth(2\r_o) b_o^2+5 h_1^2-12 b_o N_f\right) \r^{1/2}}{384 b_o^2}+\co(\r),\NO\\
e^{4(\f-\f_o)}&=&1-\frac{4  \left(b_1-N_f\right) \r^{1/2}}{h_1}
\NO\\
&&+\frac{2 \left(18 b_1^2 b_o+2 N_f \left(-h_1^2+9 b_o N_f\right)-b_1 \left(h_1^2+36 b_o N_f\right)\right) \rho }{3 \left(b_o h_1^2\right)}+\co(\r^{3/2}),\NO\\
a &=&\frac{\sinh^{-1}(2\r_o)}{-1+\coth(2\r_o)}+\frac{\sinh^{-1}(2\r_o) h_1 \r^{1/2}}{(-1+\coth(2\r_o))^2 b_o}+\co(\r).
\eea
In the limit $\r_o\to-\infty$ (type {\bf A}) these expansions reproduce 
the expansions of eq. (3.10) in \cite{Casero:2007jj}.
Here we have generalized this IR solution to the type \bN backgrounds.

Let us finally consider the case $\r_{IR}=\r_o$. Note that in this case $Q$ has a pole at $\r_o$ unless
$Q_o=-(2N_c-N_f)\left(\r_o+\frac12\right)$. Setting then, without loss of generality, $\r_o=0$, we have
\be\label{q-exp}
Q=(2N_c-N_f)\left(\frac23\r^2-\frac{8}{45}\r^4+\frac{64}{945}\r^6+\co(\r^8)\right).
\ee
Since we are looking 
for solutions such that $H$ or $G$ go to zero as $\r\to 0$, it follows that
in this case {\em both} $H$ and $G$ will go to zero as $\r\to 0$ and hence this will be a type III solution,
which we will consider below.

\begin{flushleft}
\underline{\bf Type III:}
\end{flushleft}

Finally we look for IR solutions for which $H\to 0$ and $G\to 0$ in the IR. We again first consider 
$\r_{IR}>\r_o$ and we take $\r_{IR}=0$. In terms of the expansion 
(\ref{Q-exp}) this requires that $b_0=0$. We then find, 
\bea\label{III}
P=h_1\r^{1/3}-\frac{9N_f}{5}\r-\frac{2h_1}{3}\coth(2\r_o)\r^{4/3}-\frac{1}{175h_1}\left(50b_1^2-18N_f^2\right)\r^{5/3}
+\co(\r^2),
\eea
where $h_1\neq 0$ is an arbitrary constant. Evaluating the rest of the background functions we find
\bea
e^{2h}&=&-\frac{1}{4} h_1 \tanh(2\r_o) \rho ^{1/3}+\frac{1}{20} \tanh(2\r_o) \left(9 N_f+5 b_1 \tanh(2\r_o)\right) \rho \NO\\
&&+\frac{1}{6} \left(1+\frac{3}{\cosh^{2}(2\r_o)}\right) h_1 \rho ^{4/3}+\co(\r^{5/3}),\NO\\
e^{2g}&=&-\coth(2\r_o) h_1 \rho ^{1/3}+\left(-b_1+\frac{9}{5} \coth(2\r_o) N_f\right) \rho \NO\\
&&+\frac{1}{3} (-5+\cosh(2\r_o)) \sinh^{-2}(2\r_o) h_1 \rho ^{4/3}+\co(\r^{5/3}),\NO\\
Y&=&\frac{h_1}{24 \rho ^{2/3}}-\frac{N_f}{10}-\frac{1}{9} \coth(2\r_o) h_1\rho ^{1/3}
+\frac{\left(-25 b_1^2+9 N_f^2\right) \rho ^{2/3}}{420 h_1}+\co(\r),\NO\\
e^{4(\f-\f_o)}&=&1+\frac{6  N_f \rho ^{2/3}}{h_1}+\frac{3 \left(5 b_1^2+39 N_f^2\right) \rho ^{4/3}}{5 h_1^2}+\co(\r^{5/3}),\NO\\
a &=&\frac{1}{\cosh(2\r_o)}-\frac{ b_1 \tanh^2(2\r_o) \rho ^{2/3}}{h_1\sinh(2\r_o)}+\frac{2 \tanh^2(2\r_o)}{\sinh(2\r_o)} \rho+\co(\r^{4/3}).
\eea
This expansion generalizes the type \bA  III expansion of eq.~(3.12) in  \cite{Casero:2007jj} to the type \bN backgrounds. 
Indeed, eq.~(\ref{III}) reproduces this expansion in the type \bA limit 
$\r_o\to-\infty$.  

Coming finally back to the case $\r_{IR}=\r_o$, we have seen that one should set $Q_o=-(2N_c-N_f)\left(\r_o+\frac12\right)$
to avoid the pole in $Q$, which takes the form ($\r_o=0$) (\ref{q-exp}). Looking for solutions with $P\to 0$ 
as $\r\to 0$, and noting that $P$ should go to zero slower than $Q$ to ensure that both $H$ and $G$ remain
positive, we find that there is a solution provided $N_f=0$. Namely,
\be
P= h_1 \r+ \frac{4 h_1}{15}\left(1-\frac{4 N_c^2}{h_1^2}\right)\r^3
+\frac{16 h_1}{525}\left(1-\frac{4N_c^2}{3h_1^2}-\frac{32N_c^4}{3h_1^4}\right)\r^5+\co(\r^7),
\ee
where $h_1$ is again an arbitrary constant. This gives,

\bea
e^{2h}&=&\frac{h_1 \rho ^2}{2}+\frac{4}{45} \left(-6 h_1+15 N_c-\frac{16 N_c^2}{h_1}\right) \rho ^4+\co(\r^6),\NO\\
e^{2g}&=&\frac{h_1}{2}+\frac{4}{15} \left(3 h_1-5 N_c-\frac{2 N_c^2}{h_1}\right) \rho ^2+\frac{8 \left(3 h_1^4+70 h_1^3 N_c-144 h_1^2 N_c^2-32 N_c^4\right) \rho ^4}{1575 h_1^3}\NO\\
&&+\co(\r^6),\NO\\
Y&=&\frac{h_1}{8}+\frac{\left(h_1^2-4 N_c^2\right) \rho ^2}{10 h_1}+\frac{\left(6 h_1^4-8 h_1^2 N_c^2-64 N_c^4\right) \rho ^4}{315 h_1^3}+\co(\r^6),\NO\\
e^{4(\f-\f_o)}&=&1+\frac{64  N_c^2 \rho ^2}{9 h_1^2}+\frac{128 N_c^2 \left(-15 h_1^2+124 N_c^2\right) \rho ^4}{405 h_1^4}+\co(\r^6),\\
a &=&1+\left(-2+\frac{8 N_c}{3 h_1}\right) \rho ^2+\frac{2 \left(75 h_1^3-232 h_1^2 N_c+160 h_1 N_c^2+64 N_c^3\right) \rho ^4}{45 h_1^3}+\co(\r^6).\NO
\eea
This is a new IR solution with a `good singularity' for the type \bN backgrounds in the {\it 
flavorless} case. It should be interesting to study its dynamics.  
\subsection{Solutions of the BPS equations as RG flows}
The solutions described in \cite{Casero:2006pt,Casero:2007jj} are a 
special case of the set of possible solutions to 
the BPS equations, given the ansatz in eq.~(\ref{metric}). We have 
seen that more general asymptotics are possible, for large 
values of the radial coordinate $\rho$, the metric elements 
can grow as $H\sim G\sim Y \sim e^{4\rho/3}$. Another 
possibility is that one of the elements of the metric can collapse 
at a finite value of the radial coordinate, 
ending the space. In order to study how generic these 
solutions are;
we can make a qualitative 
analysis of the system of BPS equations as a four-dimensional 
dynamical system. In the type \bA case the system is simpler and reduces 
to 
three dimensions. We would like to interpret each solution 
as describing a different RG flow in the space of holographic dual theories.

We will work with the $\rho$-dependent functions 
$P=N_c p$, $Q= N_c q$ and $u=(p^2-q^2) Y/N_c$ and 
use the parameter $x\equiv N_f/N_c$. Using the first order equations 
in (\ref{BPS-eqs}), with the constraint (\ref{constraint}) imposed, 
we can rewrite the BPS equations as,
\begin{eqnarray}\label{RGBPSnonabelian}
\notag p' & = & -x +  {8 u\over p^2-q^2}, \\
\notag q' & = & {2-x\over \cosh\tau} -2 q \sinh\tau \tanh\tau, \\
\notag u' & = & 4 u\left[\cosh\tau -{ x p\over p^2-q^2} - { q \over p^2-q^2} \left( {2-x\over \cosh\tau} -2 q \sinh\tau \tanh\tau \right)\right], \\
\notag \tau' & = & -2 \sinh\tau, \\
\notag \phi' & = &  {x p\over p^2-q^2}+ { q\over p^2-q^2}\left({2-x\over \cosh\tau} -2 q \sinh\tau \tanh\tau \right).\\
\end{eqnarray}
Let us fix $1<x\leq 2$,\footnote{
The analysis for $x > 2$ can be repeated using Seiberg 
duality $N_c\to N_f-N_c$, $H\leftrightarrow G$ that 
in this notation corresponds to $p\to (x-1) p$, $q\to -(x-1)q$, 
$u\to (x-1)^3 u$ and then use $\tilde x = x/(x-1)$, so 
$1<\tilde x\leq 2$. }
since the equations are independent of $\phi$, 
the solutions can be described as trajectories 
in the $(p,q,u,\tau)$ space. The trajectories are tangent to the vector 
field defined by the values of $(p',q',u',\tau')$. The BPS type \bA 
equations correspond to $\tau=0$. That is a co-dimension one fixed subspace in the $\tau$ direction. 

For physical solutions 
all the functions (appearing as warp factors in the metric) should be 
positive, 
so $p^2\geq q^2$. When $\tau>0$, $\tau'<0$, 
so type \bN solutions flow from a finite or infinite value 
of $\tau$ in 
the IR ($\rho=\rho_{IR}$)
to $\tau=0$ in the UV ($\rho \to \infty$). This is another way to see 
that types \bA and \bN solutions have the same UV  
asymptotics. At fixed values of 
$\tau > 0$ and $u$ there are three relevant curves in the $(p,q)$ plane
\begin{eqnarray}\label{RGattractors}
\notag p' = 0 \; \Rightarrow & p^2-q^2=\frac{8 u}{x}, & {\rm if}\ \ p^2-q^2 >  {8 u\over x},\ \ {\rm then}\ \  p'<0,\\
 q' = 0 \; \Rightarrow & q={2-x\over (2\sinh^2\tau)}, & {\rm if}\ \ q > 
{2-x\over (2\sinh^2\tau)},\ \ {\rm then}\ \ q'<0,
\end{eqnarray}
and the curve $u'=0$
\begin{equation}\label{RGu}
 \left(p-{x\over 2 \cosh\tau}\right)^2-(1-2\tanh^2\tau) \left(q+{2-x\over 2 (1-\sinh^2\tau)} \right)^2 = {x^2\over 4\cosh^2\tau} - {(2-x)^2\over 4 \cosh^2\tau (1-\sinh^2\tau) }\,.
\end{equation}
where $u'>0$ above this curve (larger values of $p$). The curves $q'=0$ and $p'=0$ attract the flow in the $(p,q)$ plane. The curve $q'=0$ is independent of $u$ and $p$ and it disappears for \bA solutions ($\tau\to 0$). However if $x=2$ in the \bA solution, then $q'=0$ exactly. The curve $p'=0$ is independent of $\tau$ and is a hyperbola in the $(p,\,q)$ plane,
asymptoting the lines $p=\pm q$. 
When $u\to 0$, the hyperbola approaches these 
lines. The curve $u'=0$ can be different conical 
curves depending on the sign of the coefficient of the 
$q^2$ term and therefore on the value of $\tau$. 
For $2\tanh^2\tau =1$, the curve is a parabola passing through $(p=x/\sqrt{2},\,q=0)$ and $(p=0,\,q=0)$. For larger values of $\tau$, the  curve is an ellipse, passing through $(p=x/\cosh\tau,\,q=0)$ and $(p=0,\,q=0)$. When $\tau\to \infty$, the ellipse collapses to the point $(p=0,q=0)$, so there is no region where $u'<0$. 

For smaller values of $\tau$, the curve is a hyperbola. For low values of $\tau$, one of the branches passes through $(p=0,\,q=0)$ but lies below the $p=\pm q>0$ lines, so it does not affect the physical solutions. The other branch passes through the point $(p=x/\cosh\tau,\,q=0)$. As we increase $\tau$, the two branches come closer until they merge at 
\beq\label{merging}
\sinh^2\tau = {4(x-1)\over x^2},
\eeq
and form a new hyperbola where now the relevant branch passes through the points $(p=0,\,q=0)$ and $(p=x/\cosh\tau,\,q=0)$.

\begin{figure}[!htbp]
\begin{tabular}{cc}
\includegraphics[width=6cm, height=6cm]{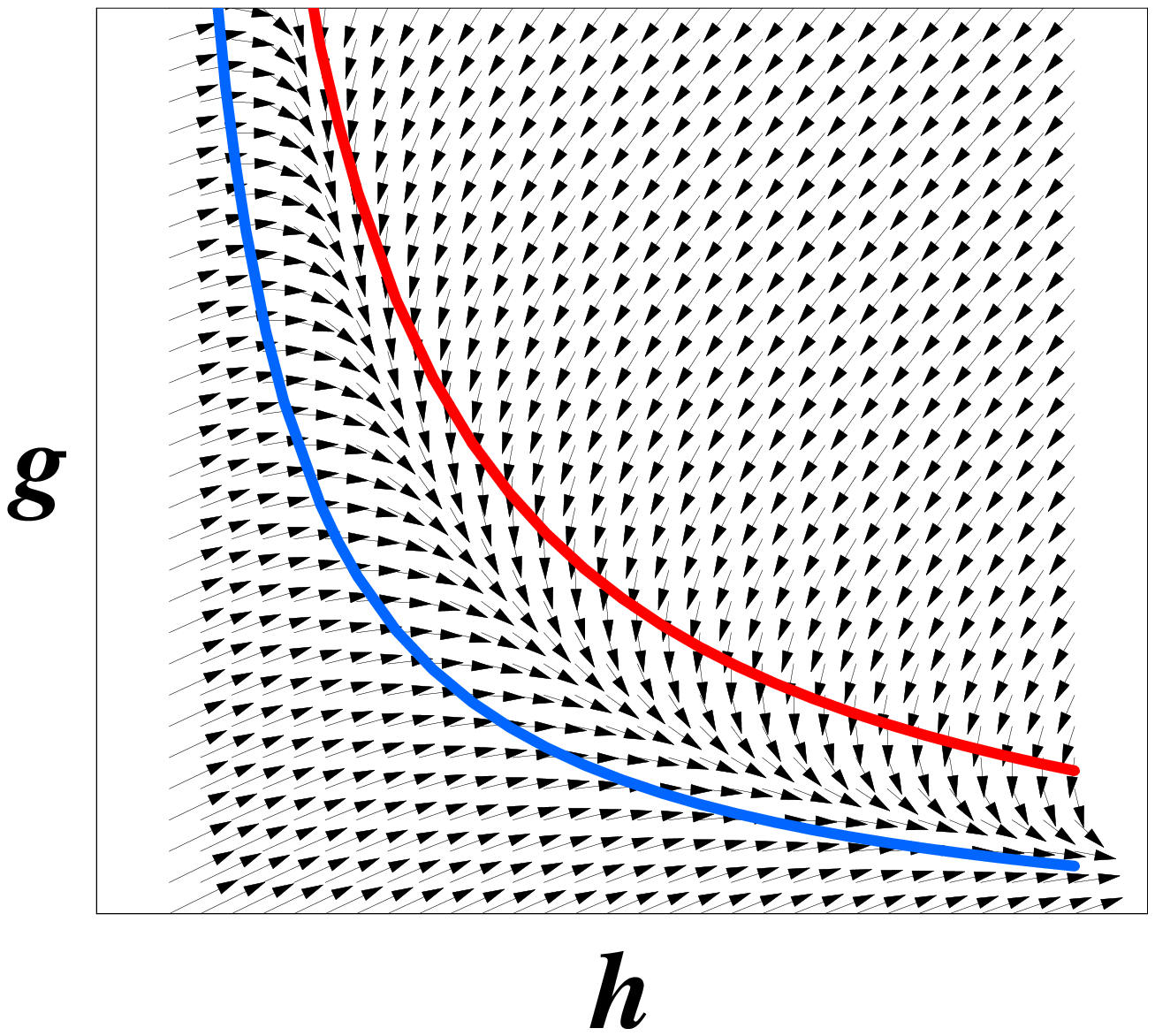} & \includegraphics[width=6cm ,height=6cm]{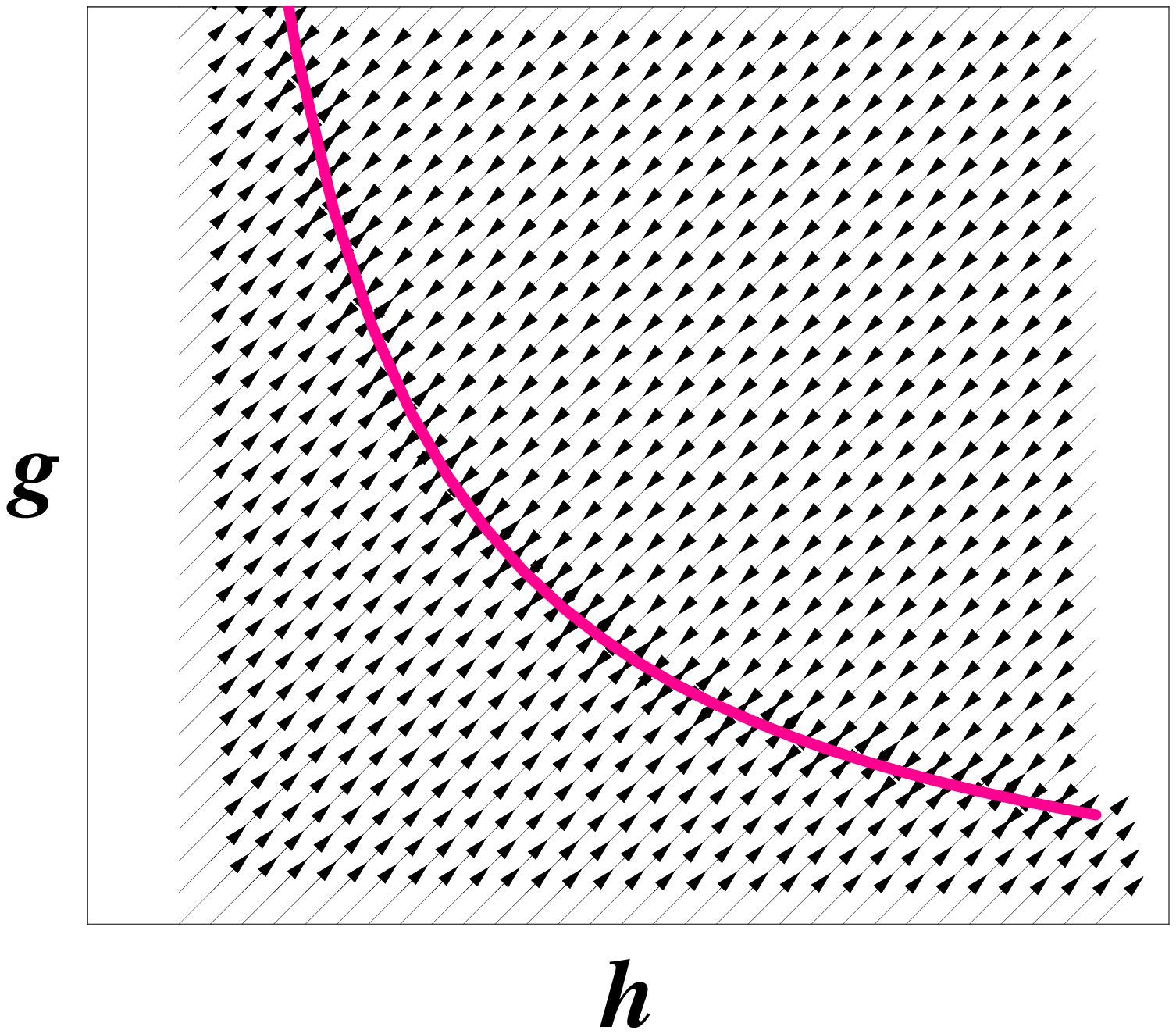} \\
(a) & (b)
\end{tabular}
\centering
\caption{\label{fig:RGflows1}\small We represent the flow in 
the $(p,q)$ plane at fixed $u$ and $\tau=0$ (type \bA solutions). The 
physical 
region is above the black lines $p=\pm q >0$. (a) The RG flow  for $N_c< N_f< 2N_c$. The $p'=0$ line (red) acts as an attractor for the flow, the dark lines are the limits of the attractor region $h'=0$ and $g'=0$. The $u'=0$ line (blue) has also been represented. (b) The RG flow for $N_f=2N_c$. The $p'=0$ (red) and $u'=0$ (blue) lines have been represented. There are two fixed points of the flow at the intersection of the two curves.}
\end{figure}

\underline{For solutions of type \bA} ($\tau=0$), when $N_f=2 N_c$ 
($x=2$) the $p'=0$ curve is a fixed line for 
the flow in the plane defined by constant $u$ 
(Fig.~\ref{fig:RGflows1}). The points where this 
curve intersects the curve $u'=0$ are fixed points 
of the three-dimensional system, and they correspond 
to the exact $N_f= 2 N_c$ solution found in 
\cite{Casero:2006pt,Casero:2007jj}. In terms of the original 
functions appearing in the metric $h=(p+q)/4$, $g=(p-q)/4$, 
the set of fixed points form a curve in the $(h,g,u)$ space that 
can be parametrized as
\begin{equation}\label{fixedline}
\left(h,g,u\right)=\left( h ,{h\over 4 h -1} , {4 h^2 \over 4 h-1}\right), \ \ \infty > h > 1/4\,.
\end{equation}
The fixed line only exists for $N_f=2 N_c$. In the notation of \cite{Casero:2006pt,Casero:2007jj}, the line of fixed points is parametrized by  $\xi=1/h$ (see eq.~(\ref{self-dual-1})).

We can easily extend the analysis for a smaller number of flavors $N_f\leq N_c$. There are two main effects as $x\to 0$. The first is the lift of the attractor region (\ref{RGattractors}) towards infinity, so it disappears completely when $x=0$, and then $p'>0$ on the $(p,q)$ plane. The second is the modification of the $u'=0$ curve. When $x=1$, the merging of the two branches of the low $\tau$ hyperbola occurs at $\tau=0$. Then, the $u'=0$ curve is a straight line $p=q+1$ on the physical region and also at the border $p=-q>0$. The region $p>q+1$ corresponds to $u'>0$, while $q+1>p>q>0$ corresponds to $u'<0$. When $x<1$, the $u'=0$, $\tau=0$ curve is a hyperbola passing through the points $(p=0,q=0)$ and $(p=x,q=0)$ and asymptoting $p=q+1$.

In Section \ref{sec:asymp} we have presented the possible asymptotic behavior of solutions as a function of the radial coordinate $\rho$. We will comment on them under the perspective of flows in the dynamical system defined by (\ref{RGBPSnonabelian}).

Let us start with the UV behavior $\rho\to \infty$. As we have already commented, both \bN flow towards the $\tau=0$ fixed plane, where \bA solutions live, so they have common asymptotics. There are two possible classes of asymptotics (see Section~\ref{UV-asymptotics}). 
\begin{itemize}
\item For class II all functions grow exponentially, so this corresponds to a situation where the flow is in the $u'>0$ region and the attractor $p'=0$ moves towards infinity. The attractor $q'=0$ will move towards positive or negative values of $q$ depending on the value of $N_f$. The flow will try to reach the intersection of both. This is a very generic situation.

\item Class I solutions, on the other hand, correspond to cases where the flow is in the $u'<0$ region, so the $p'=0$ attractor is dragged towards the lines $p=\pm q$. Generically, the flow would hit the lines and the solution will end at a finite value of the radial coordinate, so the corresponding solutions will have a boundary. Class I solutions are very special, because they correspond to a situation where $u'\to 0^-$ asymptotically, so they are just marginal solutions between class II solutions and solutions with boundary. In the special case of $N_f=2 N_c$ there is no flow in the $q$ direction, so the flow will hit the fixed line $p'=0$, $u'=0$. Notice that this is still a very special situation, usually the flow will miss the fixed line and go to a class II solution.
\end{itemize}
Let us examine now the IR behavior. Depending on the solution we can have $\rho\to 0$ (for either $\rho_0<0$ or $\rho_0=0$) or $\rho\to-\infty$. The first case corresponds to type II and III, while the last corresponds to type I (see Section~\ref{IR-asymptotics}). In terms of the dynamical flow, the meaning is the following
\begin{itemize}
\item Type I: It corresponds to flows coming from infinity in the $(p,q)$ plane, starting at the fixed subspace $u=0$. Notice that they are always in the $u'>0$ region for \bN solutions or for \bA solutions when $N_f>N_c$. However, when $N_f\leq N_c$, the $u'=0$ curve on the $(p,q)$ plane is different and can go to infinity. When $N_f= 2N_c$ it is also possible to have flow starting at the fixed line $p'=0$, $u'=0$, an example of this is the exact solution (\ref{self-dual-2}), that then runs into class II UV asymptotics. Notice that this is possible because the fixed line is not an attractor in all directions, it rather corresponds to a `saddle point' in the flow.

\item Type II and III: If $\rho_0<0$, they correspond to flows starting on the $p=\pm q$ lines, with type III corresponding to the point $p=q=0$. If $\rho_0=0$, then solutions have to start on the $q'=0$ line ($q=0$) at $\tau =\infty$.
\end{itemize}
A final comment is on the enlargement of the $u'<0$ region for $N_f\leq N_c$. Flows on that region are more likely to hit the $p=q$ line, thus corresponding to a solution with boundary. From the field theory perspective this may correspond to the runaway behavior of the $\cN=1$ SQCD theory when an ADS superpotential is generated. In the theories we consider here there is also a quartic superpotential, so the theory can show a different behavior.

\section{Physics of the new solutions}\label{Physics}
\setcounter{equation}{0}
In this Section we will study some aspects of the non-perturbative 
physics encoded by the new solutions we presented in Section 
\ref{newsolutions}. We will start with general comments, then 
we concentrate on aspects that involve the 
IR physics, hence using the IR (small $\rho$) expansions derived in 
Section \ref{newsolutions}, and then we will move into some observables computed 
with the UV expansions ($\rho\to\infty$). Some of the computations  we 
will perform to calculate 
non-perturbative effects, have been 
thoroughly checked in the past, using different backgrounds with or 
without flavors; in those cases, we will be brief in our presentation.
In contrast we will give details when new material is presented.

\subsection{General Comments on the Field Theory}
Let us first study an aspect that does not refer to the UV or the IR 
of the QFT, that is Seiberg duality \cite{Seiberg:1994pq}.

It is by now well understood that for the particular field theories in the 
class that we are considering see eq.~(\ref{ourtheory}), Seiberg duality is 
not just the infrared equivalence of two theories, but the duality is 
valid all along the flow. This peculiar behavior is due to the presence of 
the quartic term in the quarks. Basically the argument is that when 
Seiberg dualized,
\beq
W\sim \kappa (Q\tilde{Q)^2}= \kappa MM \to W_{dual}\sim \kappa 
MM+\frac{1}{\mu} \tilde{q}M q,
\eeq
where $M$ is the meson superfield and $q,\;\tilde{q}$ are the dual 
quarks. Now, the presence of the $M^2$ term-a mass term- allows us to 
integrate out the meson, leaving us with a superpotential of the form
\beq
W_{dual}\sim \frac{1}{\kappa}(q\tilde{q})^2,
\eeq
then, the theory is Seiberg dual to itself. This, by the way, is at the 
root of the working of the Klebanov-Strassler duality cascade. See the 
lectures 
\cite{Strassler:2005qs} for  lucid explanations.

How is the previous discussion reflected by our backgrounds?
This was discussed at length in \cite{Casero:2007jj}. Here, we can briefly 
mention that our generalized BPS eqs.~(\ref{BPS-eqs}) and our Hamiltonian 
eq.~(\ref{hamiltonian}) do indeed show
an interesting symmetry. Indeed,
\beq
P\to P,\;\; Q \to -Q,\;\; \tau \to \tau,\;\; Y \to Y,\;\;\sigma \to 
-\sigma,\;\;N_c \to N_f-N_c,
\label{seibergr}
\eeq
is an invariance of the equations, the Hamiltonian and the Hamilton-Jacobi 
principal function in eq.~(\ref{hamiltonjacobi}). This implies that
if we find one solution, we have by replacement in eq.~(\ref{seibergr}) 
found another one with the correct relations between color and flavor 
groups.

The $U(1)$ R-symmetry of the QFT is associated with translations in the 
angle $\psi$. The breaking of the symmetry will be briefly discussed 
later, 
but this point together with the matching of global anomalies does not 
present any subtlety aside from what was discussed in 
\cite{Casero:2007jj}.

As was proposed in the paper
\cite{Apreda:2001qb} the gaugino bilinear is related to the function 
$b(\r)$ that appears in the RR three form. We will keep that 
identification that will allow us to write an energy-radius relation (in 
the UV) 
\be
e^{\frac{2\rho}{3}}=\frac{\mu}{\Lambda} ,\;\;\;\; (UV,\;\rho\to\infty).
\eeq
Using the
information above, one can easily obtain that the solutions whose UV 
asymptotics is given in eqs.~(\ref{UV-II-A})-(\ref{UV-II-N})-(\ref{gggg}) 
represent the 
field theory described 
in Section \ref{stft}, once a dimension six operator is added. The 
addition 
of this irrelevant term in the Lagrangian 
(like in a related example of \cite{Intriligator:1999ai,Skenderis:2006di}) 
dramatically changes the UV of 
the field theory leading to a solution that is `away' from the near 
horizon of the D5 branes. We believe that this irrelevant operator
is related to the gauge sector, for instance
$O_6=(\W_\alpha \W^\alpha)^{3/2}\sim (F_{\mu\nu})^3$, or 
$O_6 \sim (D_\mu F_{\nu\rho})^2$,
though there could be also operators related to the KK multiplets like 
$O_6=|\Phi_k|^6$, etc. Our proposal that the irrelevant term is made out of 
operators transforming in the adjoint of the gauge group is due to the 
fact that the asymptotic 
eqs.~(\ref{UV-II-A})-(\ref{UV-II-N})-(\ref{gggg}) exist also 
in the case of $N_f=0$. The field theory aspects in the flavor-less 
case were studied and the solution was (numerically) found in 
Section 8 of \cite{Casero:2006pt}.

Let us proceed by studying aspects of the IR of the field theory, encoded 
in the backgrounds of Section \ref{newsolutions}.

\subsection{Physics in the Infrared}
As stated above, we will start by studying non-perturbative effects
that the backgrounds encode in the small radius region 
 ($\rho\to 0$)\footnote{For convenience we are taking the IR region to be 
at $\r \to 0$, 
but as explained above, in general it is at $\r \to \r_{IR}$}. We 
will also comment on Wilson, 't Hooft and dyon loops, domain walls and the behavior 
of the quartic coupling in the field theory described in 
eq.~(\ref{ourtheory}).  Finally we will give a general analysis to find what are the conditions 
that determine if an object is screened in the dual theory.

\subsubsection{Enhancement of the flavor group}\label{enhanced}
In this section we describe something that applies equally to the 
solutions discussed here and the ones in 
\cite{Casero:2006pt}
and \cite{Casero:2007jj}. We can ask the following question: what is the 
dual flavor group?
The first answer may be that the group is $SU(N_f)$, but a better 
analysis shows that actually the flavor branes are separated-this is an 
effect of the smearing-so, a more likely answer is that the group is 
$U(1)^{N_f}$. Many dynamical aspects will not depend on this and it may 
happen that if the distance between branes is vanishing (for example in 
the IR) then, the strings stretching between flavor branes become massless 
and the $SU(N_f)$ symmetry is recovered in that regime of energies.
The fact that the F1-strings stretching between flavor branes have a mass
has important consequences regarding the existence-or not- of diagrams 
that could likely correct the BI-WZ action that we used for the flavor 
branes. 

To analyze this, we will proceed as follows. We will compute the volume of 
the space $\Sigma_4$ on which we smear the flavor branes. We will divide 
this by $N_f$ and associate a given ``volume per flavor brane''. We will 
then propose that the distance between flavor branes is given by the 
fourth 
root of such `volume per flavor brane' and estimate the mass of the 
F1-strings as the string tension times this quantity.

We also need to estimate the proper energy $E_{proper}$ of an object in the bulk, given  
the energy of the dual object in the field theory $E_{QFT}$. The relation is
\beq\label{eq:propenergy}
E_{QFT}=\sqrt{g_{tt}} E_{proper}=e^{\phi/2}E_{proper}.
\eeq
Due to the dependence of the dilaton on the radial position, if we keep 
the QFT energy fixed, then the proper energy of the mode must vary. If the proper energy grows very fast in the IR, 
it could be possible to excite strings between different branes even if they are at a finite separation, leading to an enhancement of 
the flavor group to $SU(N_f)$.

Let us see the computation.
We focus first on type {\bf A} solutions. In the string frame, the induced 
metric on the four cycle $\Sigma_4$ is,
\beq
ds_{\Sigma_4}^2=e^{\phi}\Big[H(\rho)
(d\theta^2 + \sin^2\theta d\varphi^2)
+G(\rho) (d\tilde\theta^2 +\sin^2\tilde\theta
d\tilde\varphi^2) + Y(\r)(\cos\theta 
d\varphi+\cos\tilde{\theta}d\tilde{\varphi})^2\Big].
\label{sigma4A}
\eeq
So, the volume to be computed is
\bea
& & V_4=\int d\theta d\tilde{\theta}d\varphi d\tilde{\varphi}  
\sqrt{g_{\Sigma_4}}=\nonumber\\
& & 4\pi^2 e^{2\phi}\sqrt{HG}\int d\theta d\tilde{\theta} 
\sqrt{HG\sin^2\theta\sin^2\tilde{\theta} +YG\cos^2\theta 
\sin^2\tilde{\theta} + HY \cos^2\tilde{\theta}\sin^2{\theta}}.
\label{volsigma4}
\eea
So, we need to look for the values of this integrals in the IR. Indeed, in 
the UV, is clear from the asymptotic expansions, that this integral will 
indeed diverge, making the mass of the F1s to diverge. The group is then 
broken to $U(1)^{N_f}$ in the UV.
In the IR, the analysis depends on the type of solution we consider. Let 
us first analyze this for the three types of IR type {\bf A} solutions of 
the $SU(N_c)$ dual background in \cite{Casero:2007jj}. They were called 
type 
I, II, and III IR behaviors. In type I and II, the previous volume is 
constant, indicating that there will be a fixed separation between 
flavor branes. In the type II case, the dilaton goes to a constant, so at very low energies, the strings between 
branes will decouple, breaking the flavor group to $U(1)^{N_f}$. 
In type I the dilaton vanishes, so the proper energy grows unbounded in the IR 
and the group gets enhanced to $SU(N_f)$. In type 
III, we observe that the volume vanishes, also enhancing the flavor group.

A similar analysis can be done for the 
type {\bf N} solutions, using the determinant of the induced metric 
in the far IR.
 Given the expansion for type {\bf N} solutions in
Section \ref{newsolutions}, 
we obtain that for type I and II the volume diverges at
$\rho=0$ and the dilaton is constant. This implies that non-diagonal strings are very massive and 
 do not contribute to processes in the IR physics. The 
flavor group is then broken to $U(1)^{N_f}$ by the solutions.
This is basically the reason why in the Introduction we proposed different 
couplings between quark superfields and adjoint KK modes see 
eq.~(\ref{oooo})-the difference may be just a phase factor for different 
positions of branes in $\Sigma_4$. In type III solutions on the other 
hand, the volume 
vanishes in the IR and the flavor group gets enhanced to $SU(N_f)$.
\subsubsection{String-like objects}\label{sec:stringy}
It is known that the holographic dual 
to the Wilson loop corresponds to a fundamental string describing 
the 4d Wilson loop in the Minkowski directions, 
exploring the radial directions and ending on a brane at spatial infinity 
\cite{Maldacena:1998im}, 
while in the type of backgrounds that are not S-duality invariant, the 
't Hooft loop is computed by 
a D3 brane wrapping a topologically trivial two-cycle. These 
string-like
objects should have only a finite length due to the 
screening from fundamental matter. 

In the case of type {\bf A} backgrounds, it was shown that depending
on the IR of the solution, different vacua of the QFT present distinctive 
behavior for Wilson and 't Hooft loops, see Sections 4.4 and 4.5 of the 
paper \cite{Casero:2007jj} for a detailed analysis.
Indeed (always in the case of type {\bf A} backgrounds) it was seen that in
type I solutions the Wilson loop tension 
vanishes, 
while it is finite for type II and III. On the other hand, the 
tension of the 't Hooft loop vanishes only for type III solutions. 
Therefore, type I solutions correspond to a Higgs phase, 
while type III correspond to confined phases. 
In type II both quarks and monopoles are confined, 
so it should correspond to a oblique confined phase, 
where dyons with both electric and magnetic charge should be free. 
\paragraph{Wilson loops}
In a QFT with fundamental matter, the QCD-string can break, basically due 
to pair creation. This effect is offset if one is working in the quenched 
approximation ($\frac{N_f}{N_c}\sim 0$). Our backgrounds do indeed capture 
screening effects as was seen in \cite{Casero:2006pt,Casero:2007jj}, by the existence of maximal length for the 
quark-antiquark separation. Nevertheless one can get an {\it intuitive} 
idea about the way a quark pair interacts by looking at the QCD string 
tension in the far IR. This is computed in the string duals by the value 
of the quantity (see  \cite{Sonnenschein:1999if} for a summary)
\beq
T_{QCD}= \sqrt{g_{tt}g_{xx}}|_{\rho_{IR}}.
\eeq
We can see that this quantity is non-zero for the three types of IR 
asymptotics of type {\bf N} backgrounds. Indeed, it is just proportional 
to the value of $e^{\phi(\rho_{IR})}$; this indicates that at low 
energies, the pair feels a linear potential\footnote{Of course, at even 
longer distances, screening takes place and the potential changes.}. 
Contrast this with the type {\bf A} backgrounds where for asymptotics of 
type I the force was vanishing. The main difference between type {\bf A} 
and type {\bf N} backgrounds resides in the existence of the functions 
$a(\rho),\; b(\rho)$ that as was discussed in the literature 
\cite{Apreda:2001qb}, can be 
associated with the formation of a gaugino condensate.
This seems to be at the root of the different behavior for the QCD string 
tension. 

Let us now study another gauge invariant observable, the 't Hooft loop.

\paragraph{'t Hooft loops}
As discussed previously in the literature, for backgrounds that are not 
invariant under S-duality like the present ones, the 't Hooft loop is 
computed via a D3 brane that wraps a two cycle in the internal space.
The manifold on which  the D3 is proposed to extend is,
\beq
\Sigma_4=[x,\; t,\;\theta=\tilde{\theta},\;\varphi=2\pi-\tilde{\varphi},\; 
\psi=\pi,\; \rho(x)].
\eeq
The induced metric for a D3 on that manifold is
\beq
ds_{ind}^2= e^{\phi}\Big[-dt^2 +dx^2(1+e^{2k}\rho'^2) 
+(e^{2h} +\frac{e^{2g}}{4}(a-1)^2)(d\theta^2+\sin^2\theta d\varphi^2)   
\Big].
\eeq
After integrating over the internal manifold, the `effective string' 
(representing the 't Hooft loop) has an action,
\beq
S_{eff}=4\pi T_{D3}\int dx e^{\phi}(e^{2h} +\frac{e^{2g}}{4}(a-1)^2) 
\sqrt{(1+e^{2k}\rho'^2)}= \int dx T_{eff}\sqrt{(1+e^{2k}\rho'^2)}.
\eeq
This effective tension $T_{eff}$ plays a similar role to that of the QCD-string 
tension in the previous subsection. This object can be studied in the  
IR (again, keeping in mind that the 't Hooft loop will also present 
screening effects at even lower energies) to get an intuitive idea of 
the force, at low energies, between a pair of monopoles.

We obtain using the type {\bf N} IR expansions that for type I and type 
III backgrounds the effective tension vanishes, leaving the monopoles 
free. In contrast, for type II backgrounds the effective tension is 
constant, leading to a linear potential between them.

It is a natural next step, to compute the tension between a pair of 
dyons. We are 
not aware of a concrete computation to calculate this effect, then we will 
propose one.
So, we propose that the dyon-anti-dyon 
is represented by a D3 brane extended on $(t,x)$ wrapping the 
topologically trivial cycle, 
\beq
\theta=\tilde{\theta},\;\;\;\varphi=2\pi-\tilde{\varphi}\,,
\label{cycle}
\eeq
and that is charged by the presence 
of a world-volume electric field $F_{tx}$. We will show that for a 
particular 
value of the electric charge, the tension of the dyon loop 
indeed vanishes, in agreement with the oblique confinement picture. We 
will do this in the case of the simpler type {\bf A} backgrounds for 
illustrative purposes.

\paragraph{Computing dyon-antidyon loops}\label{sec:dyons}
Since, as stated above, we believe this is new material, we will be more 
detailed.  We  study the 
proposal in type {\bf A} and then (briefly) in type {\bf N} 
backgrounds.
In type {\bf A} configurations, the induced 
metric, RR three form  and 
world-volume fields on the D3 brane 
described  around eq. (\ref{cycle}) is (the brane 
also extends in the radial direction $\rho(x)$),
\bea
& & ds_{ind}^2= e^{\phi}\Big[ -dt^2+(1+4Y\rho'^2)dx^2 +(H+G)(d\theta^2
+\sin^2\theta
d\varphi^2) \Big],\nonumber\\
& & F_{tx}=\partial_t A_x, \nonumber\\
& & F_3=d\Big[\frac{(2N_c-N_f)}{4}(\psi-\psi_0)\sin\theta d\theta\wedge
d\varphi  \Big].
\label{d3configuration}
\eea
We then compute the Born-Infeld part of the action,
\beq
\det[g_{ab}+2\pi F_{ab}]^{1/2}=e^{2\phi}(H+G)\sin\theta
\sqrt{1+4Y\rho'^2 -4\pi^2 e^{-2\phi}F_{tx}^2}.
\eeq
After integrating over the sphere
directions and considering 
the WZ term that appears because of the
presence of a nontrivial $C_2$ on the worldvolume, the action for the D3 
brane is, 
\beq
S=-4\pi T_{D3} \Big[\int dx e^{\phi}(H+G)\sqrt{1+4Y\rho'^2 -4\pi^2
e^{-2\phi}F_{tx}^2} -\Theta F_{tx}   \Big],
\label{actiond3}
\eeq
where $\Theta$ is given by $
\Theta=\frac{2N_c-N_f}{4}(\psi-\psi_0)
$.
We then compute the eq. of motion for the gauge field, obtaining
\beq
F_{tx}=-\frac{e^{\phi}(\Theta-\Theta_0) \sqrt{1+4 Y \rho'^2}}{2\pi
\sqrt{4\pi^2 (H+G)^2
+(\Theta-\Theta_0)^2}},
\eeq
where $\Theta_0$ is an integration constant. 
Replacing back into the action (\ref{actiond3}), we get
\beq
S=-4\pi \int dx \; T_{eff}\sqrt{1+4Y\rho'^2},
\eeq
which is the action for an `effective string' that is what we consider to
be the QCD string (for dyons), with tension given by
\beq
T_{eff}=2\pi T_{D3} e^{\phi}\frac{\Big[ (H+G)^2 +\Theta(\Theta-\Theta_0)
\Big]}{\sqrt{4\pi^2 (H+G)^2+(\Theta-\Theta_0)^2}}.
\eeq
We can see that the dyon-anti-dyon tension depends on the type of 
solutions we
deal with. Always thinking in the case of type {\bf A} configurations and 
using the asymptotic expansions derived in \cite{Casero:2007jj}, we 
see 
that for type I the tension is 
always finite, so the dyons are confined. 
For type III the tension vanishes only when 
$\Theta_0=\Theta$ (which means that $F_{tx}=0$) 
or $\Theta=0$. In both cases,
the dyon has no electric charge and becomes a monopole. 
For type II theories, in the IR $H+G\to |C|$ (constant). 
If we choose $\Theta_0$ conveniently, then the 
tension of the dyon vanishes. So, for each of the type 
II solutions and its Seiberg dual, 
a dyon with a precise value of the electric charge is free. 
Notice that a shift in $\Theta$ changes 
the value of the electric charge of the free dyon, 
so the oblique confined phase is different, 
this is what we expect from field theory.

Having finished with the possible string-like objects (to which we will 
come back for a more general analysis in Section 
\ref{sec:break}), we 
would like to 
study now higher dimensional defects.  

\subsubsection{Domain-wall tensions}
A domain wall in these backgrounds is represented by a D5 brane that wraps 
(in a SUSY preserving way) a three-cycle and extends in three of the four 
Minkowski directions, 
preserving $SO(1,2)$. The proposed manifold on which the five brane 
extends was
given in \cite{Casero:2006pt} is
\be
\Sigma_6=[t,x_1,x_2,\theta,\varphi,\psi].
\ee
The effective tension of this (2+1)-dimensional object-that is after 
integration over the internal three manifold- is
\beq
T_{eff,DW}=16\pi^2
e^{2\phi+k}(e^{2h}+ \frac{e^{2g} a^2}{4}).
\eeq
The IR behavior of this quantity is such that the effective tension is 
constant for type I and type III solutions, while is divergent for type II 
asymptotics. This means (in the case of type II solutions) that the 
domain-wall like object has to sit at 
a finite value of the radial coordinate, where its tension is minimized.
We move now to the study of our last IR observable, the value of the 
quartic coupling in the different asymptotics.

\subsubsection{Quartic coupling}
In the paper \cite{Casero:2007jj} a possible definition for the quartic 
coupling $\kappa$ between quark superfields in the QFT has been proposed.
The definition is just based on dimensional analysis and has the right UV beta function,
but does not transform correctly under Seiberg duality. We would like to propose
here a different definition with the right properties
\beq\label{eq:quartic}
\log \kappa^2 =\frac{Vol[\tilde S^3]-Vol[S^3]}{Vol[S^1]\times Vol[S^2]}=-\frac{e^{2 h}+\frac{e^{2 g}}{4}(a^2-1) }{e^{2 h}+\frac{e^{2 g}}{4}(a-1)^2}= -\frac{Q }{P}e^\tau
\eeq
For type \bA backgrounds $\log \kappa^2=(G-H)/(G+H)$. Seiberg duality interchanges the two three spheres in the metric, that in (\ref{eq:quartic}) is seen as $\kappa\to \kappa'=1/\kappa$. In self-dual configurations $H=G$, $\kappa=\kappa'=1$. Let us stress again that this choice is based on some general properties but we do not have a proof of it being the actual quartic coupling of the dual field theory.

Analyzing the behavior of this quantity in the case of type \bA and {\bf N} 
backgrounds, we see that $\kappa$ always goes to a constant value in the IR or UV. In type \bA III, \bN I and \bN III infrareds and in class II ultraviolets, it flows to the self-dual value $\kappa=1$.

\subsubsection{Screening as string breaking}\label{sec:break}
In the papers \cite{Casero:2006pt}
and \cite{Casero:2007jj} it was explicitly shown (using the dual 
backgrounds)
that due to the presence of the fundamental matter,
different `loops' (Wilson, 't Hooft, dyon)
may present a maximal length, indicating screening.
In this Section, we would like to present a general study whose outcome 
is a {\it sufficient } condition for which a finite maximal length will 
occur (or not). Our analysis will not depend on a particular asymptotics.
Let us emphasize that, while the existence of a maximal length
implies screening, the existence of an infinite length does not imply 
confining behavior. We are considering the ``connected'' Wilson loop and 
its energy must be compared with the possible ``disconnected'' solution 
indicating creation of pairs and screening-for another example where the 
same comment applies see \cite{Bigazzi:2008gd}. The energetically 
favorable solution will be physically realized.

Let us then  examine the issue of the maximal length of string-like 
objects 
described by a string/brane extending on $t$, $x$ 
and wrapping the internal space. 
The bulk radial coordinate is $\rho$, and the profile 
of this object is described by the function $\rho(x)$.
If the metric is
\beq
d s^2 = e^{\phi(\rho)}
(- dt^2 + d{\bf x}^2 + 4 Y(\rho) d\rho^2 )+ 
g_{\alpha\beta}(\rho,y)d y^\alpha dy^\beta,
\eeq
with $g_{\alpha \beta}$ the metric of the internal 
space, then the Lagrangian coming from the Born-Infeld-Wess-Zumino 
action, after integration over the internal directions $y^\alpha$, will
have the general 
form,
\beq
{\cal L} = T(\rho) \sqrt{1+4 Y \rho'^2},
\eeq
where $T(\rho)$ is the effective tension of the brane that depends 
in general on the dilaton and the components of the metric in the 
directions of the internal space that 
the brane is wrapping. The usual analysis goes as follows:
we first define the Hamiltonian, that  is a conserved quantity
\beq
{\cal H}={\partial {\cal L}\over \partial \rho'} 
\rho' - {\cal L} = -{T \over \sqrt{1+4 Y \rho'^2}} \equiv - T_0.
\label{zzzz}
\eeq
If for some value of 
$\rho=\rho_0$ \footnote{Notice that in this Section $\r_0$ 
denotes the turning point of the string-like object and should not be 
confused with the quantity defined earlier 
$\r_o$ that is the lowest value of the coordinate $\r$ in the background. 
The context will make this clear.} the profile ends smoothly 
$\rho'=0$, then $T_0 =T(\rho_0)$. 
From this expression, it is straightforward to find
\beq\label{eq:dxdr}
{d x\over d \rho}= {2 \sqrt{Y} T_0\over \sqrt{T^2-T_0^2}}.
\eeq
In order to have a sensible (real-valued solution), 
we need that $T >T_0$ over all the brane 
(we are assuming that all quantities are positive). So,
\begin{itemize}
\item If the tension is a growing function of the radial position $T'(\rho)\geq 0^+$ as $\rho\to \rho_0$, then the solution exist for $\rho>\rho_0$.
\item If the tension is a decreasing function of the radial position $T'(\rho) \leq 0^-$ as $\rho\to \rho_0$, 
then the solution exist for $\rho<\rho_0$.
\end{itemize}
Physically, this means that the object 
extends towards values where its tension is smaller. In 
the backgrounds we are considering we have 
usually dilaton factors that grow exponentially in the 
tension as $\rho\to \infty$, 
hence the branes will `hang' from the UV towards the IR. 
However, close to the singularity, 
some functions of the metric grow and some functions decrease, 
so in principle there could be objects that cannot go beyond 
some finite value of the radial coordinate, and there 
could be objects that extend 
from the singularity towards the UV, although they cannot go arbitrarily far. 
When this happens, the objects will have a finite tension, the 
region where $T'=0$ acting as an effective IR wall for them. 
Let us analyze these cases in detail.

The simpler case is when there is no 
IR wall and the far IR is at $\rho= 0$, 
so we can use the asymptotic expansions. Let us 
assume that the tension of the object vanishes 
in the IR as some power $T\sim \rho^b$, $b>0$. Also, we assume
that the function $Y$ also has some power-like 
behavior, but can vanish, be a constant 
or blow up like $Y\sim \rho^a$. Then, the equation 
of motion in eq.~(\ref{zzzz}) 
reads,
\beq
{d x\over d \rho}\sim {\rho^{a/2} \rho_0^b\over \sqrt{\rho^{2 b}-\rho_0^{2 b}}}.
\eeq
Close 
to the tip, $\rho \simeq \rho_0(1+\epsilon)$. 
We can integrate trivially on $\epsilon$ 
to find a square root behavior, so the profile at the tip looks like
\beq
x(\rho)\sim \rho_0^{(1+a)/2} \sqrt{\rho-\rho_0}.
\eeq
The criterion we will adopt to quickly evaluate the `length' of the 
putative loop is the following: the 
factor in front of the square root determines how ``open'' is the 
configuration. If the configuration is to reach infinite length 
in the $x$ direction when the object explores the 
far IR of the dual theory, then this factor 
should diverge when $\rho_0\to 0$. 
In the case at hand, this implies that $a<-1$. This is indeed the case of 
the AdS string, where 
$T\sim \rho^2$ and $Y\sim \rho^{-4}$.

The next possibility is to consider an object that has finite 
tension, so that $T\sim T_*+c\rho^b$, $b>0$, $c>0$ 
(this is necessary to have $T'>0$) when $\rho\to 0$. 
Now the equation of motion is a bit different
\beq
{d x\over d \rho}\sim {T_* \rho^{a/2} \over \sqrt{(T_*+c\rho^b)^2-(T_*+c \rho_0^b)^2}} \sim {T_*^{1/2} \rho^{a/2} \over \sqrt{\rho^b-\rho_0^b}}.
\eeq
The profile at the tip now looks like
\beq
x(\rho)\sim T_*^{1/2} \rho_0^{(1+a-b)/2} \sqrt{\rho-\rho_0}.
\eeq
For example, we can analyze the previous expression for the case of the 
flavorless solution ($N_f=0$) in eq.~(\ref{nf=0solu}), and check that this 
gives 
$a=0,b=2$ implying an infinite length. As another example, we analyze 
the IR of type {\bf A} backgrounds. In 
order to have an object that can extend to infinite length, we should 
have that $b>1/2$ in type II and $b>1/3$ in type III. 
Notice that the monopoles and dyons have 
$b=1/2$ and $b=1/3$, so they are screened objects.

Suppose now, that we have an object that feels an effective IR wall at 
$\rho=\rho_*$ where the tension is minimized, so $T'(\rho_*)=0$ 
and $T''(\rho_*)>0$. We are assuming that the background 
at this point is smooth, and the functions of the metric 
do not vanish, so $Y$ and the tension admits Taylor expansions
\bea
&& T=T_* +T_2 (\rho-\rho_*)^2+\dots, \quad  T_2 >0, \NO\\
&& Y=Y_*+Y_1 (\rho-\rho_*) +\dots.
\eea
We take a configuration where the object ends smoothly 
at some point $\rho_0>\rho_*$ and consider the 
limit $\rho_0\to \rho_*$. It is 
convenient to use the coordinate $r=\rho-\rho_*$ and $r_0=\rho_0-\rho_*$ 
so eq.~(\ref{eq:dxdr}) becomes,
\beq
{d x\over d r} \sim {\sqrt{Y_*} T_*\over \sqrt{(T_*+T_2 r^2)^2-(T_*+T_2 r_0^2)^2}} \sim \sqrt{Y_* T_* \over T_2}{1\over \sqrt{r^2-r_0^2}}.
\eeq
Close to the tip, $r \simeq r_0(1+\epsilon)$. We 
can integrate trivially on $\epsilon$ to find 
a square root behavior, so the profile at the tip looks like
\beq
x \sim \sqrt{Y_* T_* \over T_2} r_0^{-1/2} \sqrt{r-r_0}.
\eeq
Therefore, when the tip of the object comes close 
to the IR wall $r_0\to 0$, the length diverges.

It should be interesting to extend the analysis of this Section 
to the case in which we have 
massive quarks; because it could provide more 
hints on the first order phase transition-in terms of the mass- that 
occurs for the connected part of the Wilson loop, see 
\cite{Bigazzi:2008gd}.
Also, it may be interesting to use the results of this Section to
improve models of AdS/QCD to incorporate screening effects.
To close, let us mention that it should be nice to provide
a link between the formalism developed here and the ideas presented 
in \cite{Armoni:2008jy}.

Let us now move to study aspects of the UV of the field theory, using the 
large $\rho$ region of the solutions.
\subsection{Physics in the Ultraviolet}
Let us briefly study how some aspects of the UV field theory are captured 
by the backgrounds. R-symmetry anomalies will be 
treated in a sketchy way. The results are similar to those in 
\cite{Casero:2006pt} and \cite{Casero:2007jj}.
The study of beta functions will be more detailed because it will 
shed light on a long-standing puzzle.

The way the R-symmetry anomaly will work, as explained at length in 
\cite{Casero:2006pt}
and \cite{Casero:2007jj} does not change here. Basically this is because 
it only depends on the form of the RR field $F_3$ when restricted to a 
particular manifold-see eq.~(\ref{cycle})-considered at large values of the 
radial coordinate. 
The result is not changed when switching between  the different UV 
solutions we found in Section \ref{newsolutions}. 

Let us 
then concentrate on the beta function of the field theory as computed by 
the two different UV expansions. We will present details, as this 
will teach us something 
interesting about the field theory and its UV completion.

\subsubsection{Beta function}
A definition of the gauge coupling in the dual 
QFT in terms of type IIB objects was given in 
\cite{DiVecchia:2002ks}. Using that definition, we have an expression for 
the gauge coupling that reads,
\beq\label{gauge-coupling}
\frac{8\pi^2}{g^2}=2(e^{2h} +\frac{e^{2g}}{4}(a-1)^2)=e^{-\t}P.
\eeq
As mentioned above, the relation between the scale of the QFT and the 
radial coordinate was also
proposed in \cite{Apreda:2001qb}
and  \cite{DiVecchia:2002ks}. This relation 
comes by associating 
the functions $a(\rho)$ and/or $b(\rho)$ with the VEV of the gaugino 
condensate. One gets, 
\beq
e^{\frac{2\rho}{3}}=\frac{\mu}{\Lambda} ,\;\;\;\; (UV,\;\rho\to\infty).
\label{uvir}
\eeq
We have found two types of UV behavior. One of them was 
already studied in \cite{Casero:2006pt}
and \cite{Casero:2007jj}. The other, for the solutions that asymptote to 
the conifold, was given in 
eqs.~(\ref{UV-II-A})-(\ref{UV-II-N})-(\ref{gggg}). For both 
asymptotics, the relation in eq.~(\ref{uvir}) above holds.

This allows us to compute the behavior of the beta function in the UV of 
the QFT.
The results are that for the UV expansions presented in 
\cite{Casero:2006pt} and \cite{Casero:2007jj} we get (for $N_f<2N_c$ as 
explained 
in those papers) 
that
\beq
\beta_{\frac{8\pi^2}{g^2}}=\frac{3}{2}[2N_c -N_f],
\eeq
that is the result expected from the NSVZ beta function, in the case in 
which the anomalous dimension of the quark superfield is 
$\gamma_Q=-\frac{1}{2}$. See \cite{Casero:2007jj} for various consistency 
checks on 
this value of the anomalous dimension.

Now, this is the puzzle: how is it possible that
a theory with so many extra fields (in the UV, the extra modes should be 
integrated-in) presents a beta function that  adjusts so well to the 
exact treatment of NSVZ. The comparison with the result obtained 
using the other possible UV solution will illuminate this point. 
This also applies to the case $N_f=0$ and 
we believe it answers the criticism to the papers 
\cite{DiVecchia:2002ks}.

\paragraph{The beta function for a 6d theory}
The set-ups
described in the Introduction 
are the holographic duals of the six-dimensional 
field theories living on the D5 branes that 
wrap a 2-cycle of the deformed conifold. 
The dimensionful gauge coupling can be read from the 
D5 brane action $g_6^2= \alpha' g_s$, and one 
can define the dimensionless coupling
\beq
\tilde{g_6}^2 =  \mu^2 g_6^2 Z^2(\mu)\,,
\eeq
where $\mu$ is some mass scale and 
$Z(\mu)$ is the renormalization factor 
of the 6d coupling, analogous to a mass renormalization factor. 
The theory is perturbatively non-renormalizable, 
in the sense that radiative corrections 
can generate an infinite set of irrelevant operators 
as the theory flows to the UV. From a four-dimensional perspective, 
we can define a (dimensionless) coupling
\beq
g_4^2 = {1\over L^2} g_6^2 Z^2(\mu)\,,
\eeq
where $L^2\sim N_c \alpha'$ is the size of the 2-cycle. 
In this case, the beta function is
\beq
\beta_{\frac{8\pi^2}{ g_4^2}}=-\frac{16\pi^2}{g_4^2} 
\frac{d \log Z}{d\log \mu}\equiv  -\frac{16\pi^2}{g_4^2} \gamma_{\tilde g}\,,
\eeq
At one-loop in perturbation theory, the 
infinite tower of extra states (KK modes) would generate 
a power-like running of the coupling in the UV
\beq
g_4^2 \sim {1 \over (L\mu)^2}\,.
\eeq
If we integrate out the KK modes, then an effective coupling for 
an irrelevant operator of dimension 
six would be generated, also modifying the running of 
$g_4$. As the energy scale becomes of the order of the Kaluza-Klein mass, more 
irrelevant operators would enter, making the theory non-renormalizable.

However, a six dimensional theory could be renormalizable at a 
non-perturbative level. This is possible if there is a 
UV fixed point, as was  shown 
in \cite{Seiberg:1996qx}. 
Looking at the beta function (see Kazakov's paper in 
ref.~\cite{Seiberg:1996qx})
\beq
\beta_{\frac{8\pi^2}{\tilde g_6^2}}=
-\frac{16\pi^2}{\tilde g_6^2} (1+\gamma_{\tilde g})\,,
\eeq
we see that a UV fixed point corresponds to an 
anomalous dimension $\gamma_{\tilde g}=-1$. 
On the other hand, the four dimensional coupling would have 
the same running as the one-loop perturbative result, 
suggesting that only a dimension six operator is generated, 
instead of an infinite set of irrelevant operators. We can take this 
as an indication that the UV theory is truly six dimensional, 
where such operator is marginal. 
It has been suggested that the marginal operator respects 
scale invariance but not the full conformal invariance of the theory. 

We can now see how this picture emerges in the holographic description.
Using the prescription of \cite{DiVecchia:2002ks}, 
we can see that the renormalization factor 
$Z^2$ is proportional to the inverse size of the cycle. 
In the UV, the relation between the radial 
coordinate $\rho$ and the scale $\mu$ is given as explained above by 
eq.~(\ref{uvir}). Then, for the solutions asymptoting to 
the conifold,  eqs.~(\ref{UV-II-A})-(\ref{UV-II-N})-(\ref{gggg}), 
we have
\beq
\log Z=-\frac{2}{3}\rho = -\log \frac{\mu}{\Lambda} \ \Rightarrow \ \gamma_{\tilde g}=-1\,.
\eeq
So the growing solutions correspond to 
six dimensional theories flowing towards a UV fixed point. 
Let us examine the metric in the far UV, keeping 
only the leading asymptotics and using the coordinate $2r= 3e^{2\rho/3}$, 
the background in eq.~(\ref{nonabmetric424}) with the asymptotics 
(\ref{UV-II-A})-(\ref{UV-II-N})-(\ref{gggg}) will read,
\beq
ds^2= e^{\phi_0}\big[dx_{1,3}^2  + \frac{4c_+}{3}(dr^2 + r^2 T^{1,1})
\big]
\label{minkconi}
\eeq
There is a scaling symmetry; we can rescale the coordinates
$
r\to \lambda r, \ \ x^\mu \to \lambda x^\mu\,.
$
This changes the metric only by a conformal 
factor, that can be absorbed in the dilaton. 
Notice that the RG flow is independent of the value of the 
dilaton. There is also a self-similarity relation 
between different RG flows in the UV, since the transformation
$r\to \lambda r,\;\;\; c_+\to\lambda^{-2}c_+\,,
$ 
leaves the metric invariant.

The special solutions with UV given in \cite{Casero:2007jj} -see 
eqs.~(\ref{UV-e})-(\ref{UV-c})- have a 
rather different behavior. The renormalization factor goes as
\beq
\log Z\sim -\frac{1}{2}
\log \rho \sim -\frac{1}{2}\log\log \frac{\mu}{\Lambda}\,.
\eeq
This gives the anomalous dimensions and gauge couplings
\beq
\gamma_{\tilde g}= -{1\over 2 \log \frac{\mu}{\Lambda}}\ \Rightarrow \ g_4^2 \sim {1\over \log\frac{\mu}{\Lambda}}, \ \ {\tilde g}_6^2 \sim\, {\mu^2 \over \log\frac{\mu}{\Lambda}}\,.
\eeq
The running of the $g_4$ coupling 
is exactly the one of a four dimensional 
${\cal N}=1$ theory. This is quite surprising, 
since a six dimensional theory is showing a 
completely four dimensional behavior. 
From the string theory point of view, 
the six-dimensional solutions (\ref{minkconi}) 
look as a decompactification limit and actually 
the metric coincides with the metric `outside' 
the near-horizon region of the 5-branes 
\footnote{As predicted to happen in a  different context in the papers 
\cite{Intriligator:1999ai,Skenderis:2006di}}.
Notice that the dilaton goes to a constant, so the energy 
of probes in the gravitational background is simply proportional 
to the energy in the field theory-see eq.~(\ref{eq:propenergy})- 
and the extra dimensions can be easily explored.

On the other hand, in the `four dimensional solutions' 
eqs.~(\ref{UV-e})-(\ref{UV-c}) 
the dilaton factor in the proper energy formula (\ref{eq:propenergy}) grows exponentially with the radial coordinate, 
while the relative size of the compact dimensions grows only linearly. 
Then, for an object with fixed energy in the field theory, 
the proper energy in the string dual is exponentially 
suppressed at larger values of the radial coordinate 
in such a way that the wavelength will grow much 
faster than the size of the compact dimensions. Therefore, 
it is very difficult to explore the extra dimensions 
and the theory remains four-dimensional even in the ultraviolet. 

In the dual field theory, the interpretation is that these 
backgrounds correspond to RG flows where the six dimensional irrelevant 
operator is not generated, so they are renormalizable 
not only in the six dimensional sense but also in a four dimensional sense.
All these comments are consistent and provide a different view on the 
findings in 
\cite{Gursoy:2005cn}, 
where it was shown 
that the beta function of the unflavored background was unaffected by the 
KK modes. For a 
field theory argument that complements this one, see Appendix 
\ref{appendixsup}.

Let us now analyze another contribution of this paper. The extension 
of these set-ups to the case of orthogonal groups.

\section{The case of $SO(N_c)$ gauge group}\label{sons}
\setcounter{equation}{0}
In this Section we address some aspects of the version 
of SQCD described in Section \ref{qft}, but in the case in which the gauge 
group is
$SO(N_c)$.
The best places where the field theory is described are, for the pure SQCD
case in \cite{Intriligator:1995id} and the paper \cite{Carlino:2001ya}
for the theory that is more like
the one we are interested in. For a 
recent analysis of a field theory related to ours in non-critical strings 
set-up see \cite{Armoni:2008gg}. 

The $N=1$ theory with orthogonal group has a rich phase structure, with fixed infrared conformal points for ${3\over 2} (N_c-2)<N_f < 3(N_c-2)$ and a free magnetic dual description for $N_c-2<N_f \leq {3\over 2} (N_c-2)$. For $N_f < N_c-2$ there is a dynamically generated superpotential of the ADS type, associated to gaugino condensation. For $N_f=N_c-4$ and $N_f=N_c-3$ there is also an extra branch with no superpotential. In the field theories described by the holographic duals 
we study, this picture is modified by the presence of a classical superpotential coming from the reduction from the six dimensional theory.

The dual geometry is based on the same 5-brane construction wrapping a 2-cycle of the resolved 
conifold. In order to have an orthogonal gauge group, we introduce an orientifold 
O5 plane parallel to the 5-branes. This setup was studied in 
\cite{Gomis:2001xw}; the case of symplectic 
groups was also addressed there. The orientifold has the effect of making a 
reflection on the orthogonal directions, 
so the $S^3$ surrounding the 5-branes becomes a projective space $S^3\to RP^3$. 
When we go from the brane picture to the geometric picture, this is shown in the periodicity of the coordinate $\psi\sim \psi+ 2 \pi$. The gauge group and the 
three form flux in the geometry depend on the type of orientifold. We can classify 
them according to their five-brane charge and discrete torsions for the NS three-form $H_3$ 
and the RR one-form $F_1$ 
\cite{Witten:1998xy,Hanany:2000fq}. If the number of D-branes is $2 n$, 
then we have,
$$
\begin{array}{|c|c|c|c|c|}
\hline {\rm orientifold} & {\rm charge} & (H_3,F_1) & {\rm group} & F_3 \\
\hline O5^- & -1 & (0,0) & SO(2 n) & 2 n -2\\
\hline O5^+ & +1 & (\half ,0)& Sp(2 n) & 2n+2  \\
\hline \widetilde{O5}^- & -\half & (0,\half) & SO(2 n+1) & 2n - 1\\
\hline \widetilde{O5}^+ & +1 & (\half,\half) & Sp(2 n) & 2n+2 \\ 
\hline 
\end{array}
$$
The geometry of the orientifolded construction is essentially the same, 
with even the same Killing spinors. The only differences are the change of topology of a 
three-cycle of the internal space and the amount of $F_3$ form flux. 

In the following and to avoid a cluttered notation, 
we will concentrate on type {\bf A}
backgrounds. It will be clear that the important features 
do extend also to type {\bf N} configurations, in virtue of our general 
treatment of 
Section \ref{generalbackground}.  The BPS equations 
for the $SO(N_c)$ case can be read directly 
from the $SU(N_c)$ case from eqs. (\ref{newbps1})-(\ref{newbps4}) by 
replacing
\be
N_c\to N_c-2\,.
\ee
So they read
\bea
\notag H'&=& \frac{1}{2} (N_c -2-N_f)  + 2 Y,
\label{newbps1so} \\
\notag G' &=& -\frac{(N_c -2)}{2} + 2Y,
\label{newbps2so} \\
\notag Y' &=&-\frac{1}{2}(N_f- N_c +2) \frac{Y}{H} -\frac{(N_c -2)}{2}
\frac{Y}{G}
-2Y^2\left(\frac{1}{H}+\frac{1}{G}\right) +4Y,
\label{newbps3so}  \\
\phi'&=&-\frac{ (N_c-2 -N_f)}{4H}   + \frac{(N_c -2)}{4G}.
\label{newbps4so}
\eea
At this stage, we observe a couple of immediate things about Seiberg
duality and R-symmetry anomalies. 
The BPS system (\ref{newbps4so}) is {\it invariant} if we
change in all the eqs.,
\beq
N_c \to N_f -N_c+4,\;\;\; H\to G,\;\;\; G\to H,
\label{zzzzz}
\eeq
leaving all the other functions and $N_f$ untouched. In 
our eq.~(\ref{seibergr}) and in the paper \cite{Casero:2007jj} a similar 
change 
to 
eq.~(\ref{zzzzz}) was identified as Seiberg duality 
(for the group $SU(N_c)$). In the case of orthogonal gauge groups, the 
change is as in eq.~(\ref{zzzzz}) and coincides with what is known 
from \cite{Intriligator:1995id}.

Using the RR three form and restricting it to 
a topologically trivial two cycle (on which $F_3=dC_2$), 
we can compute the anomaly of the R-symmetry, that is initially 
identified with changes in the angle $\psi$. 
Indeed, one can see that as previously 
explained in \cite{Casero:2007jj} the partition function 
for a Euclidean D1 that wraps the topologically trivial cycle 
(\ref{cycle})  is given by $
Z=Z_{BI} e^{\frac{i}{2\pi}\int C_2}
$,
where $Z_{BI}$ is the Born-Infeld part of the partition 
function. This implies that the partition function is invariant if,
$ \psi\to \psi+ \frac{2k\pi}{2N_c-4 -N_f}
$,
hence showing that the R-symmetry is broken 
to $Z_{2N_c-N_f-4}$. The extension to type {\bf N} backgrounds is 
immediate. In this case, spontaneous breaking of R-symmetry imply that there are $2N_c-N_f-4$ vacua.

The new and interesting thing is that, although the geometry is pretty 
much the same, the change in topology in 
the duals to theories with orthogonal groups allows new kind of objects 
corresponding to branes wrapping cycles of the internal space. We follow 
the discussion in \cite{Witten:1998xy}, adapting it to our case. The 
internal space has topology $RP^3\times S^2$, and due to the O5 
orientifolding in the brane construction, D1 and D5 branes can be wrapped 
in $RP^3$ according to the untwisted homology, while F1 strings, NS5 and 
D3 
branes should be wrapped according to the twisted one. The non-trivial 
homology groups are $H_0(RP^3,Z)=H_3(RP^3,Z)=Z$, $H_1(RP^3,Z)=Z_2$, while 
the twisted ones are $H_0(RP^3,\tilde Z)=H_2(RP^3,\tilde Z)=Z_2$. There 
could be further constraints on the kind of wrapping that is allowed due 
to the discrete torsions of the different types of orientifold. Let us 
list the new possibilities of wrapping cycles inside the three-fold. 
\begin{itemize}
\item[a)] A string on $RP^2 \subset RP^3$: this gives a 
point-like object localized in space and time. 
Two objects of this kind can annihilate each other, 
since they are classified by the group $H_2(RP^3,\tilde Z)=Z_2$.
\item[b)] A D1 brane on $RP^1\subset RP^3$: this gives a particle in the non-compact space. This particle should be able to annihilate with another of the same kind, since it is classified by a $Z_2$ group.
\item[c)] A D3 brane on $RP^2\subset  RP^3$: a string-like object in the non-compact space, also classified by $Z_2$.
\item[d)] A D3 brane on $RP^2\times S^2$: a point-like object, classified by $Z_2$. By $S^2$ we mean any 2-cycle, it could be contractible.
\item[e)] A D5 brane on $RP^1\subset RP^3$: an object with four spatial dimensions, classified by $Z_2$.
\item[f)] A D5 brane on $RP^1\times S^2$: a domain wall-like object, 
classified by $Z_2$.
\item[g)] An NS5 brane on $RP^2\subset RP^3$: an object with three spatial dimensions, classified by $Z_2$.
\item[h)] An NS5 brane on $RP^2\times S^2$: a string-like object, classified by $Z_2$.
\end{itemize}
The most interesting objects are b), c)/h) and f), that have some of the properties to be the duals of the gauge theory Pfaffian, spinor Wilson loop and domain walls.

Discrete fluxes can impose a constraint on the allowed branes from the following considerations: imagine we have a brane wrapping a cycle in the internal space that contains a subcycle with discrete flux on it. In our case the possibilities are to have a $RP^2$ or $RP^1$ subcycle. We could consider a brane or string wrapping this subcycle, that due to the discrete flux, will be seen as an instantonic contribution to the wavefunction of the larger brane, changing its sign. For instance, wrapping a string on a $RP^2$ subcycle will give a factor
$$
e^{i\int_{RP^2} B_2} = -1\,.
$$
This would rule out the wrapping brane as a good state of the theory unless it can be compensated by a twisted bundle of the gauge field on the brane,
$$
e^{i\int_{RP^2}\widetilde  F_2}=-1\,.
$$
However, the possible twisted bundles of the gauge field are determined by the twisted cohomology of the cycle the brane is wrapping, and not by the twisted cohomology of the complete internal space. They do not need to be the same, so it could happen that it is not possible to construct a twisted gauge bundle on the subcycle. In this case, the change of sign of the wavefunction cannot be compensated and the wrapping brane is ruled out.

In the case at hand, the discrete fluxes do not impose constraints on the branes, since the twisted cohomologies of $RP^2$ and $RP^1$ are non-zero. This makes it more difficult to identify the right objects, compared to the $AdS_5\times RP^5$ case, where the topological restrictions coming from the discrete torsions distinguish objects that exist only in orthogonal $SO(2n)$ or $SO(2n+1)$ theories.

\subsection{Spinor Wilson loop}
In orthogonal theories it is possible to introduce external 
sources in the spinor representation. These sources are not 
screened by the matter content of the theory, so 
the dual object should be able to extend to infinite length. 
Spinor charges do not exist in unitary theories, so the 
spinor Wilson loop should correspond to a 
D3 on $RP^2$. 
We will see that this ``no screening criterion'' allows to confirm that 
this D3 brane is the right object.

We then study our possible candidates for a spinor Wilson loop represented 
as a 
D3 brane wrapping $RP^2$. A possible $RP^2$ could be the cycle 
\beq
\begin{array}{lll}
\theta=\tilde\theta, & \varphi=\tilde \varphi=2\pi-\psi/2 & {\rm if}\ 0\leq\theta<\pi/2,\\
\theta=\tilde\theta, & \varphi=\tilde \varphi=\psi/2 & {\rm if}\  \pi/2\leq \theta<\pi.
\end{array}
\eeq
We are taking 
here $\psi$ to be defined over $[0,4\pi)$ but with 
the identification $\psi \sim 4\pi-\psi$. 
Notice that the cycle is non-orientable. 

For the clarity of this exercise, we will restrict ourselves to the 
case in which our branes probe a type {\bf A} background. In this 
case, the induced metric will be 
\begin{equation}
ds^2_{D3} = e^{\phi}\left(-dt^2 +(1+4 Y \rho'^2)d x^2+(H+G)d\Omega_2^2 + 4 Y(1\mp \cos\theta)^2 d\varphi^2 \right)\,.
\end{equation}
The effective tension is then
\begin{equation}
T_{D3}(\rho) =4\pi T_3  e^{\phi} \sqrt{H+G}\int_0^{\pi/2}d\theta  \sqrt{(H+G) \sin^2\theta+4 Y (1-\cos\theta)^2}\,.
\end{equation}
The integration can be done analytically, the final result is 
\begin{equation}
T_{D3}(\rho) =  4\pi T_3  e^{\phi} (H+G)\left(2-\sqrt{1+y} +  {2 y\over \sqrt{1-y}}\log\left[ \sqrt{2}{1+
\sqrt{1- y}\over \sqrt{1+y}+\sqrt{1-y}}\right] \right). 
\end{equation}
Where we have defined $y\equiv 4 Y/(H+G)$.
We compute numerically the value 
of the tension for this particular D3, using the expansions given in 
\cite{Casero:2007jj}. 
It turns out that the tension of the D3 brane always has a minimum at 
a finite value of $\rho$.
As we have explained in Section~\ref{sec:break}, this means that 
the D3 brane is not screened in any 
background, so it is a good candidate for a spinor Wilson loop. 

The fact that 
this object exists in the backgrounds dual to theories with orthogonal 
gauge group and does not exist for unitary groups is a very stringent 
dynamical check of 
our proposal and of the whole set-up.

\begin{figure}[htb!]
\centering
{\tiny 
\begin{tabular}{ccc}
\includegraphics[scale=0.65]{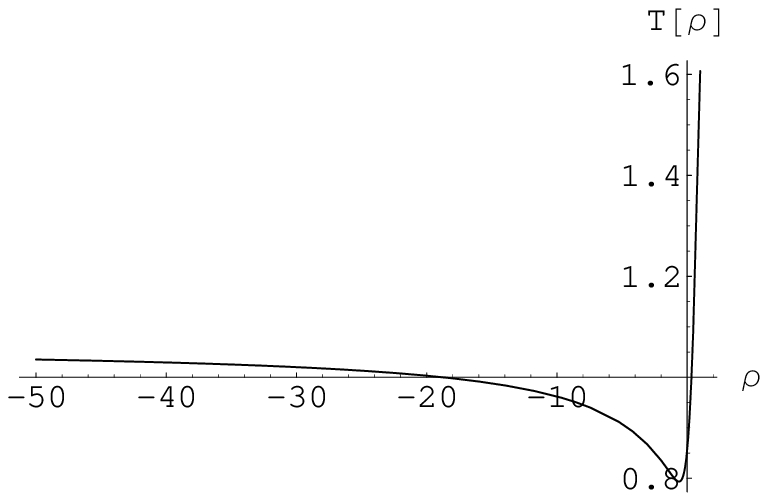} &  \includegraphics[scale=0.65]{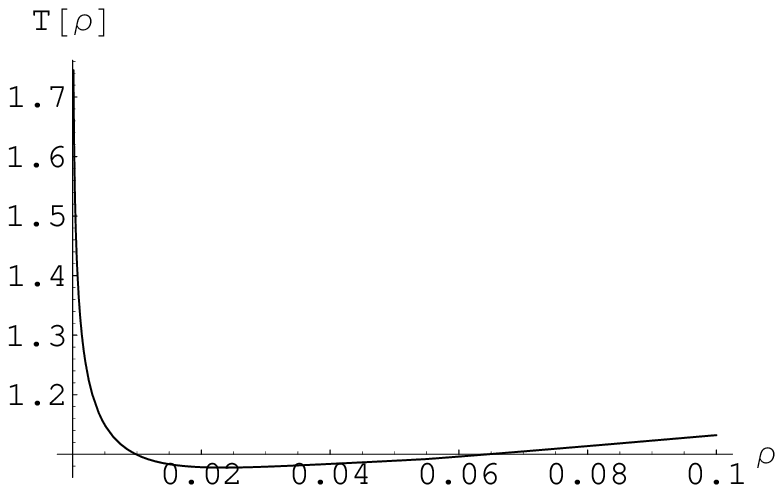} & \includegraphics[scale=0.65]{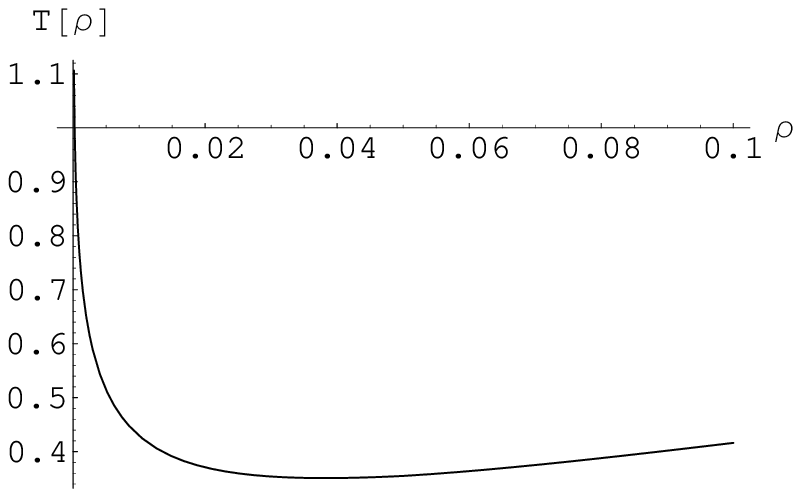} \\
type I D3 tension &  type II D3 tension & type III D3 tension 
\end{tabular}}
\caption{\small The tension of the D3 brane wrapping a $RP^2$ cycle has a minimum as a function of $\rho$ for the three types of \bA backgrounds, so the holographic dual object is not screened in any of them.}
\end{figure}

\section{General comments, criticism and conclusions}\label{gcc}
\setcounter{equation}{0}
In this Section we would like to present the outcome of many discussions 
we have had with different colleagues who asked challenging questions 
regarding various aspects of this line of research.

The first comment concerns the use of flavor branes represented by the 
Born-Infeld-Wess-Zumino Action in contrast with the color branes, encoded 
in the type IIB supergravity fields, in other words, we seem to treat 
differently the $N_c$ ``color branes'' and the $N_f$ flavor branes. The 
reason for this asymmetric treatment is that the two types of symmetries 
(flavor and gauge) 
are actually different. As is well known, a  flavor symmetry is a symmetry 
of the physical spectrum (a `real' symmetry), that can be broken; while 
the previous two properties do not apply to a gauge symmetry, that is just
a way of indicating a redundant description.
So, the main point is that an arbitrary flavor symmetry group $SU(N_f)$ 
needs extra degrees of freedom to be added to the type IIB Action, this is 
what the Born-Infeld-Wess-Zumino Action is doing. This is the proposal of 
\cite{Karch:2002sh}.

As a consequence, a fluctuation of the fields in the action of 
eq.~(\ref{accion})
would imply a fluctuation of the gauge field on the brane, that is dual to 
a meson with flavor quantum numbers, showing the existence 
of meson-glueball 
interactions.  

\begin{figure}[htb!]
\centering
\includegraphics[scale=0.5]{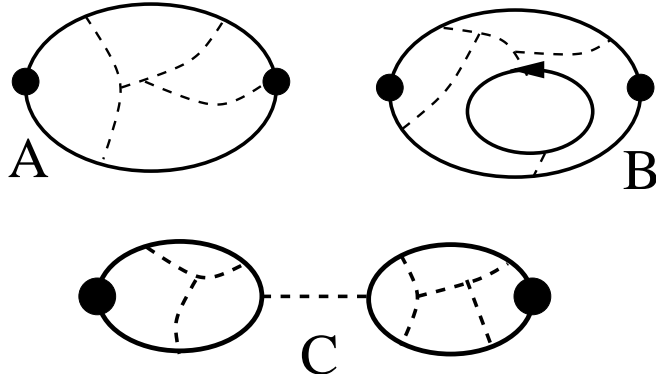} 
\caption{\small Diagram for a meson propagator, with two 
insertions of the meson operator ($n=2$) shown as thick 
points on the boundaries. The dashed lines are gluons 
that fill the diagram in the large $N_c$ limit and the 
thick lines are quarks. A) Planar diagram with no internal quark loops 
($h=0$, $w=0$), the scaling is $\sim 1$. 
B) Planar diagram with an internal quark loop $h=0$, $w=1$, 
the scaling is $\sim N_f/N_c$. 
C) Non-planar diagram with no internal quark loops $h=0$, $w=0$, $b=2$, 
the scaling is $\sim 1/N_c $.}
\label{diagrams}
\end{figure}

The reader may wonder what are the limitations of the line of research 
represented by \cite{Karch:2002sh} and papers that followed. Since the 
$N_f$ `flavor branes' are 
not deforming the spacetime (they are very few with respect to the color branes 
$\frac{N_f}{N_c}\sim 0$)
it is clear that some physical effect will not be reproduced 
by the String configuration. This can be seen  diagrammatically in the 
dual field theory.
Indeed, consider for example the scattering of two mesons in a theory with 
$N_f$ flavors and $N_c$ colors. A formula for the kinematical factor was 
produced in \cite{Capella:1992yb}  for the scattering of $n$ mesons, 
considering diagrams with $w$ internal fermion loops (windows), $h$ 
non-planar handles and $b$ boundaries,
\beq
<B_1....B_n> \sim \left(\frac{N_f}{N_c}\right)^w  N_c^{(2-\frac{n}{2} -2h -b)}.
\label{scatk}
\eeq
Consider the case of scattering of two mesons $n=2$. We see that diagrams 
like the first one in Figure \ref{diagrams} ($w=h=0, b=1,n=2$) scales like a constant 
$N_c^0\sim 1$, the 
second diagram (with $w=1,h=0,b=1,n=2$) scales like $\frac{N_f}{N_c}$, 
while the third one (with $w=0, h=0,b=2,n=2$, that is non-planar) goes 
like $N_c^{-1}$.

There are two possible scalings that can be taken to make contact with 
String theory:
\begin{itemize}
\item {The one introduced by 't Hooft \cite{'t 
Hooft:1973jz} where
$\lambda =g^2 N_c$ is kept fixed while 
\beq
g\to 0,\; N_c\to \infty,\; 
N_f=fixed.
\label{thscaling}
\eeq
In 't Hooft's scaling (\ref{thscaling}) the first diagram of the figure is 
dominant, the second one is suppressed like $N_c^{-1}$ 
due to the quark loop and the third one, 
being non-planar is also suppressed like $N_c^{-1}$.
We observe  a suppression of both 
planar (with quark loops) and non-planar diagrams. It may be desirable to 
first sum up all planar diagrams and then 
consider non-planar ones as correction, this is achieved by the so called 
Veneziano expansion.}
\item{Veneziano \cite{Veneziano:1976wm}
introduced another expansion ( called the topological expansion ) where 
the scaling is 
\beq
g\to 
0,\; N_c\to \infty,\;
N_f\to \infty,\; \frac{N_f}{N_c}=fixed.
\label{vscaling}
\eeq
Obviously here we also impose 
that $\lambda=g^2 N_c$ is fixed. In this scaling, the first two diagrams 
in the figure contribute at the same order and the one violating the OZI 
rule is suppressed.} 
\end{itemize}
The connection between these expansions and the way of adding flavors 
described above goes as follows. When we add probe branes to the 
background (hence not considering its backreaction), we are working in 
the 
't Hooft scaling (\ref{thscaling}). The discussion 
above shows that some processes, those 
in which diagrams scaling like $\frac{N_f}{N_c}$ are of importance (like 
screening, for example), will 
not be well captured by the `probe approximation'.

We actually would like to impose the Veneziano scaling (\ref{vscaling}), 
that considers diagrams of the first 
two types, and hence  
is more likely to capture more physics. The way of doing 
this is by `backreacting with the flavor branes'. That is to say, consider 
solutions to the eqs. of motion derived from the action (\ref{accion}).

Let us now comment  on the effects of the smearing of flavor branes. As we 
hinted along the 
paper and more concretely in Section \ref{enhanced}, the smearing seems to 
break the $SU(N_f)$ flavor symmetry of the localized 
brane system to $U(1)^{N_f}$. This will obviously have influence on the 
dynamics, but also many aspects will remain unaffected (like global 
anomalies and their matching in dual descriptions, beta functions, 
Seiberg 
duality, etc). We would like 
to comment a bit more on the following related fact: if $N_f$ is 
bigger than $N_c$ 
and we take the numbers such that we do not loose asymptotic freedom, it 
may be expected that there will be diagrams correcting the Born-Infeld 
action, 
that will be weighted by $g_s N_f\sim \frac{N_f}{N_c}>1$. So, in 
principle, one should not trust the Born-Infeld-Wess-Zumino action, unless 
there is a way of 
understanding that such corrections will not change the tree-level form
of the action. Here is where the smearing comes to help. The strings 
between flavor branes will typically be massive as we showed in 
Section \ref{enhanced}. Even more, in the 
far 
UV, the separation 
between flavor branes is large (because the `internal four manifold 
$\Sigma_4$' is 
very large). In the IR (typically) something similar happens, with the 
`internal 
manifold' having a fixed size and with low energies to excite the 
non-diagonal strings, see Section \ref{enhanced} for details. We believe 
that this makes the above mentioned 
corrections to be very suppressed or non-existent (a very similar 
argument was presented in \cite{Bigazzi:2008zt}). As an aside, the fact 
that the 
system 
is SUSY suggests that even in the case of localized (in contrast to 
smeared) flavor branes the diagrams mediated by the non-diagonal strings 
will cancel each other (or the system would lose stability and be 
non-SUSY). The complementary possibility that would exclude what we are 
arguing 
here is that there is no {\it weakly coupled } String dual to a 
field theory with $N_f \sim N_c$.

A second point related to the smearing is: can we consider the smearing as 
an approximation
when we wrote above, that for example in the UV the distance between 
flavor branes grows unbounded? In which sense can we consider a `continuous 
distribution' of flavor branes? Perhaps, in this sense, the smearing 
should be considered as the s-wave, in a multipole expansion of the true 
solution. Nevertheless, if we insist on the superpotential of the field 
theory with flavors to have the 
same symmetries that the unflavored theory has (see Section 5.1 of 
\cite{Casero:2006pt}) the smearing will be necessary. In the case in which 
the superpotential breaks the isometries of the original unflavored 
theory, the dual string solutions should depend-aside from the radial 
coordinate-on angles, hence partial differential eqs. will be describing 
the dynamics, making the problem more complicated. 

Another point on which we would like to comment is related to the T-dual 
version of our set-up. As is well known (see for example \cite{Giveon:1998sr}),
the IIA version of our type of field theories is represented by an array 
of NS5, D6 and D4 branes. The positions of the D6 branes are arbitrary, 
one can have them coincident or separate them (at the only cost of 
breaking the $SU(N_f)$ symmetry). In our type IIB construction we 
perform the 
smearing discussed above to find a simple solution. This seems to 
point to a particular distribution of six branes in the T-dual picture. 
Nevertheless, it may be possible to find solutions where non-uniform 
configurations of flavor branes are smeared (an example in a 
different model is in the paper \cite{Bigazzi:2008zt}). The different 
smearings 
should correspond to different distributions of $D6 $ branes in the T-dual 
version. The influence (or not) of the different smearings over the IR of 
the solution should teach us about features of the field theory and the 
stringy completion.

Finally an issue that is occasionally raised is the possible existence of 
tadpoles. This question mostly arises when the flavor branes are D7's as 
it happens 
in some of the models discussed in \cite{varios}. One must find a (closed) 
cycle  inside $\Sigma_p$-where the flavor branes are pointlike objects-to 
compute the integral $\int F_{RR}$ and this should be zero to avoid 
tadpoles. One can see that such manifold does not exist or that integral 
is zero in the case of D5 branes considered in this paper or 
in the models using D7's flavor branes on the conifold.
\subsection{Conclusions}
In this paper we have presented a unified way of working with different 
type IIB String backgrounds conjectured to be dual to a version of $N=1$ 
SQCD. We have presented various new solutions, including a new exact solution.
We also studied different field theory 
aspects of the new solutions, providing among other new things a 
systematic criterion for string breaking (screening) and a new 
understanding of UV properties in the case in which the theory is deformed 
by an irrelevant operator. We also extended the treatment to the 
case of $SO(N_c)$ gauge groups, we checked that many immediate properties 
are satisfied. Interestingly enough, we proposed a candidate object to 
compute 
the Wilson loop in the spinorial representation. This object, shows 
confinement as predicted on general grounds.

It should be nice to extend our studies in the following directions:
\begin{itemize}
\item{To extend our analysis in Section \ref{generalbackground} to the 
case of massive 
flavors and to construct duals to SQCD using  type IIA backgrounds 
\cite{Atiyah:2000zz}.}
\item{To explore new gauge theory aspects of the new solutions presented 
in Section \ref{newsolutions} and the new String-inspired objects 
mentioned in Section \ref{sons}}.
\item{To find extensions of the set-up presented in this paper to study 
different versions of 
SQCD, with different superpotentials, in different vacua, etc.}
\item{To apply this line of research to possible strongly coupled field 
theories in Physics beyond the Standard Model.}
\end{itemize}
We leave these for future work.

\section{Acknowledgments} We benefited from discussions 
with various colleagues, whom we gratefully thank. A partial list 
includes: Adi Armoni, Ofer Aharony, Francesco Bigazzi, Aldo Cotrone,  Alex 
Buchel, Amit Giveon, David Kutasov, Prem Kumar,  Asad Naqvi, Alfonso 
Ramallo, Angel Paredes, Diana Vaman and Pedro Silva.

\appendix
\renewcommand{\thesection}{\Alph{section}}
\renewcommand{\theequation}{\Alph{section}.\arabic{equation}}

\section{Appendix: Some aspects of the QFT}
\label{appendixsup}
\setcounter{equation}{0}

In this appendix we comment on some aspects of the field theory 
described in Section \ref{qft}.

Let us recall that the field theory contains an $N=1$ massless vector 
multiplet $\W_\alpha=(\lambda, A_\mu)$ plus a tower of massive KK chiral 
fields 
denoted by $\Phi_k=(\phi_k,\psi_k)$
and massive vector multiplets $W_k$.
\footnote{A massive vector superfield has the degrees of freedom of a 
vector superfield together with the ones of a chiral multiplet. That is,
\bea
& & V^0= \frac{1}{2}A^0 +\sqrt{2}[\theta \bar{\psi} +\bar{\theta}\psi] 
-\theta\sigma^\mu\bar{\theta} v_\mu +\theta^2 F+\bar{\theta}^2 F^*+ 
\nonumber\\
& & 
i\theta^2\bar{\theta}[\bar{\lambda} +\frac{1}{\sqrt{2}} 
\sigma^\mu \partial_\mu \psi]- i\bar{\theta}^2\theta [{\lambda} 
+\frac{1}{\sqrt{2}}
\sigma^\mu\partial_\mu \bar{\psi}] +\theta^2\bar{\theta}^2 (D+\nabla^2 
A^0).
\eea
From here we can construct, as usual, the field strength $W_k$.}

It is useful to give an assignment of R-charges to the fields in a 
massive vector  multiplet,
\beq
R[V^0]=R[A]=R[v_\mu]=R[D]=0,\;\;R[\lambda]=-R[\psi]=1,\;\;\; 
R[F]=-R[F^*]=-2.
\label{rchargesmassive}\eeq

The  reader should wonder 
how is it possible that a theory like (\ref{unflav}),
so different 
to N=1 SYM, 
can reproduce so many non-perturbative results of SYM (obtained using the 
dual description, that is the
string background 
in \cite{Maldacena:2000yy}). Indeed, it was shown in 
\cite{Gursoy:2005cn}, by using the string dual description of the field 
theory,
that many of the observables are such that the contribution
of the KK modes is zero. 

The question can be posed like this: is there  a QFT way of 
understanding why the 
KK modes do not contribute to typical quantities like the beta function, 
the R symmetry 
anomaly, etc?

For anomalies, one can take the simple view that only massless fields will 
contribute to them. In this case, the only massless fermion is a Majorana 
spinor, that is the gaugino of N=1 SYM. So, this would produce the same 
anomaly of $U(1)_R\to Z_{2N_c}$. But if we compute things at very high 
energies, many of the KK modes can be considered massless, so, they 
should indeed count towards  the anomaly computation. In other words, 
since the 
anomaly coefficient is invariant under RG flows, we may wonder how it 
happens that all the KK fermions do not contribute. 

This can be understood by analyzing the R-charge assignment of the KK 
fields. We assign R-charge $R[\Phi_k]=1$ to the massive chiral multiplets, 
then the fermions in the multiplets
will have $R[\psi_{k}]=0$, so, they will not contribute to any triangle 
involving the R-charge. For the massive vector multiplets, we have that 
the two fermions inside them have R-charges $R[\lambda_k]=-R[\psi_k]=1$, 
then the multiplet cancels within itself. This may be the reason why we 
get the `correct' R-symmetry pattern (that is the one of N=1 SYM).

Notice that with the previous R-charge assignment, it is clear that if 
the superpotential (like the one proposed in Section \ref{qft}) has 
interactions of three 
superfields, or higher order interactions with the massive vectors, the 
R-symmetry is explicitly broken, because the coupling constants are 
charged. This could be thought as spontaneous symmetry breaking. But we 
should still have an `effective Lagrangian', that is what the 
supergravity solution provides, that matches the anomalies.

Now, let us focus on beta functions. The question here is how is it 
possible that the papers \cite{DiVecchia:2002ks}
 get the correct NSVZ beta function. One 
possibility is that all the 
massive fields have in the far UV (when they could be considered massless) 
anomalous dimensions such that they do not contribute to the beta 
function.
Indeed, the NSVZ result for the Wilsonian beta function,
\beq
\beta_g=-\Big[3N_c +N_{k}(1-\gamma_k)  \Big],
\eeq
shows that if the anomalous dimensions of each massive field behaves in 
the UV as
\beq
\gamma_k\sim 1+ O(m_k/E),
\eeq
then, the field will not contribute to the anomalous dimension. 

Let us now consider the theory after the quark $Q$ and anti-quark 
superfields $\tilde{Q}$
have been added to the theory.
The superpotential is then,
\beq
{\cal W}=  \sum_{ijk, abc}z_{ijk}^{abc} \Phi_i^{ab} \Phi_j^{bc} 
\Phi_k^{ca} + 
\sum_k \hat{f}(\Phi_k)W_{k,\alpha} W_{k}^\alpha +\sum_k \mu_k 
|\Phi_k^{ab}|^2 
+\sum_{r,ab}\kappa_{ij,(r)}\tilde{Q}_{i}^{a} \Phi_{r}^{ab} Q_{j}^{b},
\eeq
and  the F-term eqs. for the massive KK modes will be,
\beq
2\mu_p \Phi_p^{ab} +\Big[3!\sum_{c,jk} z_{pjk}^{abc} 
\Phi^{ac}_j\Phi^{cb}_k 
+\frac{\partial \hat{f}(\Phi)}{\partial \Phi_{p}^{ab} }W_kW_k + \sum_{ab}
\kappa_{ij,(p)}\bar{Q}^{a,i}Q^{b,j}
   \Big] =0,
\eeq
and for the massive vector,
\beq
\partial_{W_\alpha}\W =\hat{f} W_k=0.
\eeq
So, we impose that there is a solution of the form
\beq
W_k=0,\;\;\;\Phi_p^{ab}=-\frac{\kappa_{(p),ij}}{2\mu_p} 
\tilde{Q}^{a}_i Q^{b}_j =\frac{\kappa_{p,ij}}{\mu_p} 
M^{ij}\delta^{a,b}=-\frac{\kappa_p}{2\mu_p}M\delta^{ab},
\label{solutionss}
\eeq
where we used that the constants $z_{ijk}^{abc}$ are antisymmetric in both 
sets of 
indexes, and the $W_k=0$ in the vacuum. 
When replacing this into the superpotential, we find an 
effective 
superpotential that is quartic in the quarks or quadratic in the meson 
fields 
\beq
W_{eff}\sim -\sum_p \frac{\kappa_{(p),ij}^2}{2\mu_{p}^2}(\bar{Q_i}Q_j)^2
\sim \frac{\kappa^2}{2\mu}M^2.
\label{weff2}
\eeq
In Section \ref{qft}, we used that all the $\kappa_{p}$ are equal, because 
we made no 
distinction between quarks due to the uniform smearing (likely, there is 
a phase in the coupling for each quark that we 
are neglecting here!). In the coefficient $\mu$ we are counting the sum of 
all the masses 
that have been integrated out, with its respective degeneracy.

\section{Appendix: Expansions in integration constants}
\label{moduli-expansions}
\setcounter{equation}{0}

\subsection{Large $c_+$ expansion}

In this appendix we construct systematically the $c_+$ expansions of both type \bA and type \bN solutions
and show that the former asymptote to the conifold, while the latter to the deformed conifold. As we will see,
these expansions resum certain terms in the class II UV expansions (\ref{UV-II-A})-(\ref{UV-II-N}).

Let us start by observing that eq.~(\ref{master}) for $P$ can be written
in the form 
\be\label{master-c+}
\pa_\r\left(s(P^2-Q^2)(P'+N_f)\right)+4s(P'+N_f)(N_fP+QQ')=0,
\ee 
where
\be
s(\r)=\left\{\begin{matrix} e^{-4\r}, & \r_o\to-\infty, \\
							\sinh^2\t, & \r_o>-\infty. \end{matrix}\right.
\ee
Integrating (\ref{master-c+}) twice we obtain
\bea\label{master-c+-integrated}
&&P^3-3Q^2P+3\int d\r\left(2QQ'P+N_f(P^2-Q^2)\right)\NO\\
&&+12\int d\r s^{-1}\int d\r s (P'+N_f)(N_fP+QQ')=c_-+4c_+^3\int d\r s^{-1},
\eea
where $c_\pm$ are arbitrary integration constants. Now 
\be
4\int d\r s^{-1}=\left\{\begin{matrix} e^{4\r}, & \r_o\to-\infty, \\
							\frac12\left(\sinh(4(\r-\r_o))-4(\r-\r_o)\right), & \r_o>-\infty, \end{matrix}\right.
\ee
and so in both cases $\int d\r s^{-1}\sim e^{4\r}$ as $\r\to\infty$. Since $Q$ only goes linearly with $\r$, it
follows from (\ref{master-c+-integrated}) that solving (\ref{master-c+-integrated}) in a large $c_+$ expansion
should precisely reproduce the class II UV expansions (\ref{UV-II-A}) and (\ref{UV-II-N}), but it should 
resum certain terms associated with $s(\r)$ in the type \bN case. Inserting an ansatz of the form
\be
P=\sum_{n=0}^\infty c_+^{1-n} P_{1-n},
\ee
in (\ref{master-c+-integrated}) we obtain
\be
P_1=\left\{\begin{matrix} e^{4\r/3} \\
							\frac{1}{2^{1/3}}\left(\sinh(4(\r-\r_o))-4(\r-\r_o)\right)^{1/3} 
 \end{matrix}\right. =
\left\{\begin{matrix} e^{4\r/3}, & \r_o\to-\infty, \\
							\sinh(2(\r-\r_o))\ck(\r-\r_o), 
& \r_o>-\infty, \end{matrix}\right.
\ee
where $\ck(\r)$ is the function introduced in Appendix C of \cite{Casero:2006pt} and corresponds to the deformed conifold
geometry. Solving recursively for the subleading terms one determines
\bea
P_0&=&-N_f P_1^{-2}\left(\int d\r P_1^2+4\int d\r s^{-1}\int d\r s P_1P_1'\right),\NO\\\NO\\
P_{-1}&=&-\frac13 P_1^{-2}\left(3P_1P_0^2+6N_f\int d\r P_1P_0+12N_f\int d\r s^{-1}\int d\r s(P_1P_0'+P_1'P_0)
\right.\NO\\
&&\left.-3Q^2P_1+6\int d\r QQ' P_1+12\int d\r s^{-1}\int d\r s(QQ'P_1'+N_f^2P_1)\right),\NO\\\NO\\
P_{-2}&=&-\frac13 P_1^{-2}\left(6P_1P_0P_{-1}+P_0^3+3N_f\int d\r (2P_1P_{-1}+P_0^2)\right.\NO\\
&&\left.
+12N_f\int d\r s^{-1}\int d\r s(P_1P_{-1}'+P_{1}'P_{-1}+P_0P_0')
-3Q^2P_0+6\int d\r QQ'P_0\right.\NO\\
&&\left.+12\int d\r s^{-1}\int d\r s (QQ' P_0'+N_f^2P_0)-3N_f\int d\r Q^2
\right.\NO\\
&&\left.+12 N_f \int d\r s^{-1}\int d\r s QQ'+c_-\right),\NO\\\NO\\
P_{-n-2}&=&-\frac13P_1^{-2}\left\{\sum_{m=1}^{n+2}\left(2P_1P_{1-m}P_{m-n-2}+\sum_{k=1}^{n-m+3}P_{1-m}P_{1-k}
P_{m+k-n-2}\right)\right.\NO\\
&&\left.+3N_f\sum_{m=0}^{n+2}\left(\int d\r P_{1-m}P_{m-n-1}+4\int d\r s^{-1}\int d\r s
P_{1-m}P_{m-n-1}'\right)-3Q^2P_{-n}\right.\NO\\
&&\left.+6\int d\r QQ' P_{-n}+12\int d\r s^{-1}\int d\r s(QQ'P_{-n}'+N_f^2P_{-n})\right\},\quad n\geq 1.
\eea
For the type \bA case these recursion relations reproduce the asymptotic expansion (\ref{UV-II-A}) and
so for these backgrounds the large $c_+$ expansion coincides with the class II asymptotic expansion. 
For the type \bN backgrounds, however, the $c_+$ expansion resums certain terms in the asymptotic expansion
(\ref{UV-II-N}). In particular, the expansion in the parenthesis multiplying $c_+$ in (\ref{UV-II-N})
is now identified with the asymptotic expansion of $\frac{1}{2^{1/3}}\left(\sinh(4(\r-\r_o))-4(\r-\r_o)\right)^{1/3}$,
after setting $\r_o=0$ and absorbing an overall constant in $c_+$.

\subsection{Small $c_+$ expansion}

In this appendix we systematically construct the most general solution of (\ref{master}) in 
the vicinity of the singular solution (\ref{bad-solution}), in a small $c_+$ expansion.   
We start by inserting an expansion of the form
\be
P=\sum_{n=0}^{\infty}c_+^{3n}P_n,
\ee
where $P_0$ is given by (\ref{bad-solution}), in (\ref{master-c+}). Solving the resulting equations recursively we find
that, provided
\bea
\Om&\equiv& P_0^2-Q^2\NO\\
&=&\left(\left(P_o-Q_o\cosh\t\right)-\frac{2N_c-N_f}{2}\left(\cosh\t-1\right)
-(N_f+(2N_c-N_f)\cosh\t)\r\right)\times\NO\\
&&\left(\left(P_o+Q_o\cosh\t\right)+\frac{2N_c-N_f}{2}\left(\cosh\t-1\right)
-(N_f-(2N_c-N_f)\cosh\t)\r\right),\phantom{-}
\eea
does not vanish identically, then
\bea
&&P_1=\int d\r s^{-1}\Om,\NO\\
&&P_n=\int d\r  s^{-1}\Om \int d\r \Om^{-2}R_n,\quad n\geq2,
\eea
where
\bea
R_n&=&-\frac13\sum_{m=1}^{n-1}\left\{\pa_\r\left[s\left(\pa_\r\left(2P_0P_mP_{n-m}+\sum_{k=1}^{n-m}P_mP_kP_{n-m-k}\right)
+3N_fP_mP_{n-m}\right)\right]\right.\NO\\
&&\left.\phantom{moremore}\rule{0.in}{0.27in}+12N_fsP_mP_{n-m}'\right\},\quad n\geq 2.
\eea

Evaluating the first few terms for the type \bA case ($\r_o\to\infty$) we find
\bea
P_0&=&-N_f\r+P_o,\NO\\
P_1&=&e^{4\r}\left(N_c(N_f-N_c)\r^2-\frac12(N_c(N_f-N_c)+N_f P_o+(2N_c-N_f)Q_o)\r\right.\NO\\
&&\left.+\frac14(P_o^2-Q_o^2)+\frac18(N_fP_o+(2N_c-N_f)Q_o+N_c(N_f-N_c))\right),\NO\\
P_2&=&\ldots,
\eea
which coincides with the type I IR expansion of \cite{Casero:2007jj} (cf. eqs. (3.7)-(3.9)). In this case 
the IR is located at $\r\to-\infty$. 

For the  type \bN case, we note that since $P_0$ contains the arbitrary constant $P_o$, we can always extend
this solution in the IR up to the smallest possible value of the radial coordinate, i.e. $\r_o$, by a suitable
shift in the constant $P_o$. However, $P$ remains finite at $\r_o$, while from (\ref{Q}) we see that 
$Q$ has a pole at $\r_o$, unless $Q_o+(2N_c-N_f)/2=0$. In order to get a well defined solution then, 
where both $H=(P+Q)/4$ and $G=(P-Q)/4$ remain positive, we must ensure that $Q$ has no pole at $\r_o$ and so we must 
set $Q_o=-(2N_c-N_f)/2$. Taking, without loss of generality, then $\r_o=0$, we obtain 
\bea\label{small-c+-expansion-N}
P_0&=&-N_f\r+P_o,\NO\\
P_1&=&\frac{1}{48}\left\{8\left((2N_c-N_f)^2+N_f^2\right)\r^3+24N_fP_o\r^2+6\left((2N_c-N_f)^2-4P_o^2\right)\r-3N_fP_o\right.\NO\\
&&\left.-\frac32\left[16N_c(N_c-N_f)\r^2+8N_fP_o\r+5(2N_c-N_f)^2-3N_f^2-4P_o^2\right]\sinh(4\r)\right.\NO\\
&&\left.+3\left[\left(3(2N_c-N_f)^2-N_f^2\right)\r+N_fP_o\right]\cosh(4\r)\right\},\NO\\
P_2&=&\ldots.
\eea
where we have chosen the integration constant in $P_1$ such that $P_1\to 0$ as $\r\to 0$.
Expanding this around $\r=0$ then one obtains
\bea\label{type-I-N}
P_0&=&-N_f\r+P_o,\NO\\
P_1&=&\frac43P_o^2\r^3-2N_fP_o\r^4+\frac{4}{5}\left(\frac{4}{3}P_o^2+N_f^2\right)\r^5+\co\left(\r^6\right),\NO\\
P_2&=&\ldots,
\eea
Identifying then $2c_+P_o^2=c_2N_c$ and $c_1=4(2N_c-N_f)/3P_o$ the expansion 
(\ref{type-I-N}) exactly reproduces the expansion (4.21) in \cite{Casero:2006pt}. The expansion (\ref{small-c+-expansion-N}),
however, resumming certain terms of this IR expansion, extends the solution to a wider range of the radial coordinate.

Finally, note that in the type \bA case ($\r_o\to-\infty$) it is possible that $\Om=0$ identically.  This happens when 
either $N_f=N_c$ and $P_o=-Q_o$ or $N_c=0$ and $P_o=Q_o$. The solution is then obtained recursively by 
setting $R_n=0$, $n\geq 2$, which gives
\bea
P_0&=&-N_f\r+P_o,\NO\\
P_1&=&e^{2\r}\sqrt{-N_f\r+P_o+N_f/4},\NO\\
P_2&=&-\frac{1}{2P_1}\int d\r e^{4\r}P_0\int d\r P_0^{-2}\pa_\r\left(e^{-4\r}P_1\pa_\r P_1^2\right),\NO\\
P_3&=&\ldots.
\eea
$P_2$ can be expressed in terms of the error function, but the integrals involved in $P_3$ and
higher orders cannot be done in closed form in general. This solution is valid for $\r>-\infty$ 
and it is a new type of IR behavior for the type \bA backgrounds.

\end{document}